\documentclass[11pt,a4paper]{article}
\pdfoutput=1
\usepackage{jheppub}
\usepackage[utf8]{inputenc}
\usepackage{enumitem}
\usepackage{bm}
\usepackage{bbm}
\usepackage{xcolor}
\usepackage{scalerel}
\usepackage{enumitem}  
\usepackage{dsfont}
\allowdisplaybreaks

\usepackage{tikz-feynman}
\tikzfeynmanset{compat=1.1.0}
\tikzfeynmanset{
graviton/.style={circle, draw=green!60, fill=green!5, very thick, minimum size=7mm}
}
\tikzfeynmanset{
codot/.style={/tikz/shape=circle,
/tikz/fill=red,/tikz/minimum size=0.1cm,/tikz/inner sep=1.8pt}
}
\tikzfeynmanset{sb/.style=
{/tikz/shape=circle,
/tikz/fill=black,
 /tikz/minimum size=0.3cm,/tikz/inner sep=1.pt
 } }
\tikzfeynmanset{myblob/.style=
{/tikz/shape=rectangle,
/tikz/fill=red,
 /tikz/minimum size=0.2cm,
 } }
\tikzset{box/.pic={\filldraw[fill=black]  (0,0) circle (2.5pt);
				   \filldraw [fill=black] (0.5,0) circle (2.5pt);
			       \draw [line width=5pt] (0,0) -- (0.5,0);}}

\makeatletter
\def\namedlabel#1#2{\begingroup
    \normalfont #2\!%
    \def\@currentlabel{#2}%
    \phantomsection\label{#1}\endgroup
}
\makeatother

\def\be{\begin{equation}}
\def\ee{\end{equation}}
\def\beal{\begin{equation}\begin{aligned}}
\def\eeal{\end{aligned}\end{equation}}
\def\nn{\nonumber}
\def\bra#1{\langle #1|}

\def\braket#1{\langle #1 \rangle}

\def \Ea{{E}}
\def \Eb{{\tilde E}}
\def\pCM{p_{\rm CM}}

\newcommand{\cSquare}{\,\rotatebox{90}{\scalebox{0.6}[0.9]{$\bowtie$}}}

\def\cM{{\cal M}}

\def\cO{{\cal O}}

\def\nn{\nonumber}

\def\Res_#1{\operatorname*{Res}_{#1}}

\def\ie{{\it i.e.} }
\def\eg{{\it e.g.} }

\newcommand{\qperp}[0]{\bm{q_{\perp}}}
\newcommand{\bperp}[0]{\bm{b_{\perp}}}

\def\eqn#1{eq.~\eqref{#1}}
\def\eqns#1#2{eqs.~\eqref{#1} and~\eqref{#2}}

\def\Fig#1{Fig.~{\ref{#1}}}

\def\rcite#1{ref.~\cite{#1}}
\def\rcites#1{refs.~\cite{#1}}

\def\vs{\varsigma}

\def\Ea{E}
\def\Eb{\tilde E}
\def\Ec{{\cal E}}
\def\Ed{\tilde {\cal E}}
\def\eperp{e_\perp}

\title{Binary Kerr black-hole scattering at 2PM \\ from quantum higher-spin Compton}

\author[a]{Lara Bohnenblust,}
\author[b,c]{Lucile Cangemi,}
\author[b,d]{Henrik Johansson,}
\author[e]{and Paolo Pichini}

\affiliation[a]{Department of Astrophysics, University of Zurich, Winterthurerstrasse 190, 8057 Zurich, Switzerland,}
\affiliation[b]{Department of Physics and Astronomy, Uppsala University,
Box 516, 75120 Uppsala, Sweden,}
\affiliation[c]{School of Mathematics and Maxwell Institute for Mathematical Sciences, \\
University of Edinburgh, EH9 3FD, UK}
\affiliation[d]{Nordita, Stockholm University and KTH Royal Institute of Technology, \\
Hannes Alfv\'{e}ns  v\"{a}g 12, 10691 Stockholm, Sweden,}
\affiliation[e]{center for Theoretical Physics, Department of Physics and Astronomy, \\
Queen Mary University of London, Mile End Road, London E1 4NS, UK.}

\emailAdd{lara.bohnenblust@uzh.ch}
\emailAdd{lucile.cangemi@ed.ac.uk}
\emailAdd{henrik.johansson@physics.uu.se}
\emailAdd{p.pichini@qmul.ac.uk}

\preprint{UUITP-31/24 \\
\phantom{~} \hfill NORDITA 2024-042}

\abstract{
Quantum higher-spin theory applied to Compton amplitudes has proven to be surprisingly useful for elucidating Kerr black hole dynamics. Here we apply the framework to compute scattering amplitudes and observables for a binary system of two rotating black holes, at second post-Minkowskian order, and to all orders in the spin-multipole expansion for certain quantities. 
Starting from the established three-point and conjectured Compton quantum amplitudes, the infinite-spin limit gives classical amplitudes that serve as building blocks that we feed into the unitarity method to construct the 2-to-2 one-loop amplitude. We give scalar box, vector box, and scalar triangle coefficients to all orders in spin, where the latter are expressed in terms of Bessel-like functions. Using the Kosower-Maybee-O'Connell formalism, the classical 2PM impulse is computed, and in parallel we work out the scattering angle and eikonal phase. We give novel all-order-in-spin formulae for certain contributions, and the remaining ones are given up to ${\cal O}(S^{11})$. Since Kerr 2PM dynamics beyond ${\cal O}(S^{\ge 5})$ is as of yet not completely settled, this work serves as a useful reference for future studies.  
}

\begin{document}
\maketitle

\section{Introduction}
\label{sec:intro}

The detection of gravitational waves from binary mergers of compact objects~\cite{LIGOScientific:2016aoc,LIGOScientific:2017vwq}, and the promises of upcoming experiments~\cite{Punturo:2010zz,LISA:2017pwj,Reitze:2019iox}, guide computational efforts towards increased analytic precision for two-body dynamics~\cite{Buonanno:1998gg,Damour:2001tu,Blanchet:2013haa}. Recently, focus on post-Minkowskian (PM) perturbation theory and hyperbolic trajectories~\cite{Damour:2017zjx}, has led to new tools that import lessons from scattering amplitudes in quantum field theory (QFT)~\cite{Bjerrum-Bohr:2018xdl,Cheung:2018wkq,Bern:2019nnu}. There are now results for 2-to-2 scattering at orders 3PM~\cite{Bern:2019nnu,Bern:2019crd,Damour:2019lcq,Cheung:2020gyp,Kalin:2020fhe,Bjerrum-Bohr:2021din,Brandhuber:2021eyq,Jakobsen:2022psy,Akpinar:2024meg}, 4PM~\cite{Bern:2021dqo,Bern:2021yeh,Dlapa:2021npj,Dlapa:2021vgp,Khalil:2022ylj,Jakobsen:2023pvx,Jakobsen:2023hig,Damgaard:2023ttc,Dlapa:2024cje} and beyond~\cite{Driesse:2024xad}. Techniques such as analytic continuation of hyperbolic results are used to learn about bound systems~\cite{Kalin:2019rwq,Kalin:2019inp,Cho:2021arx,Adamo:2022ooq,Gonzo:2023goe}, for which high-order post-Newtonian (PN) results are also available~\cite{Siemonsen:2017yux,Foffa:2019yfl,Blumlein:2019zku,Blumlein:2020znm,Levi:2019kgk,Levi:2020lfn,Levi:2022rrq,Levi:2022dqm,Mandal:2022ufb,Mandal:2022nty,Placidi:2024yld,Khalil:2021fpm,Antonelli:2020ybz,Antonelli:2020aeb,Cho:2021mqw,Bini:2023mdz}. 

An abundance of techniques has emerged for computing classical observables, which include traditional worldline effective field theory (EFT)~\cite{Goldberger:2004jt,Goldberger:2007hy,Kol:2007bc,Goldberger:2009qd,Foffa:2013qca,Foffa:2016rgu,Kalin:2020mvi} with spin effects~\cite{Porto:2005ac,Porto:2006bt,
Porto:2008tb,Porto:2008jj,Levi:2008nh,Porto:2010tr,Porto:2010zg,
Levi:2010zu,Levi:2011eq,
Porto:2012as,Levi:2014gsa,
Levi:2014sba,Levi:2015msa,Levi:2015uxa,Levi:2015ixa, 
Levi:2016ofk,Maia:2017yok, 
Maia:2017gxn,
Levi:2020kvb,Levi:2020uwu,Liu:2021zxr},
the Kosower-Maybee-O'Connell (KMOC) formalism~\cite{Maybee:2019jus,Cristofoli:2021vyo,Cristofoli:2021jas,Cristofoli:2022phh,Adamo:2022rmp}, eikonal-based approaches~\cite{DiVecchia:2023frv,KoemansCollado:2019ggb,DiVecchia:2021bdo,Heissenberg:2021tzo,Haddad:2021znf,Adamo:2021rfq,DiVecchia:2022piu,Bellazzini:2022wzv,Luna:2023uwd,Gatica:2023iws,Fernandes:2024xqr,Du:2024rkf}, wordline quantum field theory~\cite{
Mogull:2020sak,Jakobsen:2021smu, Jakobsen:2021lvp,Jakobsen:2021zvh,Comberiati:2022cpm,Ben-Shahar:2023djm,Damgaard:2023vnx}, heavy-particle effective theory~\cite{Damgaard:2019lfh,Aoude:2020onz}, approaches using the double copy~\cite{Bern:2008qj,Bern:2010ue,Bern:2019prr,Adamo:2022dcm,Bern:2022wqg,Luna:2016due,Luna:2017dtq,Shen:2018ebu,Li:2018qap,Goldberger:2019xef,Plefka:2019wyg,Bautista:2019tdr,Kim:2019jwm,Monteiro:2020plf,Haddad:2020tvs,Carrasco:2020ywq,Carrasco:2021bmu,Brandhuber:2021kpo,Gonzo:2021drq,Shi:2021qsb,Almeida:2020mrg,Wang:2022ntx,CarrilloGonzalez:2022mxx}, soft graviton theorems~\cite{Saha:2019tub,Manu:2020zxl,DiVecchia:2021ndb,Alessio:2022kwv,A:2022wsk,Alessio:2024wmz,Elkhidir:2024izo,Akhtar:2024lkk,Alessio:2024onn}, direct classical limits of QFT amplitudes~\cite{Cachazo:2017jef,Guevara:2017csg}, and twistor descriptions~\cite{Kim:2023vgb,Kim:2023aff,Kim:2024grz}. For recent reviews on gravitational waves, EFTs and scattering amplitude methods, see \rcites{Porto:2016pyg,Barack:2018yly,Levi:2018nxp,Buonanno:2022pgc,Kosower:2022yvp,Bjerrum-Bohr:2022blt,Goldberger:2022rqf,Bjerrum-Bohr:2022ows}.

Classical observables extracted from 2-to-2 scattering amplitudes, in the PM expansion, include the scattering angle, impulse, and spin-kick~\cite{Guevara:2018wpp,Bini:2018zxp,Chung:2019duq,Bjerrum-Bohr:2019kec,Cristofoli:2019neg,Goldberger:2020wbx,Kol:2021jjc,Chen:2021kxt,Bjerrum-Bohr:2021wwt,Herrmann:2021tct,Jakobsen:2022fcj,Jakobsen:2022zsx,Aoude:2022thd,Menezes:2022tcs,Damgaard:2022jem,FebresCordero:2022jts,Alessio:2023kgf,Bautista:2023szu,Akpinar:2024meg,Gonzo:2024xjk}.
Furthermore, the PM expansion has allowed the computation and understanding of radiation-reaction effects~\cite{Herrmann:2021tct,Manohar:2022dea,Kalin:2022hph,Alessio:2022kwv,Bini:2022enm,Dlapa:2022lmu,Riva:2021vnj}, higher-order waveforms~\cite{Brandhuber:2023hhy,Georgoudis:2023lgf,Georgoudis:2023ozp,Caron-Huot:2023vxl,Caron-Huot:2023ikn,Bini:2023fiz,Georgoudis:2023eke,Georgoudis:2024pdz,Bini:2024rsy,Buonanno:2024byg,Mougiakakos:2021ckm}, beyond-GR effects~\cite{Burger:2019wkq,Emond:2019crr,Cristofoli:2019ewu,Brandhuber:2019qpg,KoemansCollado:2019lnh,AccettulliHuber:2020dal,AccettulliHuber:2020oou,Carrillo-Gonzalez:2021mqj,Davis:2023zqv,Bhattacharyya:2024aeq,Bhattacharyya:2024kxj,Fransen:2024fzs,Brandhuber:2024bnz,Bautista:2024agp,Brandhuber:2024qdn,Falkowski:2024bgb}, tidal effects~\cite{Aoude:2020ygw,Kalin:2020lmz,Kim:2020dif,Bern:2020uwk,Haddad:2020que,Cheung:2020sdj,Bini:2020flp,Heissenberg:2022tsn,Mougiakakos:2022sic}, nontrivial backgrounds and/or self-dual black holes~\cite{Adamo:2020qru,Cristofoli:2020hnk,Kim:2020cvf,Crawley:2021auj,Adamo:2022qci,Adamo:2022rob,Gonzo:2022tjm,Kosmopoulos:2023bwc,Jones:2023tgz,Crawley:2023brz,Guevara:2023wlr,Adamo:2023fbj,Adamo:2023cfp,Adamo:2024oxy}, natural extensions such as quantum effects, high-energy limits, supersymmetric and string models, and Kerr-Newman black holes, mergers and other toy models~\cite{Abreu:2020lyk,Bern:2020gjj,Caron-Huot:2018ape,KoemansCollado:2018hss,Moynihan:2019bor,Parra-Martinez:2020dzs,Cristofoli:2020uzm,Kim:2022iub,Chen:2021huj,Chung:2019yfs,Monteiro:2021ztt,Jones:2022aji,Emond:2020lwi,Cangemi:2022abk,Hoogeveen:2023bqa,Bianchi:2023lrg,Gambino:2024uge,Saketh:2021sri,Akhtar:2024mbg,Aoki:2024boe}.
Comparisons to scattering computations in numerical relativity have also been done, which highlight the need for resummation~\cite{Rettegno:2023ghr,Damour:2022ybd}.

Inclusion of spin effects is vital for astrophysical black holes, and it has also proven to be interesting from a purely formal perspective. By using scattering amplitudes for massive spinning particles one can extract the spin multipoles~\cite{Vaidya:2014kza} in terms of the spin vector $S^\mu=m a^\mu$ of a Kerr black hole. In \rcite{Vines:2017hyw} the three-point amplitude was given on an exponential form to all orders in the ring radius vector $a^\mu$, and in \rcite{Arkani-Hamed:2017jhn} a family of generic spin-$s$ quantum amplitudes were given, which were later shown~\cite{Guevara:2018wpp,Chung:2018kqs} to also describe the Kerr three-point amplitude. Using unitarity and on-shell methods, the higher-spin amplitudes gave a new avenue to compute 2-to-2 scattering at 1PM~\cite{Guevara:2017csg,Guevara:2019fsj,Chung:2019duq} and conceptually treat Kerr black holes as elementary particles~\cite{Arkani-Hamed:2019ymq,Guevara:2020xjx,Aoude:2020mlg}.

Compton four-point amplitudes, corresponding to a Kerr black hole interacting with two gravitons, have been presented for the opposite-helicity~\cite{Arkani-Hamed:2017jhn} and same-helicity~\cite{Johansson:2019dnu} graviton states. The opposite-helicity Arkani-Hamed-Huang-Huang (AHH) amplitudes exhibit spurious poles for higher-spin $s>2$ states, which signals the need to resolve contact-term ambiguities. Several proposals have been put forward attempting to single out the appropriate Compton contact terms, using properties such as the high-energy limit or additional conjectured symmetries and structures~\cite{Falkowski:2020aso,Chen:2022clh,Bautista:2021wfy,Bern:2022kto,Aoude:2022trd,Saketh:2022wap,Bautista:2022wjf,Bjerrum-Bohr:2023jau,Bautista:2023sdf,Bjerrum-Bohr:2023iey,Scheopner:2023rzp,Aoude:2023vdk,Haddad:2023ylx,Azevedo:2024rrf}. The spinning Compton amplitudes for $s\le 2$ are also known to be given by the double copy~\cite{Kawai:1985xq,Bern:2008qj,Bern:2010ue} in terms of gauge-theory $s\le 1$ amplitudes~\cite{Johansson:2019dnu,Bautista:2019evw,Johansson:2015oia,Chiodaroli:2021eug,Guevara:2021yud} that are often referred to as root-Kerr theory~\cite{Arkani-Hamed:2019ymq}. 

The elementary-particle Lagrangians that underlie the well-behaved AHH amplitudes were analyzed in detail in~\rcite{Chiodaroli:2021eug}, and using tree-level unitarity constraints from higher-spin theory, the Compton family was extended up to $s=5/2$. By using the full machinery of higher-spin theory~\cite{Zinoviev:2001dt}, the AHH three-point amplitudes could also be derived from gauge-symmetry principles~\cite{Cangemi:2022bew}. Combining higher-spin gauge symmetry, a chiral-field approach~\cite{Ochirov:2022nqz}, and the appearance of symmetric homogeneous polynomials in Kerr and root-Kerr amplitudes~\cite{Cangemi:2023ysz}, lead to a proposal for the closed-form spin-$s$ family of Kerr Compton tree amplitudes~\cite{Cangemi:2023bpe}. 
Taking the classical limit of the quantum Compton amplitudes, using infinite-spin limit~\cite{Cangemi:2022abk} or coherent-spin states~\cite{Aoude:2021oqj}, gave a proposal for the all-order-in-spin tree-level Compton amplitude describing Kerr dynamics~\cite{Cangemi:2023bpe}. This was tested against explicit general-relativity calculations~\cite{Bautista:2022wjf, Bautista:2023sdf}, using black-hole perturbation theory or the Teukolsky equation. Full agreement was found for the special choice $\alpha=0$, where $\alpha$ is a bookkeeping parameter introduced in~\rcite{Bautista:2022wjf}, which tags terms related to polygamma functions that start appearing at ${\cal O}(S^5)$. Also, dissipative terms start appearing at this order, tagged by $\eta$, and they could be straightforwardly accounted for in the higher-spin Compton amplitude, again for $\alpha=0$. These non-analytic contributions, and their relation to near-zone/far-zone splitting of the solution to the Teukolsky equation, and related ambiguities, are discussed in \rcites{Bautista:2023sdf,Bautista:2024agp}. 

Related work on classical Compton amplitudes can also be found in~\rcites{Bjerrum-Bohr:2023jau,Bjerrum-Bohr:2023iey}, which gives closed-form expressions that have the same classical factorization poles as the proposal~\cite{Cangemi:2023bpe}, while the contact terms are similar but different. Proposals for closely related Compton processes in a Kerr background have also been put forward in~\rcite{Correia:2024jgr}.

In general, there have been many competing new scattering amplitude methods employed for computing state-of-the-art observables for binary systems of spinning black holes. In particular, the 2-to-2 amplitude is known in a wide range of cases: at 1PM with arbitrary spin, at 2PM up to spin ${\cal O}(S^4)$ \cite{Guevara:2018wpp,Chen:2021kxt} and ${\cal O}(S^6)$  \cite{Bautista:2023szu}, with proposed ${\cal O}(S^{8})$ extension~\cite{Chen:2024mmm}, at 3PM up to ${\cal O}(S^{2})$~\cite{Jakobsen:2022zsx,Akpinar:2024meg} and up to 4PM for ${\cal O}(S^{1})$~\cite{Jakobsen:2023ndj}; see also~\cite{Bini:2018ywr,Guevara:2019fsj,Chung:2019duq,Bern:2020buy,Chung:2020rrz,Damgaard:2022jem,Kosmopoulos:2021zoq,Jakobsen:2022fcj,Aoude:2022thd,FebresCordero:2022jts,Menezes:2022tcs,Jakobsen:2022zsx,Alessio:2023kgf,Buonanno:2024vkx,Gonzo:2024zxo}. See also recent work on spin-magnitude change~\cite{Bern:2023ity,Kim:2023drc,Alaverdian:2024spu}, dissipation and absorption with spin~\cite{Chen:2023qzo,Aoude:2023fdm}. More importantly, the radiation emitted by the binary, obtained from a five-point amplitude with an emitted graviton, was computed up to $\cO(G^3)$ for low spin orders~\cite{Jakobsen:2021lvp,Bautista:2021inx,Alessio:2022kwv,Riva:2022fru,Heissenberg:2023uvo,DeAngelis:2023lvf,Brandhuber:2023hhl,Aoude:2023dui,Bohnenblust:2023qmy}.

In this paper, we explore 2-to-2 scattering of spinning black holes and corresponding 2PM observables using the all-order-in-spin Compton amplitude proposed in~\rcite{Cangemi:2023bpe}. We develop efficient integration techniques for extracting the relevant contributions to the classical one-loop amplitude: scalar triangles, scalar and vector boxes. This requires some delicacy, since the one-loop integrand is composed of several nontrivial entire functions of three independent spin variables. While the box coefficients have closed forms in terms of exponential functions, we find after some work that the triangle integral coefficients can be expressed as integrals and derivatives of the Bessel function $J_0$ of the first kind. Our results can be contrasted to the recent 2PM work~\cite{Chen:2024bpf}, which also addressed the need for integration methods for spin-resummed entire functions that appear in Kerr scattering amplitudes. See also related results~\cite{Chen:2024mmm} that extract 2PM amplitudes and eikonal phase to relatively high spin-multipole order. 

Using the one-loop amplitude reduced to master integrals, we then explore observables at 2PM. We compute the classical impulse, scattering angle and eikonal phase up to ${\cal O}(S^{11})$. We also give various closed-form expressions to all orders in spin, both for contributions coming from boxes, such as the parallel impulse and impulse perpendicular to the scattering plane, and for the eikonal-phase contributions corresponding to triangle coefficients that originate from the pole terms of the quantum Compton amplitude. We leave the eikonal evaluation of the third triangle coefficient, corresponding to genuine contact terms, to future work.


This paper is structured as follows: In section~\ref{sec:compton}, we review the higher-spin Compton amplitude which describes gravitational perturbations of a Kerr black hole, presented by some of the authors in \rcite{Cangemi:2023bpe}. In section~\ref{sec:OneLoopDecomposition}, we study the one-loop 2-to-2 amplitude required to extract 2PM binary observables, presenting the required building blocks, the classical limit technology and the techniques to compute loop integrals. In section~\ref{sec:observables}, we review the KMOC formalism~\cite{Kosower:2018adc,Maybee:2019jus} and use it to compute classical observables from the one-loop amplitude. In particular, we compute the eikonal, the scattering angle and the impulse, and display some explicit results. We also compare our results to similar works in the literature. Lastly, we conclude in section~\ref{sec:conclusion}.

\section{Review of tree-level Kerr Compton amplitudes}
\label{sec:compton}

In this section, we review the details of the higher-spin three-point and Compton amplitudes discussed in \rcite{Cangemi:2023bpe}, and the classical black-hole amplitudes that arise in the infinite-spin limit.

To set up our notation, consider the tree-level amplitude ${\cal M}(1^s,2^s,3,\ldots, n)$ between two massive higher-spin particles and $(n-2)$ gravitons. We start by pulling out the gravitational coupling and phases,\footnote{We have suppressed a spin-dependent unphysical phase $\sim (-1)^{s}$, present due to quirks of the mostly-minus signature.}
\begin{align}
{\cal M}(1^s,2^s,3,\ldots, n)&=i\Big(\frac{\kappa}{2}\Big)^{n-2} M(1^s,2^s,\ldots, n)\,,
\end{align}
and the momenta satisfy $p_1^2=p_2^2=m^2$, $p_{i>2}^2=0$ and $\sum_i p_i^\mu=0$ (all incoming momenta). 

Then recall that the three-point higher-spin AHH amplitudes take the form 
\begin{align}
M(1^s,2^s,3^+)=2(\varepsilon_3^+ \cdot p_1)^2\frac{\braket{\bm{21}}^{2s}}{m^{2s}}\,,~~~~~~~
M(1^s,2^s,3^-)=2(\varepsilon_3^- \cdot p_1)^2\frac{[\bm{21}]^{2s}}{m^{2s}}\,,  
\end{align}
where the massive spinors include a SU(2) wavefunction $|\bm{1}\rangle=|1^a\rangle z_a$, $|\bm{1}]=|1^a] z_a$, $|\bm{2}\rangle=|2^a\rangle \bar z_a$, $|\bm{2}]=|2^a] \bar z_a$, which automatically symmetrizes over the little group indices.

Under the classical limit $\hbar\to 0$, $s\to \infty$, the $\braket{\bm{21}}^{2s}$ and $[\bm{21}]^{2s}$ factors are mapped to exponentials, giving the classical Kerr amplitudes
\begin{align} \label{Classical3pt}
M(1,2,3^\pm)=2(\varepsilon_3^\pm \cdot p_1)^2 e^{\pm p_3 \cdot a}\,, 
\end{align}
where the ring-radius vector is related to the SU(2) wavefunction as
\begin{align}\label{RingRadiusSpinS}
a^\mu = &- \frac{|a|}{m}  \bra{1^b}\sigma^\mu| 1^{c}] \, \bar z_{(b} z_{c)}\,,
\end{align}
and we normalized as $\bar z^b z_b=1$, giving that $a^2=-|a|^2$. The ring radius has units of length; it is transverse $p_1 \cdot a=0$, and related to the dimensionless spin vector as $S^\mu=m a^\mu$. 

For the tree-level Compton amplitude, there also exist candidate higher-spin amplitudes. 
The same-helicity Compton amplitude for spin-$s$ quantum states is~\cite{Johansson:2019dnu} 
\begin{align}
M(1^s,2^s,3^+,4^+)&=\frac{[34]^4 \braket{\bm{21}}^{2s}}{s_{12}t_{13} t_{14} m^{2s-4}}\,, 
\end{align}
where $s_{ij}=(p_i+p_j)^2$, $t_{ij}=2p_i\cdot p_j$.
The corresponding classical amplitude also takes the form of an exponential
\begin{align} \label{eq:clsamehel}
M(1,2,3^+,4^+)&=-\frac{[34]^4}{q^2 (p_1 \cdot q_\perp)^2} e^{q\cdot a}\,,
\end{align}
where the graviton momenta are encoded as $q=p_4+p_3$ and $q_\perp=p_4-p_3$. Note that the frequency/energy of the two graviton plane waves is $\omega=  \tfrac{p\cdot q_\perp}{2m}$.

The opposite-helicity Compton amplitude for spin-$s$ quantum states is expected to take the form~\cite{Arkani-Hamed:2017jhn,Cangemi:2023bpe},
\begin{align} \label{eq:HSCompton}
M(1^s\!,2^s\!,3^-\!\!,4^+)&= \frac{\big(\braket{\bm{1}3}[4\bm{2}]+\braket{\bm{2}3}[4\bm{1}]\big)^{2s}}{s_{12}t_{13} t_{14} \bra{3}1|4]^{2s-4}} + (\text{contact-term completion})\nn \\\
 & =  \, \frac{\langle3|1|4]^4  P_1^{(2s)}}{m^{4s} s_{12}t_{13}t_{14}}
-\frac{\braket{\bm{1}3} [{4\bm 2}] \langle3|1|4]^3 }{ m^{4s} s_{12}t_{13}} P_2^{(2s)}
+\frac{\braket{\bm{1}3} \braket{3{\bf 2}} [\bm{1}4] [4\bm{2}]}{m^{4s} 
 s_{12}} \langle3|1|4]^2 P_2^{(2s-1)} \nn \\ &
 +\frac{\braket{\bm{1}3} \braket{3{\bf 2}} [\bm{1}4] [4\bm{2}]}{m^{4s-4} s_{12}} \langle3|1|4] \langle3|\rho|4] \Big(\frac{P_2^{(2s-2)}}{m^2} {-} \braket{\bm{12}}[\bm{12}] P_4^{(2s-2)} {+}  \frac{\langle3|\rho|4]}{\langle3|1|4]} P_4^{(2s-1)}\Big) \nn \\ &
 + \frac{\braket{\bm{1}3}^2 \braket{3{\bf 2}}^2 [\bm{1}4]^2 [4\bm{2}]^2}{2m^{4s-4} }\braket{\bm{12}}[\bm{12}]
   \Big[(1+\eta) P_{5|\vs_1}^{(2s-2)} + (1-\eta) P_{5|\vs_2}^{(2s-2)} \Big]
 + \alpha\, C^{(s)}_\alpha \,,
\end{align}
where the first-line expression is the AHH Compton amplitude~\cite{Arkani-Hamed:2017jhn}, which is known to need a completion of contact terms to cancel out the spurious pole coming from the $\langle3|1|4]$ denominator. The last equality gives the Compton amplitude found in \rcite{Cangemi:2023bpe}, which gives the completion, up to free contact terms controlled by the parameter $\alpha$, and $\eta$ controls dissipative terms. The $\alpha$ and $\eta$ were introduced as bookkeeping parameters in~\rcite{Bautista:2022wjf}.\footnote{We work under the assumption $\alpha=0, \eta=\pm1$, although it is currently unclear which contact-term choice best describes the {\it tree-level} Compton amplitude beyond fourth order in spin~\cite{Bautista:2022wjf, Bautista:2023sdf,Cangemi:2023bpe}.} 

In \eqn{eq:HSCompton}, the $P^{(k)}_{n}$ are complete homogeneous symmetric polynomials, which can be written as
\begin{equation}
\label{PolyDef}
P_n^{(k)} := \sum_{i=1}^n \frac{\vs_i^k}{\prod_{j \neq i}^n (\vs_i-\vs_j)} = \sum_{\sum l_i=k-n+1}\!\vs_1^{l_1}\!\cdots \vs_n^{l_n}\,,
\end{equation}
with globally assigned spin-dependent variables
\beal
\label{VarSigmaDef}
\vs_1 & := \bra{\bm{1}}4|\bm{2}] + m[\bm{21}] , \qquad~\;\,\quad
\vs_3 := m \braket{\bm{21}} , \\*
\vs_2 & := - \bra{\bm{2}}4|\bm{1}] + m [\bm{21}] , \qquad \quad
\vs_4 := m [\bm{21}] .
\eeal
Polynomials with more than four variables are subject to the limit $P_{n|\vs_i}^{(k)}:=\lim_{\vs_n \to \vs_i} P_{n}^{(k)}$.
Finally, the $\rho^\mu$-vector in \eqn{eq:HSCompton} is defined as $\rho^\mu:=\frac{1}{2}(\langle {\bm 2}|\sigma^\mu | {\bm 1}]+\langle {\bm 1}|\sigma^\mu | {\bm 2}])$.

After taking the classical limit, the opposite-helicity Compton amplitude can be expressed in terms of four classical spin-dependent variables, abbreviated as
\beal \label{eq:clComptonVars}
x &: = a \cdot q_\perp , \qquad\;\,\qquad
y := a \cdot q , \\
z & := |a| v \cdot q_\perp , \qquad \quad
w := \frac{\langle 3|a|4]}{\langle 3|v|4]} v \cdot q_\perp ,
\eeal
where the four-velocity of the black hole is identified with the first particle, $p_1^\mu= m v^\mu$, with $v^2=1$. In these variables, the classical opposite-helicity Compton amplitude is
\begin{align}
\label{eq:classicalcompton}
{M}(\bm{1},\bm{2},3^-\!,4^+) = -\frac{\langle 3|1|4]^4}{q^2 (p_1 \cdot q_\perp)^2}  
\Big(&e^x \cosh z - w \, e^x {\rm sinhc}\, z + \tfrac{w^2 - z^2}{2} E + (w^2 - z^2) (x-w) \tilde{E} \nn \\ &
- \frac{(w^2-z^2)^2}{2\xi} \big({\cal E} +\eta \,\tilde {\cal E}\big) \Big)+ \alpha\, C_\alpha^{(\infty)} ,
\end{align}
where $\xi=(v\cdot q_\perp)^2/q^2$ is the optical parameter, ${\rm sinhc}\, z:= z^{-1}\sinh z$ is an even function, and we make use of four entire functions
\begin{align} \label{eq:clEfns}
\Ea(x,y,z)&=\frac{e^y - e^x \cosh z + (x - y) e^x \, {\rm sinhc}\,z}{(x - y)^2 - z^2}~+~ (y\to -y)\,, \nn  \\
\Eb(x,y,z)&=\frac{1}{2y}\frac{e^y - e^x \cosh z + (x - y) e^x \, {\rm sinhc}\,z}{(x - y)^2 - z^2}~+~ (y\to -y) \,, \\
&\quad \Ec(x,y,z)= \frac{\partial \Eb}{\partial x}\,, \qquad \qquad \Ed(x,y,z)= \frac{\partial \Eb}{\partial z}\,. \nn
\end{align} 
Note that if we had instead defined the velocity in terms of the second particle, $p_2^\mu= -m v^\mu$, the above classical formulae would be unchanged, since the minor difference in definition is automatically projected out in the variable definitions. The classical Compton amplitude matches the black hole perturbation theory results of~\rcites{Bautista:2022wjf,private} for the choice $\alpha=0$, and is compatible with~\rcite{Bautista:2023sdf}.

It is useful to consider the Compton amplitudes evaluated on one of the massive poles, corresponding to $t_{14}= (p_1+p_4)^2 - m_1^2 \to 0$. In this limit it factorizes into two three-point amplitudes
\begin{align}
M_{t_{14}=0}(1^s\!,2^s\!,3,4)&= t_{14} M(1^s\!,2^s\!,3,4)\big|_{t_{14=0}}= \sum_{\rm states}M(1^s\!,p^s\!,4) M(-p^s\!,2^s\!,3)\,, \nn \\
M_{t_{14}=0}(1^s\!,2^s\!,3^-,4^+)&= \frac{\big(\braket{\bm{1}3}[4\bm{2}]+\braket{\bm{2}3}[4\bm{1}]\big)^{2s}}{s_{12}t_{13} \bra{3}1|4]^{2s-4}}\Bigg|_{t_{14=0}} =\frac{\langle3|1|4]^4  \vs_1^{2s}}{m^{4s} s_{12}t_{13}}\Bigg|_{t_{14=0}} \,.
\end{align}
The contact terms do not contribute and all information is captured by the AHH amplitude, given on the first line of \eqn{eq:HSCompton}. In the classical limit, the factorized amplitudes ${M}_{t_{14}=0}$ have the following compact forms,
\beal
\label{eq:classicalComptonCut}
{M}_{t_{14}=0}(\bm{1},\bm{2},3^+\!,4^+)  = -\frac{m^4 [3 4]^4}{q^4} e^{y} \,, \quad {M}_{t_{14}=0}(\bm{1},\bm{2},3^-\!,4^+) = -\frac{\langle 3|1|4]^4}{q^4} e^{x} \,.
\eeal 
The amplitudes above are only valid for $t_{14}=0$ which changes the classical scaling of the spin-dependent variables in \eqn{eq:clComptonVars}. The details are discussed in appendix C of \rcite{Cangemi:2023bpe} for the Compton amplitudes in the analogous $\sqrt{\mathrm{Kerr}}$ gauge theory.  
\section{Classical 2-to-2 scattering for Kerr binary}
\label{sec:OneLoopDecomposition}

We are interested in 2-to-2 scattering of two spinning black holes, which at $L$ loop orders takes the general form     
\begin{equation}
\mathcal{M}^{(L)} (1^{s_1}2^{s_2}\to 3^{s_2}4^{s_1})=  (-i)^L\Big(\frac{\kappa^2}{4}\Big)^{L+1} \int \frac{d^{DL}\ell}{(2\pi)^{DL}} \, {\cal I}^{s_1 s_2}(p_1,p_2,q,\ell_i)\,,
\end{equation}
where ${\cal I}$ denotes the integrand, which depends on $L$ independent loop momenta $\ell_i$ and the transfer momentum $q^\mu$; the measure is $d^{DL}\ell=\prod_{i=1}^Ld^{D}\ell_i$.  The external momenta satisfy
\begin{align}
q^\mu=p_2-p_3=p_4-p_1\,, ~~~~
 p_1^2=p_4^2=m^2_1\,,~~~~  p_2^2=p_3^2=m^2_2\,,~~~(p_1+p_2)^2=s=E^2\,.
\end{align}
For classical scattering we will assume that the loop and exchange momenta exhibit soft scaling $\ell^\mu \sim q^\mu \sim \hbar$, and that the quantum spins $s_1$ and $s_2$ are approaching infinity, so that classical ring-radius vectors $a_1^\mu = S_1^\mu/m_1$ and $a_2^\mu= S_2^\mu/m_2$ are the appropriate variables,
\begin{align}
{\cal I}^{s_1 s_2}(p_1,p_2,q,\ell_i) ~ \stackrel{\hbar \to 0}{\longrightarrow} ~  {\cal I}(p_1,p_2,q,\ell_i, a_1,a_2)\,,
\end{align}
The formulae that relate quantum and classical spin are analogous to~\eqn{RingRadiusSpinS}, and we will not need the details. 
We take the ring-radius vectors to satisfy
\begin{align}
a_i \cdot p_i=0\,,~~~~ a_i^\mu  \sim \frac{1}{\hbar}\,,~~~~~~~~ (i=1,2)
\end{align}
which implies that each spin multipole needs to be accompanied with corresponding loop or transfer momenta such that the products, $a^\mu q^\nu \sim  a^\mu \ell^\nu \sim 1$, are invariant under scaling. 

Let us illustrate the above considerations with the warm-up case of tree-level 2-to-2 scattering. For $L=0$  there is no integration, and the relevant contribution to the amplitude is obtained by sewing together two three-point amplitudes using a BCFW-shifted graviton momentum:
\be
q^\mu\rightarrow q_{\rm null}^\mu = q^\mu + i \eperp^\mu\,,~~~~~~~~ e_\perp^\mu:= \frac{\epsilon^{\mu \nu \rho \sigma} p_{1\nu}p_{2\rho} q_\sigma}{m_1m_2\sqrt{\sigma^2-1}}\,.
\ee
The factorization pole contribution is then given by the sum over graviton helicities
\be \label{Quantum2to2pole}
\sum_\pm \!\frac{M(1^{s_1}4^{s_1}q^\pm_{\rm null}) M(-q^\mp_{\rm null} 2^{s_2}3^{s_2})}{q^2}\!= \frac{m_1^{2-2s_1}m_2^{2-2s_2}}{q^2}\Big( \! [\bm {41}]^{2s_1} \langle \bm {32}\rangle^{2s_2} e^{2 \zeta} 
{+}\langle \bm {41}\rangle^{2s_1}[\bm {32}]^{2s_2} e^{\!-2 \zeta}\!\Big),
\ee
where $\zeta$ is the relative rapidity\footnote{For simplicity, we assume that the rapidity is positive, \ie $\zeta=|\zeta_2-\zeta_1|$ in terms of the individual black-hole rapidities.} of the black holes, related to the kinematics through
\begin{equation}
\sigma:= \frac{p_1 \cdot p_2}{m_1 m_2} = \cosh \zeta \,,~~~~~~\sqrt{\sigma^2-1} = \sinh \zeta \,, 
\end{equation}
and $\sigma$ is the relative Lorentz factor. 

However, the factorization has two independent kinematic branches $q_{\rm null}^\mu = q^\mu \pm i \eperp^\mu$, and the above formula \eqref{Quantum2to2pole} only holds on the positive branch. The negative branch is given by a similar formula except that square and angle brackets are interchanged. To smoothly connect the two branches, we instead use the following quantum higher-spin amplitude
\begin{align} \label{Quantum2to2}
M^{(0)} (1^{s_1}2^{s_2}\to 3^{s_2}4^{s_1}) =& \frac{m_1^{2-4s_1}m_2^{2-4s_2}}{q^2}\sum_{\pm}
\Big(p_1 {\cdot} \rho_1 \pm \frac{1}{2} i \bar \rho_1 {\cdot}  \eperp \Big)^{2s_1}\!\Big(p_2 {\cdot} \rho_2 \pm \frac{1}{2} i \bar \rho_2 {\cdot}   \eperp\Big)^{2s_2}\! e^{\pm 2 \zeta} \nn \\
&\hskip 3cm \null +{\cal O}(q^0)\,,
\end{align}
where the sum is again over the graviton helicities, but the chiralities of the spinors are now uncommitted, due to the use of the spin-dependent vectors
\begin{align}
\rho_1^\mu =\frac{1}{2}(\langle {\bm 4}|\sigma^\mu | {\bm 1}]+\langle {\bm 1}|\sigma^\mu | {\bm 4}])\,,~~~~\bar \rho_1^\mu =\frac{1}{2}(\langle {\bm 4}|\sigma^\mu | {\bm 1}]-\langle {\bm 1}|\sigma^\mu | {\bm 4}])\,, \nn \\
\rho_2^\mu =\frac{1}{2}(\langle {\bm 3}|\sigma^\mu | {\bm 2}]+\langle {\bm 2}|\sigma^\mu | {\bm 3}])\,,~~~~\bar \rho_2^\mu =\frac{1}{2}(\langle {\bm 3}|\sigma^\mu | {\bm 2}]-\langle {\bm 2}|\sigma^\mu | {\bm 3}])\,, 
\end{align}
introduced in \rcite{Cangemi:2023ysz}. The amplitude \eqref{Quantum2to2} has the same massless pole as \eqn{Quantum2to2pole} on the positive branch, and it also has the correct negative branch.

The classical limit converts the spinor powers into exponentials of the ring-radius vectors. Specifically, for the above spin vectors, we have the classical map~\cite{Cangemi:2023ysz}
\be
\rho_i^\mu \to p_i^\mu\,,~~~~~  \frac{1}{2}\bar \rho_i^\mu \to    -\bar a_i^\mu \, m^2_i\,,
\ee
where $\bar a_i$ are the ring-radius vectors in the spin-1/2 representation. After converting to spin-$s_i$ representations, and taking the $s_i \to \infty$ limit, the powers become exponentials and the classical 2-to-2 tree-level result is~\cite{Guevara:2018wpp,Guevara:2019fsj,Arkani-Hamed:2019ymq}
\begin{equation} \label{ClassicalTree2to2}
M^{(0)} (12\to 34) =
\frac{m_1^2m_2^2}{q^2}\Big(e^{i a\cdot e_\perp} e^{-2 \zeta} 
+e^{-i a\cdot e_\perp}  e^{2 \zeta}\Big) = \frac{2m_1^2m_2^2}{q^2}\cosh(2\zeta-i a\cdot e_\perp) \,,
\end{equation}
where the ring-radius vectors nicely combine in $a^\mu=a_1^\mu+a_2^\mu$. This result could of course have been obtained directly from sewing the classical three-point amplitude \eqref{Classical3pt}, and hence the quantum expression \eqn{Quantum2to2} was not needed. We expect this to be valid more generally: the classical Compton amplitude should be sufficient input for the classical one-loop 2-to-2 process, and related observables. Nevertheless, it can be useful to revert back to the quantum expressions when there are doubts about the classical framework.  

\subsection{One-loop integrals for 2-to-2 scattering}

In general, we are interested in the full classical information that is contained in the loop integrand, but we are free to ignore terms that can only contribute as quantum corrections. This means that we write the one-loop integrand as a sum over two scalar boxes, two scalar triangles, and a linear-in-$\ell$ vector box, 
\begin{equation}\label{eq:masterdecomp}
    \mathcal{I} = \Big( c_{\, \square}\mathcal{I}_{\, \square }
    +c_{\cSquare}\mathcal{I}_{ \cSquare}+c_{\bigtriangleup}\mathcal{I}_{\bigtriangleup}+c_{\bigtriangledown}\mathcal{I}_{\bigtriangledown}  \Big) + \tilde c_{\square}\, \mathcal{I}_{ \square}[\ell\cdot\eperp]+\mathcal{O}(\hbar)\,.
\end{equation}
The other master integrals that we ignore (such as bubbles or tadpoles) either encode quantum contributions, or vanish after integration. We keep the vector box, as it can potentially generate classical contributions, once we perform certain operations on the integrand, needed for computing classical expectation values such as the impulse.

\begin{figure}[h]
    \centering
    \includegraphics[width=0.4\linewidth]{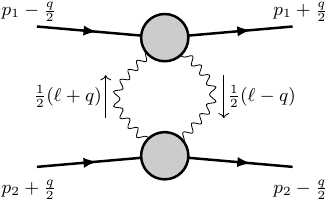}
    \caption{A convenient parametrization of the one-loop amplitude is obtained by shifting the external momenta by $\pm q/2$, and using a loop momenta that is the average of the two internal lines. The external shift will not matter in the classical limit.}
    \label{fig:paramterization}
\end{figure}

It is well established that there is a convenient re-parametrization of the external and internal momenta so to expose more symmetry, namely the one-loop process can be parametrized as shown in \Fig{fig:paramterization}.
The external momenta have been shifted $p_i\to p_i \pm q/2$, and the new $p_i$ are sometimes known in the literature as \textit{averaged} or \textit{barred} momenta. In the classical limit, the shift is insignificant and should not matter. Indeed, if we consistently work with well-behaved quantities to leading order in the $\hbar$ expansion, then the distinction between external and averaged momenta can be suppressed. Nevertheless, since the master-integral decomposition \eqref{eq:masterdecomp} uses quantum integrals as intermediate steps, we will be careful with the definitions here.

The averaged momenta $p_1, p_2$ satisfy the on-shell conditions
\begin{align}
    p_i^2 = m_i^2 - \frac{q^2}{4} = m_i^2 + \mathcal{O}(\hbar^2)\,,
\end{align}
and the last equality serves as a reminder that in the remaining part of this paper we are fully justified to ignore the correction term.  
The external on-shell conditions furthermore imposes two transversality conditions
\begin{align}
    p_i\cdot q = 0\,.
\end{align}

For later convenience, we introduce the velocities $v_1^\mu, v_2^\mu$, the vector $\eperp^\mu$ transverse to the scattering plane, the impact parameter $b^\mu$ and angular momentum $L^\mu= \frac{m_1 m_2}{E} \hat L^\mu$,  
\begin{align}
    &v_i^\mu := \frac{p_i^\mu}{m_i}\,,~~~~~~~~ \eperp^\mu: = \frac{\epsilon^{\mu \nu \rho \sigma} v_{1\nu}v_{2\rho} q_\sigma}{\sqrt{\sigma^2-1}} = (\star q)^\mu \,, \nn \\
    &\,\hat{b}^\mu :=\frac{b^\mu}{|b|} ,~~~~~~~~~ 
    \hat{L}^\mu := \frac{\epsilon^{\mu \nu \rho \sigma} v_{1\nu}v_{2\rho} \hat{b}_\sigma }{\sqrt{\sigma^2-1} }= (\star \hat b)^\mu\,,
\end{align}
where $\epsilon^{\mu \nu \rho \sigma}$ is the anti-symmetric Levi-Civita tensor with convention $\epsilon^{0123}=1$, and we have introduced a convenient Hodge-star notation for the dualization of vectors in the transverse two-dimensional space
\be \label{eq:stardef}
(\star)^{\mu}{}_{\nu}: = \frac{\epsilon^{\mu \rho \sigma}_{~~~\,\nu} v_{1\rho}v_{2\sigma} }{\sqrt{\sigma^2-1} }\,.
\ee
To be clear, some normalization properties of these vectors are 
\begin{align}
v_i^2=1\,,~~~~~ e_\perp^2=q^2=-|q|^2\,,~~~~~  \hat{b}^2=\hat{L}^2=-1 \,,~~~~~ b\cdot v_i=0\,, 
\end{align}
and recall that square-root factor is related to the rapidity $\sqrt{\sigma^2-1}=\sinh \zeta$, which we will often use whenever it is convenient. As is well known, in the classical scattering problem $q^\mu$ is not an observable momentum, instead it has to be traded via a Fourier transform for the impact parameter $b^\mu$, which measures the separation between the trajectories.

The decomposition in~\eqref{eq:masterdecomp} requires us to compute the box, crossed box and triangle integrals. 
The full quantum integrals with possible numerators $N(\ell)$ are defined as
\begin{align} \label{eq:Qintegrals}
I_{\bigtriangleup}^{\rm qu}[N(\ell)]& = \! \!\int\!\! \frac{d^D \ell}{(2\pi)^D 2^{D-4}}\frac{N(\ell)}{(\ell+q)^2(\ell-q)^2\big(\tfrac{1}{4}(\ell^2-q^2)-p_2{\cdot}\ell +i0\big)}\,, \nn \\
I_{\bigtriangledown}^{\rm qu}[N(\ell)]& = \!\! \int \!\! \frac{d^D \ell}{(2\pi)^D 2^{D-4}}\frac{N(\ell)}{(\ell+q)^2(\ell-q)^2\big(\tfrac{1}{4}(\ell^2-q^2)  +p_1{\cdot}\ell+i0\big)}\,,  \\
I_{\square}^{\rm qu}[N(\ell)] & = \!\!\int\!\! \frac{d^D \ell}{(2\pi)^D 2^{D-4}}\frac{N(\ell)}{(\ell+q)^2(\ell-q)^2\big(\tfrac{1}{4}(\ell^2-q^2)-p_2{\cdot}\ell +i0\big)\big(\tfrac{1}{4}(\ell^2-q^2) +p_1{\cdot}\ell+i0\big)},  \nn \\
I_{\cSquare}^{\rm qu}[N(\ell)] & = \!\! \int \!\!\frac{d^D \ell}{(2\pi)^D 2^{D-4}}\frac{N(\ell)}{(\ell+q)^2(\ell-q)^2\big(\tfrac{1}{4}(\ell^2-q^2)+p_2{\cdot}\ell +i0\big)\big(\tfrac{1}{4}(\ell^2-q^2) +p_1{\cdot}\ell+i0\big)}. \nn
\end{align}
For the amplitude, we can reduce all integrals to scalar masters, with numerators $N(\ell) = 1$. Keeping only the leading $\hbar$ order and using dimensional regularization with $D=4-2\epsilon$, the classical integrals evaluate to \cite{Chen:2021kxt, Herrmann:2021tct}
\begin{align}
I_{\bigtriangleup}[1]& = -\frac{i}{32 m_2 |q|}, \quad I_{\bigtriangledown}[1] = -\frac{i}{32 m_1 |q|}\,,  \nn \\
I_{\square}[1] & = \frac{i \left(-\zeta +i \pi\right)}{16\pi^2 q^2 m_1 m_2 \sqrt{\sigma^2-1}}\Big[\frac{1}{\epsilon}-\log(-q^2)  \Big]\,, \nn \\
I_{\cSquare}[1] & = \frac{i \zeta}{16\pi^2 q^2 m_1 m_2 \sqrt{\sigma^2-1}}\Big[\frac{1}{\epsilon}-\log(-q^2)  \Big]\,,
\end{align}
where the rapidity appears as $\zeta=\log(\sigma + \sqrt{\sigma^2-1})=-\log(\sigma - \sqrt{\sigma^2-1})$.
Note that we will sometimes omit the unit numerator insertion $I[1]$ when convenient, hence the absence of a numerator implies a scalar integral.

The box with numerator insertion $N(\ell)=\ell \cdot \eperp$ and $N(\ell)=(\ell \cdot \eperp)^2$ reappears when computing expectation values for observables, such as impulse. We have thus included such terms in the decomposition \eqref{eq:masterdecomp}. Let us give the relevant classical integral identities
\begin{align}\label{eq:Iins}
    I_{\square}[\ell\cdot\eperp] &= 0\,, \nn \\
    I_{\square}[(\ell\cdot\eperp)^2] &= -q^4I_{\square}\,.
\end{align}
where the latter one follows from an algebraic identity valid on the quadruple cut. 

Let us also consider the tensor-triangle integrals, which can all be reduced to scalar ones. 
Any numerator involving powers of $(\ell\cdot\eperp)$ either vanish upon integration (odd power) or is algebraically reducible (even power) on the triple cut. The remaining tensors have the following classical integral identities: 
\begin{align} \label{TensorReuctionTriangle}
I_{\bigtriangleup}[ (\ell {\cdot} v_1)^{2n}] =  \frac{(2n - 1)!!}{n!} \Big(\frac{1}{2} q^2 (\sigma^2 - 1) \Big)^n I_{\bigtriangleup}[1] \,,
\end{align}
and
\begin{align}
I_{\bigtriangledown}[ (\ell {\cdot} v_2)^{2n}] =  \frac{(2n - 1)!!}{n!} \Big(\frac{1}{2} q^2 (\sigma^2 - 1) \Big)^n I_{\bigtriangledown}[1] \,,
\end{align}
and for odd powers of $\ell {\cdot} v_i$ the triangles vanish. The above formulae automatically vanish for negative $n$, as they should.  

Finally, we need the box integral with a cut in the $s$-channel. We may evaluate this using the relation between cuts and imaginary part, giving
\begin{align} \label{eq:CutBoxInt}
    \mathrm{Cut}[I_\square] =2  \mathrm{Im}[(-i) I_\square] &=\frac{1}{\epsilon}  \frac{ 1}{8\pi  m_1 m_2 \sqrt{\sigma^2-1} } \frac{1}{ (-q^2)^{1+\epsilon}}+O(\epsilon) \nn \\
    &= -\frac{ 1}{8\pi q^2 m_1 m_2 \sqrt{\sigma^2-1}}\Big[\frac{1}{\epsilon}-\log(-q^2)  \Big]+O(\epsilon)\,,
\end{align}
where the expression on the first line, which keeps the full $\epsilon$-dependence of $q^2$, is useful when performing the Fourier transform needed for observables in impact parameter space.

\subsection{The classical one-loop triple cut}
\label{sec:triangle}

\begin{figure}[h]
    \centering
    \includegraphics[width=0.5\linewidth]{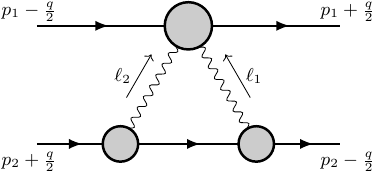}
    \caption{The triple-cut of the one-loop amplitude is constructed by sewing two Compton amplitudes and imposing the following triple-cut conditions $\ell\cdot q=0, \, 2\ell \cdot p_2 =-  q^2 =  \ell^2$. The external and internal momenta are parameterized according to the prescription given in \Fig{fig:paramterization}. In particular, the cut internal lines are parameterized as $\ell_1 = - \frac{1}{2} (\ell - q)$ and $\ell_2 = \frac{1}{2} (\ell + q)$.}
    \label{fig:triplecut}
\end{figure}

Having identified the master integrals that contribute to classical physics, we now compute the relevant classical coefficients $\{ c_{\, \square},\,  c_{\cSquare},\,  c_{\bigtriangleup},\, c_{\bigtriangledown},\,  \tilde c_{\square}\}$. All coefficients can be extracted from the triple cut of the one-loop amplitude, displayed in \Fig{fig:triplecut}. 

From now on we will work in the strict classical limit where we assume a scaling $q\sim \hbar$, $\ell\sim \hbar$ and $a_i \sim 1/\hbar$, such that the combined on-shell and cut constraints are
\begin{alignat}{2}
p_i\cdot q&=0\,, &&p_i^2 = m^2_i,~~~~~~~(i=1,2) \nn\\
p_2 \cdot \ell &= \ell\cdot q=0\,, ~~~~~&&\ell^2 =-  q^2\,.
\end{alignat}
Since the constraints have uniform scaling, they can be imposed on the classical Compton amplitudes discussed in section \ref{sec:compton}, without ambiguities or undoing of the classical limit. To make things simple, we have also aligned the tree-level notation and the one-loop cut notation, such that identifications are straightforward: for black hole 1 we have $q \to q$, $q_{\perp}\to \ell$, and for black hole 2 we have $q \to -q$, $q_{\perp}\to \ell$. 

Given that the black hole $1$ attaches to the Compton tree amplitude, we can conveniently recycle the notation from the all-order-in-spin classical Compton amplitudes in~\eqns{eq:classicalcompton}{eq:clsamehel}. Thus, we make use of the following shorthand notation for common variables depending on $a_1$:
\be
x = \ell {\cdot} a_1, ~~~
y = q{\cdot} a_1, ~~~
z = |a_1|  \ell {\cdot} v_1 \,.
\ee 
It is convenient to decompose the entire functions in \eqn{eq:clEfns} into five parts $f_i=f_i(x,y,z)$ that multiply the allowed five powers of the $w$ variable, 
\begin{align} \label{triple_cut_functions}
&f_0=e^x \cosh z -\frac{1}{2} z^2  \Ea -  x z^2 \Eb  -  z^4 (\Ec - \eta\Ed) \frac{q^2}{2 (\ell {\cdot} v_1)^2}, \nn\\ &
f_1= z^2 \Eb  - e^x \, {\rm sinhc}\,z, \nn\\ &
f_2= \frac{1}{2} \Ea + \Eb x +  z^2 (\Ec - \eta \Ed) \frac{q^2}{(\ell {\cdot} v_1)^2}, \nn\\ &
f_3=  -\Eb, \nn\\ &
f_4=  - (\Ec - \eta \Ed) \frac{q^2}{2(\ell {\cdot} v_1)^2}\,,
\end{align}
where the $x$ and $z$ now contain loop momentum. 

Since the black hole $2$ attaches to cubic vertices with an intermediate on-shell massive state, we can make use of the compact expression given in \eqn{eq:classicalComptonCut}, now with $a_2$ inserted into the exponentials. This makes sure that the triple-cut integrand can be constructed without sewing the states of the massive lines, since that would be ambiguous for classical states.   

After sewing only massless graviton states, the triple-cut integrand thus has the following four helicity contributions from $(\ell_1^\mp, \ell_2^\pm)$,
\begin{align} \label{eq:triplecut}
C^{-+}(\ell) &=\frac{e^{\ell \cdot a_2}  }{q^6 (\ell {\cdot} p_1)^2}  \sum_{n=0}^{4} f_n  (\ell {\cdot} p_1)^n \langle \ell_1|p_1|\ell_2]^{4-n}\langle \ell_1|a_1|\ell_2]^n   \langle \ell_2|p_2|\ell_1]^4  \nn\\
&=
\frac{ m_1^2 m_2^4 e^{\ell \cdot a_2} }{q^6 (\ell {\cdot} v_1)^2} \sum_{n=0}^{4} f_n \, (\ell {\cdot} v_1)^n \Big(q^2 \sigma +i \ell {\cdot} \eperp \sqrt{\sigma^2-1} \Big)^{4 - n}\Big(q^2 a_1 {\cdot}  v_2 +i \epsilon(\ell,  a_1, v_2, q)\Big)^n \,, \nn \\
C^{+-}(\ell)&= C^{-+}(-\ell) \,, \\
C^{++}(\ell)&= \frac{m_1^2 m_2^4 q^2}{ (\ell {\cdot} v_1)^2} e^{q \cdot (a_1+a_2)}\,, \nn \\
C^{--}(\ell)&= C^{++}(\ell)\Big|_{q\to -q}\,,  \nn
\end{align}
and the total cut integrand is 
\be
C(\ell)=C^{++}+C^{--}+C^{+-}+C^{-+}\,,
\ee
which is an even function of $\ell$.

Using the fact that we have a basis of four vectors $\{v_1,v_2,q,\eperp\}$, we can rewrite the Lorentz invariants involving the loop momentum and spin as 
\begin{align}
  \ell {\cdot} a_1 &= \frac{a_1 {\cdot} \eperp \,\eperp {\cdot}\ell}{q^2} +\sigma \,\frac{  a_1 {\cdot} v_2 \,\ell {\cdot} v_1}{\sigma^2-1}   , \nn \\
 \ell {\cdot} a_2 &= \frac{a_2 {\cdot} \eperp \,\eperp {\cdot} \ell }{q^2}-\frac{ a_2 {\cdot} v_1 \,\ell {\cdot} v_1}{\sigma^2-1} 
 , \nn \\
\epsilon(\ell, a_1, v_2, q)& =  \frac{\sigma \, a_1 {\cdot} v_2 \, \eperp {\cdot} \ell +
 a_1 {\cdot} \eperp \,\ell {\cdot} v_1}{
 \sqrt{\sigma^2 - 1}}\,,
\end{align}
where now the only independent invariants are $\ell {\cdot} \eperp$ and $\ell {\cdot} v_1$.

The four contributions to the triple-cut integrand $C^{\pm\pm}$ are thus functions of two Lorentz invariants depending on the loop momentum
\be
C^{\pm\pm}=C^{\pm\pm}( \ell {\cdot} \eperp, \ell {\cdot} v_1).
\ee
However, we can use the Levi-Civita identity for the even powers of the perpendicular vector
\be
(\ell {\cdot} \eperp)^2 =  -q^4+  \frac{q^2 (\ell {\cdot} v_1)^2}{\sigma^2-1}\,,
\ee
and the odd powers of $\ell {\cdot} \eperp$ will integrate to zero due to Lorentz invariance. That said, it is important to keep the odd powers in the box integral, since they will contribute classically if the integrand is multiplied by some parity-odd function. We now consider the box coefficients. 

\subsection{Extracting all-order box integral coefficients}
As is clear from \eqn{eq:triplecut}, the triple-cut integrand has both a double and a simple pole at the location $\ell {\cdot} v_1=0$. One can show that the double-pole residue corresponds to the scalar box integral, and the simple pole contributes to the vector box integral. Thus we obtain the box coefficients as the residues
\begin{align}\label{eq:defboxcoefs}
c^{\pm\pm}_{\Box}[1] & :=  \lim_{\ell {\cdot} v_1\to0} m^2_1 \frac{(\ell {\cdot} v_1)^2}{q^2} C^{\pm\pm}( \ell {\cdot} \eperp, \ell {\cdot} v_1) \nn \\
c^{\pm\pm}_{\Box}[\ell {\cdot} \eperp ] & :=  \lim_{\ell {\cdot} v_1\to0} m _1 \frac{\ell {\cdot} v_1 }{\ell {\cdot} \eperp } \frac{1}{4}\Big[C^{\pm\pm}(\ell {\cdot} \eperp, \ell {\cdot} v_1) - C^{\pm\pm}( \ell {\cdot} \eperp, -\ell {\cdot} v_1) \Big] - \big( 1 \leftrightarrow 2 \big)\,.
\end{align}
On the first line, the multiplication of $m_1^2/q^2$ gives the standard normalization for the scalar box, and on the second line the prefactor likewise gives the standard normalization for the vector box, as set by \eqns{eq:Qintegrals}{eq:Iins}. Whilst the triple-cut integrand is not symmetric under swapping particle numbers $1\leftrightarrow 2$, the residue of the double pole automatically respects the symmetry, and the vector box coefficient needs to be anti-symmetrized by hand, since $\ell \cdot e_{\perp}$ is odd. Furthermore, the difference of the two terms in the bracket is introduced to explicitly project out the double pole. 

Let us now work out the box coefficients from the different helicity contributions of the triple-cut integrand. As can be seen, the equal-helicity contributions give only boxes,
\begin{align}
c^{++}_{\square}[1]:= m_1^2 \frac{(\ell {\cdot} v_1)^2}{q^2}  C^{++} = m_1^4 m_2^4 e^{q \cdot (a_2+ a_1)}  \,, 
\end{align}
and the opposite-helicity scalar box coefficient is also easy to extract, giving
\begin{align}
c^{-+}_{\square}[1]:\!&= m_1^2 \frac{(\ell {\cdot} v_1)^2}{q^2}  C^{-+}\Big|_{\ell {\cdot} v_1\to 0} =\frac{ m_1^4 m_2^4 e^{ a \cdot \ell} }{q^8}   \Big(q^2 \sigma +i \ell {\cdot} \eperp  \sqrt{\sigma^2-1} \Big)^{4} \Big|_{\ell {\cdot} v_1\to 0}  \nn \\
&= m_1^4 m_2^4\,  e^{-i a\cdot \eperp}   \Big(\sigma + \sqrt{\sigma^2-1} \Big)^{4} = m_1^4 m_2^4\,  e^{-i a\cdot \eperp+ 4\zeta}  \,, 
\end{align}
where $a=a_1+a_2$, and we made use of the identities $a \cdot \ell = a\cdot \eperp \, \eperp \cdot \ell/ q^2$ and $\eperp \cdot \ell = -i q^2$, valid on the quadruple cut $\ell\cdot v_1=0$.
When summing over the two opposite-helicity boxes, we simply sum over the sign of the square root, giving 
\begin{align} \label{BoxCoeff1}
c^{-+}_{\square} +c^{+-}_{\square} &= 2 m_1^4 m_2^4 \Big\{(1 - 8 \sigma^2 + 8 \sigma^4) \cos(a\cdot \eperp ) - 4 i \sigma \sqrt{\sigma^2-1} (2\sigma^2-1) \sin( a\cdot \eperp)  \Big\} \nn \\
&=2 m_1^4 m_2^4  \cosh(4 \zeta-i a\cdot \eperp )\,. 
\end{align}
To sum the same-helicity boxes, we sum over $q\to -q$, giving
\be \label{BoxCoeff2}
c^{++}_{\square}+c^{--}_{\square} = 2m_1^4 m_2^4 \cosh q \cdot a \, .
\ee
This expression matches the low-order results in \rcite{Chen:2021kxt}. It is often implied that the same-helicity box coefficients are irrelevant for classical physics, however, as we demonstrate in later sections they give tangible contributions to the parallel classical impulse. 

Next, we focus on the reduction of the odd terms in $\ell {\cdot} \eperp$, which gives the vector-box coefficients.
Whereas $c^{++}_{\square}[\ell \cdot \eperp]=c^{--}_{\square}[\ell \cdot \eperp]=0$ follow trivially from the cut, the opposite-helicity case requires some work. Using the residue formula \eqref{eq:defboxcoefs}, with the intermediate steps suppressed, we similarly obtain a closed expression to all orders in spin, 
\begin{align}\label{eq:BoxIns}
(c^{-+}_{\square}+c^{+-}_{\square})[\ell \cdot \eperp] =& 2 i m_1^3 m_2^4
\Big\{\frac{1}{\sqrt{\sigma^2 - 1}}  
 \Big((4 \sigma^2-1) a_1 \cdot v_2 + 
      4 \sigma (1-2 \sigma^2) a_2 \cdot v_1\Big)  \cos( a \cdot \eperp)
 \nn \\
& +
\frac{i}{\sigma^2 - 1}  \Big(\sigma (3 - 4 \sigma^2) a_1 \cdot v_2 + (1 - 8 \sigma^2 + 8 \sigma^4) a_2 \cdot v_1\Big) \sin(  a \cdot \eperp)
\Big\} \nn \\
&- \big( 1 \leftrightarrow 2 \big)\,.
\end{align}
Note that we explicitly impose the anti-symmetrization over legs $1$ and $2$ by hand; otherwise contributions are missed, since they belong to the horizontally flipped triple cut of \Fig{fig:triplecut}. The anti-symmetrization corresponds to the flipping $\{m_1 \leftrightarrow m_2, a_1 \cdot v_2 \leftrightarrow a_2 \cdot v_1\}$ in the above equation. Note that, in our conventions, $e_{\perp}$ is even under the flip and the oddness of $\ell \cdot e_{\perp}$ originates from $\ell$. Finally, as noted above for the scalar box, re-writing \eqn{eq:BoxIns} using rapidity generates a much more compact formula
\begin{align}\label{eq:tensorboxcoef}
(c^{-+}_{\square}+c^{+-}_{\square})[\ell \cdot \eperp] =i\frac{m_1^3 m_2^3}{\sinh^2 \!\zeta} \Big(&(m_2 a_2 \cdot v_1-m_1 a_1 \cdot v_2) \sinh(4 \zeta - i a \cdot \eperp) \nn \\
- &(m_2 a_1 \cdot v_2-m_1 a_2 \cdot v_1) \sinh(3 \zeta- i a \cdot \eperp)
   \Big)  \,.
\end{align} 
As we will see below, this formula gives rise to a very simple all-order-in-spin contribution to the classical transverse impulse.

\subsection{Extracting all-order scalar triangle coefficients}

We now apply the tensor reduction to the positive powers of the loop momentum of the triple-cut integrand. This is considerably more complicated than extracting box coefficients, so we need to introduce some mathematical tools before getting to the results. 

We first note that we can use some more convenient variables. Rather than tensor numerators built using powers of $\ell \cdot v_1$ and $\eperp \cdot \ell$, we will use the following two algebraically independent factors 
\begin{align}
r:=\frac{\ell \cdot v_1}{|q| \sinh \zeta}\,,~~~~~~ \sqrt{1 + r^2}= \frac{i\eperp \cdot \ell}{q^2}\,,
\end{align}
which are both dimensionless. The triple-cut integrand is now a function
\begin{align}
C(\ell {\cdot} \eperp, \ell {\cdot} v_1)= \widetilde C(r,\sqrt{1 + r^2})\,.
\end{align}
It can be expanded as a double series
\begin{align}
\widetilde C(r,\sqrt{1 + r^2})= \sum_{n=-2}^\infty \sum_{k=0}^\infty  \widetilde C_{nk} \, r^n \,  \sqrt{1 + r^2}^k\,,
\end{align}
which can be tensor-reduced to a scalar triangle coefficient using \eqn{TensorReuctionTriangle}. We note that all odd powers in $n$ and $k$ integrate to zero, and the even powers satisfy the reduction formula
\begin{align} \label{NewTensorReductionFormula}
I_{\bigtriangleup}\big[r^{2n}  \sqrt{1 + r^2}^{2k}\big] ~~\to ~~(-1)^{n} \frac{(2 n - 1)!! (2 k - 1)!!}{(2 n +2 k)!!}I_{\bigtriangleup}[1]\,, ~~~~~~~(k \ge 0)
\end{align}
which is remarkably symmetric in $n$ and $k$. For $k=0$ we recover \eqn{TensorReuctionTriangle}, and the $k\ne0$ cases are simple to derive from \eqn{TensorReuctionTriangle}. We will not need $k<0$, but for completeness, the reduction of such negative powers is set to zero by hand.

Now let us apply this new reduction formula to a toy example, we consider an exponential function in the triple-cut integrand
\begin{align}
\widetilde C(r,\sqrt{1 + r^2}) =e^{x_1 r + x_2 i \sqrt{1 + r^2}}\,,
\end{align}
where ${\bf x}=(x_1,x_2)$ are some dummy variables (we use notation suggestive of 2D Cartesian coordinates). The scalar triangle coefficient of the toy example is then a functional transform of the exponential using the above integration rule, giving
\begin{align}
c_{\bigtriangleup} =\sum_{n=0}^\infty\sum_{k=0}^\infty (-1)^{n+k} \frac{(2 n - 1)!! (2 k - 1)!!}{(2 n +2 k)!!}\frac{x_1^{2n}}{(2n)!}\frac{x_2^{2k}}{(2k)!} =J_0\big(|{\bf x}|)\,,
\end{align}
where $|{\bf x}|=\sqrt{x_1^2+x_2^2}$. Namely, the transformed function is the zeroth-order Bessel function of the first kind,
\begin{align}
J_0\big(|{\bf x}|):=\sum_{n=0}^\infty\frac{(-1)^{n}}{(n!)^2 } \Big(\frac{|{\bf x}|}{2}\Big)^{2 n} =\sum_{n=0}^\infty\sum_{k=0}^\infty  \frac{(-1)^{n+ k}}{n! k!(n + k)! } \Big(\frac{x_1}{2}\Big)^{2 n} \Big(\frac{x_2}{2}\Big)^{2 k}\,.
\end{align}
It turns out that several terms in the Kerr triple-cut integrand are exponentials of the above form, however, with rational prefactors that depend on $r$ and $\sqrt{1 + r^2}$. These prefactors can be considered to be operators that act on the exponentials, hence after transforming the integrand we expect the full triangle coefficient to be expressible as operators acting on Bessel functions, 
\begin{align}
r \times J_0\big(|{\bf x}|)& := \frac{\partial J_0}{\partial x_1} = - \frac{x_1 J_1(|{\bf x}|)}{|{\bf x}|} \,, \nn \\
i\sqrt{1 + r^2} \times J_0\big(|{\bf x}|)& := \frac{\partial J_0}{\partial x_2}=-\frac{ x_2 J_1(|{\bf x}|)}{|{\bf x}|}\,,
\end{align}
Thus, for the repeated derivatives, it is useful to introduce the $n$th-order Bessel function of the first kind
\begin{align}
J_n\big(|{\bf x}|):=\sum_{k=0}^\infty\frac{(-1)^{k}}{k!(k+n)! } \Big(\frac{|{\bf x}|}{2}\Big)^{2 k+n}\,,
\end{align}
which satisfy $J_{-n}=(-1)^n J_n$. And for the integration operators
\begin{align}
\frac{1}{r} \times J_0\big(|{\bf x}|)& := \int dx_1 J_0\,, 
\end{align}
we define a family of new ``Bessel-like'' functions of two variables
\begin{align} \label{GeneralJdef}
J_{n,k}(x_1,x_2):=\sum_{l=-1}^\infty\sum_{m=0}^\infty (-1)^{l+m} \frac{(2 l - 1)!! (2 m - 1)!!}{(2 l +2 m)!!}\frac{x_1^{2l-n}}{(2l-n)!}\frac{x_2^{2m-k}}{(2m-k)!}\,,
\end{align}
such that $J_{0,0}(x_1,x_2)= J_0\big(|{\bf x}|)$, $J_{n,k}\big(x_1,x_2)= \partial_{x_1}^n\partial_{x_2}^k J_0\big(|{\bf x}|)$, and indefinite integration now corresponds to taking anti-derivatives
\begin{align}
\int dx_1 J_0 = J_{-1,0}\big(x_1,x_2)\,, \nn \\
\int dx_1 \int dx_1 J_0 = J_{-2,0}\big(x_1,x_2)\,.
\end{align}
We will not need more than two integrations of the $x_1$ variable, since we have at most a double pole in $r$ in the triple-cut integrand. Note that this definition of $J_{-2,0}$ requires that we are careful about the integration constants, hence we use the $l=-1$ lower bound of \eqn{GeneralJdef}. (All integration constants are correct if the summations are extended over all integers, positive and negative, but in practice we only need the $l=-1$ case.)

Next, we need to deal with the fact that the entire functions \eqref{eq:clEfns} have denominator factors. In particular, the first such expression is the entire function that makes up half of $E(x,y,z)$. Let us denote it by
\be
 \Omega(x,y,z):=\frac{e^y - e^x \cosh z + (x - y) e^x \, {\rm sinhc}\,z}{(x - y)^2 - z^2}\,,
\ee
and then $E(x,y,z)=\Omega(x,y,z)+\Omega(x,-y,z)$, and $\tilde E(x,y,z)=\frac{1}{2y}\Omega(x,y,z)-\frac{1}{2y}\Omega(x,-y,z)$.
We note that it has the simple series expansion
\be
 \Omega(x,y,z)=-\frac{e^y}{2 z}
   \sum_{n=0}^{\infty} \frac{(x - y - z)^n - (x - y + z)^n}{(n + 1)!}\,,
\ee
which removes the complicated denominator.  The $z$ in the denominator is harmless, as it is again equivalent to dividing with the $r$ variable, which can be implemented by integration.

We can further simplify the sum as
\be \label{SimplifiedContact}
 \sum_{n=0}^{\infty} \frac{(x - y - z)^n}{(n + 1)!} = \frac{e^{x - y - z}-1}{x - y - z} = \int_0^1 dt \, e^{t(x - y - z)}\,,
\ee
which exposes that it is equivalent to an unit-interval integral of an exponential function. Next, recall that the $(x\pm y\pm z)$ factors are linear functions the rescaled loop-momentum variables $r$ and $\sqrt{1+r^2}$, as 
\begin{align}
&x \to -i \sqrt{1 + r^2} a_1\cdot \eperp +  r  \sigma a_1{\cdot}  v_2   |q|  \,, \nn \\ 
&\ell\cdot a_2 \to -i \sqrt{1 + r^2} a_2\cdot \eperp -   r  a_2{\cdot}  v_1 |q|\,, \nn \\
&  y \to a_1 \cdot q\,, \nn \\
&  z \to |a_1| |q| r \sinh\zeta\,, \nn 
\end{align}
where we also included the $\ell\cdot a_2$ factor that always appears inside an overall exponential function in \eqn{eq:triplecut}. 

Thus, we are ready to study a toy example that captures the more intricate behavior of the contact terms of the triple cut. The toy integrand has the form
\be\label{Toy2integrand}
\widetilde C(r,\sqrt{1+r^2}) = e^{x_1 r+ x_2 i\sqrt{1+r^2}}\sum_{n=0}^{\infty}\frac{(y_1 r+ y_2 i\sqrt{1+r^2} +y)^n}{(n+1)!}= \int_0^1 dt \, e^{( {\bf x}+{\bf y} t)\cdot(r,\, i\sqrt{1+r^2}) + y t }\,,
\ee
where ${\bf x}=(x_1,x_2)$ and ${\bf y}=(y_1,y_2)$ are 2-vector dummy variables. In the last equality, we applied an integration over the unit interval to recover an exponential form.

After integrating out the loop momenta $r$ using the tensor reduction formula \eqref{NewTensorReductionFormula}, the function is transformed to the following triangle coefficient:
\begin{align} \label{K0def}
c_{\bigtriangleup} & =\int_0^1 dt \, J_{0}\big(|{\bf x}+{\bf y} t|\big) e^{y t} \nn \\
&=\sum_{n,k,l,j=0}^{\infty}\frac{(-1)^{n + j + k}}{(2j + l + k + 1)(n + j + k)!} 
\frac{\big(\frac{|{\bf x}|}{2}\big)^{2n}}{n!} 
\frac{\big(\frac{|{\bf y}|}{2}\big)^{2j}}{j!} 
\frac{\big(\frac{{\bf x} \cdot{\bf y}}{2}\big)^{k}}{k!} 
\frac{y^l}{l!} =: K_0({\bf x},{\bf y},y)\,,
\end{align}
where we defined it to be a new function in our arsenal. The function $K_0$ satisfies the following differential equation with a Bessel source:
\be
(y_1 \partial_{x_1}+y_2 \partial_{x_2}+y)K_0({\bf x},{\bf y},y) = J_0(|{\bf x}+{\bf y}|)e^y-J_0(|{\bf x}|)\,,
\ee
where the operator prefactor cancels out the first denominator in the summand of \eqn{K0def}, and it is equivalent to the denominator factor in \eqn{SimplifiedContact}.

Furthermore, we are interested in derivatives and integrations in the two dummy variables $x_1,x_2$, since as before this will be equivalent to multiplying/dividing with $r$ and $\sqrt{1+r^2}$ before transforming the triple-cut integrand. The derivatives can be carried out via the chain rule, but the integrations require that we introduce a new family of functions
\begin{align} \label{GeneralKdef}
 K_{n,k}({\bf x},{\bf y},y):= \!\!\!\!\!\sum_{\substack{m,j_1,j_2,j=0\\ l=-1}}^\infty \! \frac{(-1)^{l + m}(2 l - 1)!! (2 m - 1)!!}{(j + j_1 + j_2 + 1)  (2 l + 2 m)!!}
\frac{x_1^{2l-n-j_1}}{(2l-n-j_1)!}\frac{x_2^{2m-k-j_2}}{(2m-k-j_2)!}\frac{y_1^{j_1}}{j_1!} \frac{y_2^{j_2}}{j_2!} \frac{y^{j}}{j!} ,
\end{align}
%
The new functions satisfy $K_{0,0}=K_{0}$ and $K_{n,k}=\partial_{x_1}^n\partial_{x_2}^k K_{0}$, where as before the negative powers correspond to integration. Again, we will need at most two integrations in the variable $x_1$, thus we picked the lower bound $l=-1$ in order to obtain the correct integration constants. (As before, the $l,m$ sums can be extended to all integers.)

Similar to the $K_0$ function, the full family obey a corresponding differential equation with Bessel-like source,
\be
(y_1 \partial_{x_1}+y_2 \partial_{x_2}+y)K_{n,k}({\bf x},{\bf y},y) = J_{n,k}({\bf x}+{\bf y})e^y-J_{n,k}({\bf x})\,,
\ee
or, equivalently, in terms of shifted indices
\be
 y_1 K_{n+1,k}({\bf x},{\bf y},y)+y_2 K_{n,k+1}({\bf x},{\bf y},y)+y K_{n,k}({\bf x},{\bf y},y) = J_{n,k}({\bf x}+{\bf y})e^y-J_{n,k}({\bf x})\,.
\ee
It also follows that $K_{n,k}$ is an integral of $J_{n,k}$ over the unit interval
\be \label{GeneralKintegral}
K_{n,k}({\bf x},{\bf y},y)= \int_0^1 dt \, J_{n,k}\big({\bf x}+{\bf y} t\big) e^{y t} = \partial_{x_1}^n\partial_{x_2}^k \int_0^1 dt \, J_{0}\big(|{\bf x}+{\bf y} t|\big)e^{y t} \,,
\ee
and thus all are related to the Bessel $J_0$. While we have defined $K_{n,k}$ to all orders (in spin multipoles) in \eqn{GeneralKdef}, in practice it is sometimes easier to use \eqn{GeneralKintegral} by series expanding $J_{0}\big(|{\bf x}+{\bf y} t|\big)e^{y t}$ to the needed multipole order, after which the $t$-monomials can be trivially integrated, $\int_0^1 dt \, t^n = 1/(n+1)$. As before, for negative indices of $K_{n,k}$ it is best to use the formula \eqref{GeneralJdef} for $J_{n,k}$ to get the correct integration constants.  

\subsubsection*{All-order triangle coefficients:}

We can now work out the triangle coefficients coming from the first term in the all-order Compton amplitude, namely the $e^{x} \cosh z$ term in \eqn{eq:classicalcompton} (also in \eqn{triple_cut_functions}). This term is of exponential form; thus, we expect the Bessel-like functions $J_{n,k}$ of two variables to appear. 
Indeed, the all-order triangle coefficient is
\begin{align} \label{FirstTerm}
c_{\bigtriangleup,1}= \frac{m_1^2 m_2^4 \sigma^4}{\sinh^2 \zeta}
  \Big(1 + i \tanh\zeta \, \partial_2 \Big)^4  \Big(J_{-2,0}(\tau_+,\varepsilon) +J_{-2,0}(\tau_-,\varepsilon) \Big)\,,
\end{align}
where we used the shorthands $\varepsilon: =a \cdot \eperp= a_1 \cdot \eperp +a_2 \cdot \eperp$, and also
\be
\tau_\pm:=|q| \Big(\frac{ a \cdot (v_2 \sigma -v_1)}{\sinh\zeta}
      \pm |a_1|\sinh\zeta \Big)=|q|  \big(a \cdot \check v_1
      \pm |a_1|\big) \sinh\zeta \,,
\ee
where $\check v_1^\mu$ is the so-called dual velocity that has the property $\check v_1 \cdot v_1=1$ and $\check v_1 \cdot v_2=0$, see appendix~\ref{ap:FourierTransform} .
Note that in \eqn{FirstTerm} the derivatives $\partial_i$ shifts the indices $(\partial_1)^n (\partial_2)^k  J_{-2,0}=J_{-2+n,k}$. The fourth power of the operator in \eqn{FirstTerm} can be traced back to the spinor-helicity factor $\langle 3|1|4]^4$ of the tree-level Compton amplitude, or equivalently $\langle \ell_1|p_1|\ell_2]^4$ of the one-loop integrand. Note that the two helicity configurations, $C^{-+}$ and $C^{+-}$, contribute equally. This holds for all triangle coefficients since the two helicity configurations are related by loop momentum reversal $\ell\to-\ell$, and only even terms survive the tensor reduction.  

The second term in \eqn{eq:classicalcompton}, $w e^{x} {\rm sinhc}\, z$, is also of exponential form, only with a slightly more complicated $w$-prefactor, which can be implemented as an operator. Thus, we again get the all-order triangle coefficient in terms of the Bessel-like functions,
\begin{align} \label{eq:trigcoef2}
c_{\bigtriangleup,2}& = -\frac{m_1^2 m_2^4 \sigma^3 }{\sinh^2 \zeta }\frac{a {\cdot} v_2 }{ |a_1|}
\big(1 {+} i \tanh\zeta  \partial_2 \big)^3  \Big( 1  {+} 
  \frac{i \partial_2}{\tanh \zeta } {-} 
     \frac{a_1 {\cdot} \eperp}{a {\cdot} v_2} \frac{i\partial_1}{|q|} \! \Big)  \Big(\!J_{-2,0}(\tau_+,\varepsilon)-J_{-2,0}(\tau_-,\varepsilon) \!\Big), 
\end{align}
where the expression inside the second set of parentheses roughly corresponds to the $w$ factor implemented as an operator.  

All the remaining terms of the triple cut are contact terms, which for us means the terms captured by the entire functions $E,\tilde E,{\cal E},\tilde {\cal E}$; thus we set $\alpha=0$. These contact terms can be combined into a compact expression, which can be schematically written, using our Compton variables, as
\begin{align}
C^{-+} ~\sim~ \frac{e^{-y}}{4 y}   (\hat w^2 - \hat z^2) &\Big((w - x + y) +\frac{1}{2}q^2(\hat w^2 - \hat z^2)(\partial_x+\eta \partial_z)\Big)\sum_{n=0}^\infty\frac{(x + y + z)^n}{z (n+1)!} \nn \\& + \left(\begin{smallmatrix} z \to -z\\\eta \to -\eta  \end{smallmatrix}\right)+ \big(y \to -y\big)\,,
\end{align}
where the hatted variables have been stripped of $\ell \cdot v_1$ factors. Note that because of the symmetric appearance of the variables in the summand, we can use $\partial_x=\partial_y$ and $\partial_z=\partial_y-\frac{1}{z}$ so that the derivatives do not act on the loop momentum hidden inside $x$ and $z$. 

We can thus write all of the remaining triangle coefficients (for $\alpha=0$ and $\eta\neq0$) as
\begin{align}\label{eq:trigcoef3}
c_{\bigtriangleup,3}& =\frac{m_1^2 m_2^4 e^{-y}}{2 y |a_1| |q|} \frac{\mathfrak{W}^2 - \mathfrak{Z}^2}{ \sinh\zeta} \Big[ \mathfrak{U}\big( \sinh\zeta \, \mathfrak{W} \partial_1-\mathfrak{X} \mathfrak{U}\big) - \frac{1}{2}(\mathfrak{W}^2 -\mathfrak{Z}^2)\mathfrak{D}\Big] K_{-1,0}\big(\tau_2, \varepsilon_2;\tau_{1+}, \varepsilon_1;y)\nn \\
& \hskip3.5cm + \left(\begin{smallmatrix} |a_1| \to -|a_1|\\\eta \to -\eta  \end{smallmatrix}\right)+ \big(y \to -y\big)\,,
\end{align}
with derivative operators 
\begin{alignat}{3}
&\mathfrak{W}:= |q| a {\cdot} v_2\Big(1  + 
  \frac{i \partial_2}{\tanh \zeta } - 
     \frac{a_1 {\cdot} \eperp}{a {\cdot} v_2} \frac{i\partial_1}{|q|} \Big)\,, \qquad  &&\mathfrak{U}&&:= \cosh \zeta  \big(1 + i \tanh\zeta  \partial_2 \big)\,, \nn
 \\
&\mathfrak{X}:= -y  + a_1\cdot\eperp  \partial_2+ \frac{
 |q|  a{\cdot} v_2 }{\tanh \zeta} \partial_1\,,
\qquad &&\mathfrak{D}&&:= (1 + \eta) \partial_y - \frac{ \eta}{|a_1| |q|\sinh\zeta}   \partial_1^{-1}\,, \nn \\
&\mathfrak{Z}:=  |a_1| |q| \mathfrak{U}\,,  && &&
\end{alignat}
where, as before, the derivatives $\partial_1,\partial_2$ act on the two indices $(\partial_1)^n (\partial_2)^k  K_{-1,0}=K_{-1+n,k}$ (or, equivalently, first two arguments), and $\partial_y$ acts on the fifth argument.
The five arguments of the $K$ function are
\begin{align}
&y:=a_1{\cdot} q\,,~~~~~~~\tau_{1\pm}:=\frac{|q| a_1 {\cdot} v_2}{\tanh\zeta} \pm |a_1||q| \sinh\zeta~=~ |q| (a_1 {\cdot} \check v_1 \pm |a_1| )\sinh\zeta\,, 
\nn \\
&\varepsilon_i:=a_i{\cdot}\eperp\,,~~~~~\tau_2:= -\frac{|q| a_2 {\cdot} v_1 }{\sinh\zeta}~=~|q| a_2 {\cdot} \check v_1  \sinh\zeta\,,
\end{align}
and the previously introduced variables are related through $\varepsilon=\varepsilon_1+\varepsilon_2$ and $\tau_\pm= \tau_{1\pm}+\tau_{2}$.

Finally, the total triangle coefficient for the Kerr black hole, using the Compton amplitude~\eqref{eq:classicalcompton}, is given by the sum of the above three contributions,
\be
c_{\bigtriangleup}^{\rm Kerr} := c_{\bigtriangleup,1}+c_{\bigtriangleup,2}+c_{\bigtriangleup,3}\,.
\ee
Recall that these multiply the scalar triangle integral, such that the total triangle contribution to the one-loop amplitude is
\be
c_{\bigtriangleup}^{\rm Kerr} I_{\bigtriangleup}[1]+c_{\bigtriangledown}^{\rm Kerr} I_{\bigtriangledown}[1]\,,
\ee
where $I_{\bigtriangleup}[1]=-i/(32m_2|q|)$, and the upside-down triangle contribution is given by the swap of the two black holes,
\be
c_{\bigtriangledown}^{\rm Kerr} I_{\bigtriangledown}[1]:= c_{\bigtriangleup}^{\rm Kerr} I_{\bigtriangleup}[1]\Big|_{m_1\leftrightarrow m_2, a_1\leftrightarrow a_2, v_1\leftrightarrow v_2, q\to -q, \eperp \to \eperp,\zeta \to \zeta }\,.
\ee
This completes the computation of the classical one-loop amplitude for two Kerr black holes to all orders in spin.

Before moving on, let us also briefly mention the aligned-spin scenario: $a_1^\mu \propto a_2^\mu \propto L^\mu$. We note that some simplification occurs, since $a_i\cdot v_j=0$ and $\tau_{+}= - \tau_{-}=|q||a_1|\sinh \zeta$, giving
\begin{align}
c_{\bigtriangleup,1}^{\rm aligned}&= \frac{2m_1^2 m_2^4 \sigma^4}{\sinh^2 \zeta}
  \Big(1 + i \tanh\zeta \, \partial_2 \Big)^4  
J_{-2,0}(|a_1||q|\sinh \zeta,\,\varepsilon)\,, \nn \\
c_{\bigtriangleup,2}^{\rm aligned}& = \frac{2m_1^2 m_2^4 \sigma^3 }{\sinh^2 \zeta }\frac{i a_1 {\cdot} \eperp }{|a_1||q|}
\big(1 + i \tanh\zeta  \, \partial_2 \big)^3   J_{-1,0}(|a_1||q|\sinh \zeta,\,\varepsilon), 
\end{align}
where $J_{-2,k}$ is even and $J_{-1,k}$ odd in the first argument. The third triangle coefficient $c_{\bigtriangleup,3}$ also has some simplification, but it is less apparent. This matches low-order results known in the literature, \eg \cite{Chen:2021kxt, Bautista:2023szu,Bern:2022kto} and we provide expressions for $\alpha = 0$ up to $\mathcal{O}(S^{11})$ in the ancillary files.
\section{Observables for 2PM Kerr}
\label{sec:observables}

We now use some of our all-order-in-spin one-loop coefficients to compute observables at 2PM. We focus on impulse, scattering angle and eikonal phase.


\subsection{Impulse from the KMOC formalism}
The Kosower-Maybee-O'Connell formalism~\cite{Kosower:2018adc,Maybee:2019jus} conveniently evaluates classical observables from standard quantum scattering amplitudes, effectively relying on the Schwinger-Keldysh (in-in) prescription.
We start with a brief review.

The change of a classical observable $\mathcal{O}$, due to a scattering event, is found by computing the difference of the expectation values for a corresponding operator $\mathds{O}$ in the asymptotic future and asymptotic past,
\begin{equation}
\Delta\mathcal{O} = \langle \psi_\mathrm{out}|\mathds{O}|\psi_\mathrm{out}\rangle - \langle \psi_\mathrm{in}|\mathds{O}|\psi_\mathrm{in}\rangle.
\end{equation}
The out-states $|\psi_\mathrm{out}\rangle$ are related to the in-states $|\psi_\mathrm{in}\rangle$ via the $S$-matrix $|\psi_\mathrm{out}\rangle = S|\psi_\mathrm{in}\rangle$. Explicit definitions of the external wave-function packets are found in \rcite{Kosower:2018adc}.
The $S$-matrix itself
is expanded into free propagation and the interaction piece as $S=1 +i T$. In particular, the classical impulse is given by plugging in the momentum operator of the first body $\mathds{P}_1^{\mu}$,
\begin{equation} \label{eq:ImpulseDefn}
    \Delta p^{\mu}:= \Delta p_1^{\mu} = i\langle\psi_\mathrm{in}|[\mathds{P}_1^{\mu}, T]|\psi_\mathrm{in}\rangle + \langle \psi_\mathrm{in}| T^{\dagger}[\mathds{P}_1^{\mu}, T]|\psi_\mathrm{in}\rangle.
\end{equation}
In the classical limit, with appropriate wave packets, the expectation values correspond to Fourier transforms into impact-parameter space of a real and virtual kernel 
\begin{align}\label{eq:generalImpulse}
    \Delta p^{\mu} & = i\int d^{D}\!\mu \, e^{i b\cdot q}(K_\mathrm{v}^{\mu} + K_\mathrm{r}^{\mu})\,,
\end{align}
corresponding to the two terms in \eqn{eq:ImpulseDefn}, with respect to the transverse measure
\begin{align} \label{measure}
 d^{D}\!\mu  & := \frac{d^D q}{(2\pi)^{D-2}}\delta(2 p_1 \cdot q)\delta(2 p_2 \cdot q)\,.
\end{align}
The virtual kernel $K_\mathrm{v}^{\mu}$ is given by the 2-to-2 amplitude multiplied by the transfer momentum,
\begin{equation}\label{eq:virtualkernel}
\begin{split}
    K_\mathrm{v}^{\mu} &:= q^{\mu}\big\langle p_1{-}\tfrac{q}{2},p_2{+}\tfrac{q}{2}|T|p_1{+}\tfrac{q}{2},p_2{-}\tfrac{q}{2}\big\rangle
 ~   = ~~~  q^{\mu}\times \!\!\! \raisebox{-40pt}{\includegraphics[]{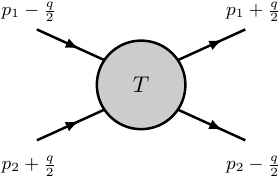}}
\end{split}\,.
\end{equation}
The real kernel $K_\mathrm{r}^{\mu}$ is given by the product of two amplitudes with all intermediate states and momenta summed/integrated over, with the appropriate transfer momenta of the first amplitude inserted, 
\begin{equation}\label{eq:realkernel}
\begin{split}
    K_\mathrm{r}^{\mu} =&  ~\mathop{\int\!\!\!\!\!\!\!\!\!\sum }_X \int \! d\Phi(l_1)d\Phi(l_2)  (l_1{-}p_1{+}\tfrac{q}{2})^{\mu} \langle p_1{-}\tfrac{q}{2},p_2{+}\tfrac{q}{2}|T^{\dagger}|l_1,l_2,{X}\rangle\langle l_1,l_2,{X}|T|p_1{+}\tfrac{q}{2},p_2{-}\tfrac{q}{2}\rangle\\
    =& -i\int d\Phi(l_1)d\Phi(l_2)\,(l_1-p_1+\tfrac{q}{2})^{\mu} \, \times \!\! \hspace{3pt} \raisebox{-40pt}{\includegraphics[]{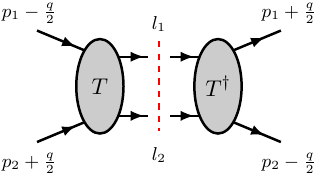}} ~+~{\cal O}(G^3).
    \end{split}
\end{equation}
Here $l_1$ and $l_2$ denote on-shell momenta of the massive states with phase-space measure $d\Phi(l_i) = d^4 l_i \, \delta(l_i^2-m_i^2)\Theta(l_i^{0})/(2\pi)^3$. The sum/integral over $X$ implies a phase-space integration over any intermediate graviton states, assuming a purely gravitational theory and no virtual black-hole states. The real kernel first contributes at ${\cal O}(G^2)$ for no intermediate gravitons, as these contribute from ${\cal O}(G^3)$. 

Computing the $n$PM impulse requires constructing the $(n{-}1)$-loop amplitude in $K_\mathrm{v}^{\mu}$, while $K_\mathrm{r}^{\mu}$ gets contributions from products of lower loop amplitudes sewn together with possible intermediate graviton states. Since in this paper we are interested in $2$PM impulse, all of the necessary contributions can be extracted from the 2-to-2 one-loop amplitude, which we already decomposed into convenient master integrals~\eqref{eq:masterdecomp}. 

For the virtual kernel, scalar triangles and scalar boxes are needed (the vector boxes integrate to zero). The scalar triangles only give rise to classical contributions, whereas the boxes contain pieces that diverge in the classical limit. The box and crossed box of scalar type start contributing at hyper-classical order $\mathcal{O}(1/\hbar)$, where they satisfy $ c_{\, \square} = c_{\cSquare}$. The next non-vanishing order for the boxes is at $\mathcal{O}(\hbar)$, and hence can be ignored. 

For the real kernel, the relevant information can be worked out from the $s$-channel cut of the master integrals. Due to the extra insertion of loop momenta in \eqn{eq:realkernel}, the reduction to scalar master integrals has to be performed again, and the otherwise vanishing vector box $I_{\square}[\ell\cdot \eperp]$ gets resurrected. Since the loop-momentum insertion increases the $\hbar$ counting, the vector boxes now only give classical contributions.  

To evaluate the real kernel \eqref{eq:realkernel}, we make the identification $\ell^{\mu} = 2 (l_1-p_1)^{\mu}$ such that the momentum insertion in front of the cut diagram becomes $(\ell+q)^{\mu}/2$, matching the parametrization in \Fig{fig:paramterization}. This relabeling of loop momenta induces overall factors of 2 that should not be forgotten in the measure of \eqn{eq:realkernel}; these factors are already seen in \eqn{eq:Qintegrals}. We can then expand $\ell^{\mu}$ into a basis of four external vectors $v_1^\mu$, $v_2^\mu$, $q^{\mu}$ and $\eperp^\mu$,
\begin{equation}\label{eq:lexpansion}
\ell^{\mu} = \frac{\sigma \ell\cdot  v_2 - \ell\cdot  v_1}{(\sigma^2-1) } v_1^{\mu} + \frac{ \sigma \ell\cdot  v_1 - \ell\cdot  v_2}{(\sigma^2-1)} v_2^{\mu}+ \frac{\ell\cdot q}{q^2}q^{\mu}+\frac{\ell \cdot \eperp}{q^2}\eperp^\mu.
\end{equation}
Based on that, it is convenient to split the impulse into the \textit{transverse} contribution (in the direction of $q^\mu$ or $\eperp^\mu$) and the \textit{parallel} contribution (in the direction of $v_1^{\mu}$ and $v_2^\mu$),
\begin{align}
    \Delta p^\mu &=\Delta p_{\parallel}^{\mu}+ \Delta p_{\perp}^{\mu} \nn \\ & =\Delta p_{\parallel}^{\mu}+ \Delta p_{q}^{\mu} +   \Delta p_{e_\perp}^{\mu}\,.
\end{align}
As shown later in this section, each term on the last line has a natural correspondence to the scalar box, scalar triangle, and vector box, respectively. 

We can also rewrite the result using a natural orthogonal basis $\{P^\mu, \pCM^\mu, \hat{b}^{\mu},\hat{L}^{\mu} \}$, where
\begin{align}\label{eq:orthobasis}
{\rm parallel}:&~~~~~P^\mu:=p_1^{\mu}+p_2^{\mu}\,,~~~~ 
\pCM^\mu := \frac{m_1 m_2}{s}\Big[ ( m_1 \sigma +  m_2) v_1^{\mu}-( m_2 \sigma +  m_1)v_2^\mu\Big]\,,\nn \\
{\rm transverse}:&~~~~~\hat{b}^{\mu} := \frac{b^{\mu}}{|b|}\,,~~~~~~~~~~~~~~~~ 
\hat{L}^{\mu}:= (\star \hat b)^{\mu} \,, 
\end{align}
which is natural for the fully integrated expressions. Since $b\cdot v_i = 0$, the unit vectors $\hat{b}^{\mu}$ and $\hat{L}^{\mu}$ are transverse to the plane spanned by $v_1$ and $v_2$. On the other hand, $\pCM^\mu$ corresponds to the spatial momentum of each body in the center-of-mass frame of the system, it is orthogonal to the total momentum of the system $P^\mu$, and its norm is $|\pCM|=m_1 m_2 \sqrt{\sigma^2-1}/E$, where the total energy is $E=\sqrt{s}= |p_1+p_2|$. Note that the parallel impulse will always be proportional to $\pCM^\mu$.  

Thus, in the above orthogonal basis, the impulse takes the form
\begin{align} \label{ScatteringAngleParametrization}
      \Delta p_{\parallel}^{\mu} &= \pCM^\mu (\cos\theta-1)\,, \nn \\
    \Delta p_{\perp}^{\mu} & = |\pCM| \sin\theta \, \big(\cos\tilde\theta \, \hat b^{\mu}+ \sin\tilde\theta \, \hat L^{\mu} \big)  \,,
\end{align}
where we have parametrized the components in terms of two unknowns, the aligned-spin scattering angle $\theta$ and a second spin-induced deflection angle $\tilde \theta$. The parametrization makes the on-shell conditions of the outgoing black holes manifest, namely $(p_1+\Delta p)^2=m_1^2$ and $(p_2-\Delta p)^2=m_2^2$. 

For later purposes, it is useful to consider how the dualized spin vector decomposes into the two transverse components,
\begin{align}
(\star a)^\mu  
=  \hat{b} \star a \, \frac{\hat{b}^{\mu}}{\hat{b}^2} + \hat{L} \star a \, \frac{\hat{L}^{\mu}}{\hat{L}^2}
=   a \cdot \hat{L} \, \hat{b}^{\mu} - a \cdot \hat{b}\, \hat{L}^{\mu}\,.
\end{align} 
If the spin is aligned with the orbital angular momentum, $a^\mu = |a| \hat L^\mu$, then we have $a \cdot \hat{b}=0$ and $a \cdot \hat{L}=-|a|$, so that $(\star a)^\mu=|a| \, \hat{b}^{\mu}$. In the case of aligned-spin kinematics, where both spin vectors $a_{1,2}^\mu$ are aligned with $L^\mu$, by total-angular-momentum conservation the impulse must be confined to the $(\pCM^\mu,\hat{b}^\mu)$-plane, such that the second angle will vanish, $\tilde \theta=0$.

\subsubsection*{${\bm 1}$PM impulse}
Before moving on to loop level, we first compute the $1$PM impulse, which only involves the virtual kernel \eqref{eq:virtualkernel}. Using the tree amplitude in \eqn{ClassicalTree2to2}, the Fourier integrals are simple, see appendix~\ref{ap:FourierTransform}, and the impulse can be compactly expressed as  
\begin{align}\label{eq:1PMimpulse}
    \Delta p^{\mu} =  \Delta p^{\mu}_\perp &= \frac{i \kappa^2}{4} m_1^2 m_2^2\int d^{D}\!\mu \, e^{i b\cdot q} \frac{q^\mu}{q^2}\Big(e^{-i e_\perp \cdot a+2\zeta }+e^{i e_\perp \cdot a-2\zeta}\Big) \nn \\
    &= -\frac{G m_1 m_2}{\sqrt{\sigma^2-1}}\bigg[ e^{2\zeta} \frac{(b+\star a)^\mu}{(b+\star a)^2} + e^{-2\zeta} \frac{(b-\star a)^\mu}{(b-\star a)^2} \bigg] \, .
\end{align} 
This result is well known, \eg it agrees with~\rcite{Vines:2017hyw}. Since the virtual kernel is proportional to transverse vectors, at this order there is no parallel contribution to the impulse.

\subsubsection*{Transverse impulse at $2$PM}
At $2$PM the full impulse has a contribution from both the virtual and real kernels. However, since the virtual kernel is in the $q$-direction, only $\Delta p_{q}^{\mu}$ receives a contribution from both kernels.

The decomposition of the classical one-loop integrand \eqref{eq:masterdecomp} allows us to effectively assume that the two exchanged gravitons are on shell.
In the real kernel, the cut conditions impose $\ell\cdot q \rightarrow 0$ such that the loop momentum $\ell^\mu$ has no contribution in the $q^\mu$ direction, as per \eqn{eq:lexpansion}. Therefore, in computing $\Delta p_{q}^{\mu}$, we can replace the prefactor of the cut $(\ell^\mu+q^\mu)/2 \to q^\mu/2$ and the transverse impulse in the $q$-direction therefore simplifies to
\begin{align}
    \Delta p_{q}^{\mu} = &\left( \frac{\kappa}{2}\right)^4 \int d^{D}\!\mu \, e^{i b\cdot q}q^{\mu}\Big( c_{\, \square}(I_{\, \square }+I_{ \cSquare})+c_{\bigtriangleup}I_{\bigtriangleup}+c_{\bigtriangledown}I_{\bigtriangledown} - i \frac{c_{\, \square}}{2} \mathrm{Cut}[I_{\, \square }]\Big ) \nn \\
    =&\left( \frac{\kappa}{2}\right)^4\int d^{D}\!\mu \, e^{i b\cdot q}q^{\mu}\Big( c_{\, \square}(I_{\, \square }+I_{ \cSquare} - i \mathrm{Im}[(-i)I_{\, \square }])+c_{\bigtriangleup}I_{\bigtriangleup}+c_{\bigtriangledown}I_{\bigtriangledown}  \Big)\\
    =&\left( \frac{\kappa}{2}\right)^4 \int d^{D}\!\mu \, e^{i b\cdot q}q^{\mu}\Big(c_{\bigtriangleup}I_{\bigtriangleup}+c_{\bigtriangledown}I_{\bigtriangledown}  \Big)\,.\nn
\end{align}
Thus all hyper-classical contributions from the scalar boxes cancel with the cut term, and this impulse contribution is fully determined by the triangles. We defer the explicit details of the remaining Fourier integral of the triangles to the next subsection, where we compute the closely related eikonal phase. The Fourier integrals are simple to perform at any given order in the spin-multipole expansion; see appendix~\ref{ap:FourierTransform} as well as the final result in the ancillary file. However, computing the integral in spin-resummed form involves Fourier-transforming the Bessel-like functions that appear in the triangle coefficients, which is somewhat challenging.

Next, the transverse impulse also receives a contribution in the direction $\eperp^\mu$, generated by the real kernel. This contribution only appears for spinning black holes, where spin-induced precession forces point out of the initial plane of scattering. 
The coefficient of $e_{\perp}^{\mu}$ in the loop momentum decomposition \eqref{eq:lexpansion} cannot be reduced by means of cut conditions as done with the coefficient of $q^{\mu}$ in the computation of $\Delta p_{q}^{\mu}$. Instead, we have to observe how the reduction to the basis of master integrals changes with an insertion of $\ell \cdot \eperp$. Its effect on the decomposition in \eqn{eq:masterdecomp} is to map 
\begin{align}
    I_\square[\ell\cdot\eperp] &\rightarrow I_\square[(\ell\cdot\eperp)^2]=-q^4 I_\square\,.
\end{align}
The corresponding transverse contribution is thus
\begin{align}
      \Delta p_{e_\perp}^{\mu} &=\frac{i}{2} \left (\frac{\kappa}{2}\right)^4 \int d^{D}\!\mu \, e^{i b\cdot q}  \, q^2 \eperp^\mu \, \tilde c_{\square}\mathrm{Cut}[I_{\, \square }]\,.
\end{align} 
The tensor box coefficient ${\tilde c}_{\Box} := (c^{-+}_{\square}+c^{+-}_{\square})[\ell \cdot \eperp] $ is given in \eqn{eq:tensorboxcoef}, and it indeed vanishes in the spinless limit. Recall that while the loop integral $\mathrm{Cut}[I_{\, \square }]$ has a $1/\epsilon$-pole \eqref{eq:CutBoxInt}, this cancels in the above formula when the Fourier transform is consistently treated in dimensional regularization; for details, see appendix~\ref{ap:FourierTransform}. The contribution to the transverse impulse can now be computed to all spin-multipole orders, plugging in \eqn{eq:BoxIns} we obtain
\begin{align} \label{eq:ImpulseEperp}
   \Delta p_{e_\perp}^{\mu} =& \frac{G^2 m_1 m_2}{\left(\sigma ^2-1\right)^{2}} 
   \Big( a\cdot (p_1-p_2) \, 
   \Big[ e^{3\zeta}  \frac{ (\star b-\Pi a)^\mu }{(b+\star a)^4 }-e^{-3\zeta}  \frac{(\star b+ \Pi a)^\mu}{(b-\star a)^4 }\Big]\nn\\
   &\hskip2cm + a\cdot(m_2 v_1 - m_1 v_2)\,\Big[ e^{4\zeta}\frac{ (\star b-\Pi a)^\mu}{(b+\star a)^4 }-e^{-4\zeta} \frac{ (\star b+\Pi a)^\mu}{(b-\star a)^4 }\Big]\Big)\,,
\end{align}
where we used that $(\star b\pm\Pi a)= \star ( b\mp \star a)$, and the transverse projector is given by
\be
(\Pi)^\mu{}_\nu = - (\star^2)^\mu{}_\nu= \delta^\mu_\nu- v^\mu_i M^{ij}v_{j\nu}\,,
\ee
where $M^{ij}$ is the inverse Gram matrix of velocities, $(M^{-1})_{ij}= v_i\cdot v_j$. It projects to the $2$-dimensional subspace transverse to $v_1$ and $v_2$, see appendix~\ref{ap:FourierTransform} for the explicit form of $\Pi$. As discussed in section~\ref{sec:comparison}, this new all-order result reproduces known low-order results in the literature.

Note that we can slightly massage \eqn{eq:ImpulseEperp} to give an alternative presentation of this impulse, which makes use of the center-of-mass momentum $\pCM^\mu$, and the rapidity exponentials are now uniformly fourth powers. The alternative formula is
\begin{align}
\Delta p_{e_\perp}^{\mu} =\frac{G^2 m_1 m_2}{\sinh^3\zeta} 
  \Big[&  a\cdot (p_1 - p_2)  \Big(e^{4\zeta}  \frac{ (\star b-\Pi a)^\mu }{(b+\star a)^4 } + e^{-4\zeta}  \frac{ (\star b+\Pi a)^\mu }{(b-\star a)^4 }\Big) \nn \\
   &- 
 \frac{s \, a \cdot \pCM}{m_1 m_2 \sinh\zeta} \Big(e^{4\zeta}  \frac{ (\star b-\Pi a)^\mu }{(b+\star a)^4 } - e^{-4\zeta}  \frac{ (\star b+\Pi a)^\mu }{(b-\star a)^4 }\Big)
 \Big]\,.
\end{align}
Finally, we note that in the aligned-spin scenario, where $v_i \cdot a_j=0$ for any $i,j \in \{1,2\}$,  $\Delta p_{e_{\perp}}^{\mu}$ vanishes, which is natural since the absence of spin precession gives trajectories (orbits) that do not wobble, instead they are confined to the spatial plane orthogonal to the angular momentum. 
\subsubsection*{Parallel impulse at ${\bm 2}$PM}
For impulse directions parallel to the initial black-hole velocities $v_i^\mu$, only the real kernel, and hence the cut of the box, contributes. Concerning the insertion of $(\ell+q)^{\mu}/2$, the second and third terms in \eqn{eq:lexpansion} are relevant.  Since the cut conditions allow us to replace $v_1\cdot \ell\rightarrow q^2/(2  m_1)$ and $v_2\cdot \ell\rightarrow -q^2/(2  m_2)$ the integrand reduction is left unchanged. 
The parallel impulse is thus becomes
\begin{equation}
      \Delta p_{\parallel}^{\mu} =  i\left( \frac{\kappa}{2}\right)^4\int d^{D}\!\mu \, e^{i b\cdot q} \frac{  q^2 s}{4(\sigma^2-1) m_1^2  m_2^2}\pCM^\mu c_{\, \square}\mathrm{Cut}[I_{\, \square }]\,,
\end{equation}
where the center-of-mass momentum $\pCM^\mu$ was introduced in \eqn{eq:orthobasis}. Note that this impulse is orthogonal to the total momenta (total energy) $P^\mu=(p_1+p_2)^\mu$, with $s=P^2$, that defines the rest frame of the system. Hence, the parallel impulse does not cause a shift in the energy (time) direction of the rest frame, as expected.  

We can now work out the parallel impulse to all orders in spin, using \eqns{BoxCoeff1}{BoxCoeff2} the result is
\begin{align}
   \Delta p^\mu_{\parallel} =& \frac{G^2 s\, \pCM^\mu}{2\left(\sigma ^2-1\right)^2} \Big( \frac{1}{(b+i\Pi a)^2}
   + \frac{1}{(b-i \Pi a)^2}
   +\frac{e^{4\zeta}}{(b+\star a)^2}
   + \frac{e^{-4\zeta}}{(b-\star a)^2} \Big)\,.
\end{align}
Note that the first two terms come from the same-helicity Compton amplitudes, and the last two terms come from the opposite-helicity Compton amplitudes. Thus, contrary to commonplace statements, the same-helicity Compton amplitudes are not automatically quantum contributions.
While this all-order result for the parallel impulse is new, low-order pieces of this result have been previously given in the literature, \eg \cite{Bautista:2023szu}. More details of how our results compare to the literature are given in section~\ref{sec:comparison}.

Note that for aligned spin the parallel impulse simplifies to
\begin{align}
   \Delta p^\mu_{\parallel,{\rm aligned}} =& -\frac{G^2 s\, \pCM^\mu}{2\left(\sigma ^2-1\right)^2} \Big( \frac{2}{|b|^2-|a|^2}
   +\frac{e^{4\zeta}}{(|b|+|a|)^2}
   + \frac{e^{-4\zeta}}{(|b|-|a|)^2} \Big)\nn\\
   &= -\frac{G^2 s\, \pCM^\mu}{2\left(\sigma ^2-1\right)^2} \Big(
   \frac{e^{2\zeta}}{|b|+|a|}
   + \frac{e^{-2\zeta}}{|b|-|a|} \Big)^2
   \,,
\end{align}
which is the square of the 1PM impulse for aligned spin, up to overall simple factors. Indeed, the two impulse contributions are kinematically constrained, through \eqn{ScatteringAngleParametrization}, to be related as
\be
|\Delta p_{\parallel,{\rm aligned}}^{\rm 2PM}|= \frac{|\Delta p^{\rm 1PM}_{\rm aligned}|^2 }{2|\pCM|}\,,
\ee
and we recall $|\pCM|=m_1 m_2\sqrt{\sigma^2-1}/E$ and $E=\sqrt{s}$. 

More generally, for non-aligned spin, one can check that the following relation holds for our results: 
\be
2 \pCM \cdot \Delta p^{\rm 2PM}_\parallel= -(\Delta p^{\rm 1PM})^2\,,
\ee
which also follows from the kinematic constraints imposed by \eqn{ScatteringAngleParametrization}.

\subsubsection*{Explicit low spin-multipole results for ${\bm 2}$PM impulse}
For reference, we print here some results up to $\mathcal{O}(S^1)$ for each of the contributions,
\begin{align}
    \Delta p_{q,\,S^0}^\mu &=  \frac{3 \pi G^2 m_1^2 m_2}{4 |b|^2 \sqrt{\sigma^2-1}}(5 \sigma^2-1)\hat{b}^\mu - \left(1 \leftrightarrow 2\right)\,, \nn \\
    \Delta p_{q,\,S^1}^\mu &= \frac{\pi G^2 }{4 |b|^3 (\sigma^2-1)}\sigma(3 - 5 \sigma^2)\Big(2(4 a_1\cdot\hat{L}+3 a_2 \cdot \hat{L})\hat{b}^\mu+(4 a_1\cdot \hat{b}+ 3 a_2\cdot \hat{b})\hat{L}^\mu   \Big) \nn \\
    &- \left(1 \leftrightarrow 2\right)\,, \nn \\
    \Delta p_{\eperp,\,S^0}^\mu &= 0\,,\nn \\
    \Delta p_{\eperp,\,S^1}^\mu &= \frac{2 G^2}{|b|^3 (\sigma^2   -1)^{3/2}}\Big(4\sigma(1-2\sigma^2)a \cdot v_2+(4 \sigma^2 -1)a\cdot v_1  \Big)\hat{L}^\mu - \left(1 \leftrightarrow 2\right)\,, \nn \\
     \Delta p_{\parallel,\,S^0}^\mu &= -\frac{2 G^2 s}{|b|^2(\sigma^2-1)^2}(2\sigma^2-1)^2 \pCM^\mu\,, \nn \\
     \Delta p_{\parallel,\,S^1}^\mu &= \frac{8 G^2 s }{|b|^3 (\sigma^2-1)^{3/2}}\sigma(2\sigma^2-1)a\cdot\hat{L}\,\pCM^\mu\,.
\end{align}
Further explicit results for the impulse up to ${\cal O}(S^{11})$ are provided in the ancillary files of this paper. 
\subsection{Eikonal and scattering angle}
In the case of scattering of two spinless bodies, the motion is confined to a plane. Therefore, the scattering event is fully specified by the scattering angle $\theta$ (see \eg \cite{Bjerrum-Bohr:2018xdl}).
Although the orbit is not planar for generic spinning bodies, restricting to the aligned-spin scenario maintains planarity,
\be
a_1^\mu\propto a_2^\mu \propto L^\mu\,,~~~~~~ v_i\cdot a_j=0\,,~~~~~~  b\cdot a_i=0\,.
\ee
In this case, the scattering process is fully specified by the scattering angle $\theta$, which is given by the transverse impulse through~\eqn{ScatteringAngleParametrization},
\begin{equation}
    \sin{\theta} = -\frac{\Delta p_\perp \cdot \hat b}{|\pCM|} = - \frac{E \, \Delta p \cdot \hat{b}}{m_1 m_2\sqrt{\sigma^2-1}}\, ,
\end{equation}
and one can use $\sin{\theta}=\theta +{\cal O}(\theta^3)$ for all terms that are not iterations, such as the triangle contributions. 

An alternative way to compute $\theta$ is through the eikonal approach. At 1PM and 2PM, the eikonal phase $\chi$ is given by the Fourier-transformed aligned-spin amplitudes \cite{Bern:2020buy}, not including iteration terms such as boxes,
\begin{align}
    \chi_{\mathrm{1PM}} &= \int d^{D}\!\mu \, e^{i q\cdot b} \cM^{(0)}(q)\,, \nn \\
    \chi_{\mathrm{2PM}} &= \int d^{D}\!\mu \, e^{i q\cdot b} \cM^{(1)}_{\bigtriangleup+\bigtriangledown}(q)\,,
\end{align}
where the measure is given in \eqn{measure}. The scattering angle can be extracted through a derivative with respect to the impact parameter \cite{Kosmopoulos:2021zoq},
\begin{equation}
    \theta_{n\mathrm{PM}} = \frac{E}{ m_1  m_2 \sqrt{\sigma^2-1}}\frac{\partial}{\partial|b|}\chi_{n\mathrm{PM}}\,.
\end{equation}

We can work out the eikonal phase at 1PM using \eqn{ClassicalTree2to2}, it takes the simple form
\begin{align}\label{eq:1PMeikonal}
    \chi_{\mathrm{1PM}} &= \frac{\kappa^2}{4} m_1^2 m_2^2\int d^{D}\!\mu \,\frac{ e^{i b\cdot q} }{q^2}\Big(e^{-i e_\perp \cdot a+2\zeta }+e^{i e_\perp \cdot a-2\zeta}\Big) \nn \\
    &= -\frac{G m_1 m_2}{\sqrt{\sigma^2-1}}\bigg[ e^{2\zeta} \log |b+\star a| + e^{-2\zeta} \log |b-\star a|+ \dots \bigg]\,,
\end{align} 
where the ellipsis are (divergent) terms that are independent of $b^\mu$, and hence irrelevant. The $1$PM impulse \eqref{eq:1PMimpulse} then also follows via $\Delta p^{\mu}_{\mathrm{1PM}} = \partial\chi_{\mathrm{1PM}}/\partial b_{\mu}$, which kills the divergent terms.

At 2PM, we will give the full eikonal result using spin-multiple expanded formulae. However, it is interesting to first Fourier transform two of our simpler all-order-in-spin triangle coefficients, and inspect the partial result. The first triangle coefficient \eqref{FirstTerm} integrates to
\begin{align}
 \chi_{\mathrm{2PM},1}&:= -i \left(\frac{\kappa}{2} \right)^{4} \int d^{D}\!\mu \, e^{i q\cdot b} c_{\bigtriangleup,1} \mathcal{I}_{\bigtriangleup} \nn\\
 &=\frac{G^2 \pi m_1 m_2^2 \sigma^4}{2  |b|\sinh^3 \zeta} \Big(1 + i \tanh\zeta \, \partial_2 \Big)^4 \sum_{\pm}{\cal J}_{-2,0}\Big(\frac{\tilde\tau_\pm^2}{2|b|^2},\frac{a\star b}{|b|^2},\frac{(\star a)^2}{2|b|^2}\Big)\,.
\end{align}
where ${\cal J}_{n,k}$ is the Fourier transform of the Bessel-like function $J_{n,k}$. Using  ${\bf x}=(x_1,x_2,x_3)$, it can be written as 
\begin{align}
{\cal J}_{n,k}({\bf x})&:= |b| \int d^{D}\mu e^{i b \cdot q} |q|^{-1} J_{n,k}(|q|\tilde{\tau}, a \star q)\\
&=i^{n-k}\!\!\!\!\sum_{\substack{l=-1\\m,j=0}}^{\infty}\!\!\!\! \frac{(2 l {-} 1)!!(2 m + 2 j - 1)!! (2 (2m {+}  j {+} l{-}k)-n {-} 1)!!}{ (2 l {+} 2 m {+}  2 j)!!} \frac{x_1^{l-n/2}}{(l{-}n/2)!}\frac{x_2^{2m-k}}{(2m{-}k)!}\frac{x_3^j}{j!}, \nn
\end{align}
and we will only use $n=-2$ (the transformed $J_{n,k}$ vanish for odd $n$) and $k\geq 0$. Similarly to $J_{n,k}$, we use operators to indicate shifted indices $\partial_2^k{\cal J}_{-2,0}={\cal J}_{-2,k}$; these no longer correspond to derivatives of the arguments. 
The variable $\tilde\tau_\pm$ is defined analogously to $\tau_\pm= |q|\tilde\tau_\pm $, such that
\be
\tilde\tau_\pm:= ( a \cdot \check v_1 
      \pm |a_1|)\sinh\zeta \,.
\ee

For the $c_{\bigtriangleup,2}$ triangle coefficient \eqref{eq:trigcoef2}, one has to be more careful with the differential operators that act on $J_{n,k}$, since one of these prefactors now contains $q$. One can in principle introduce some new integral/differential operators, along the lines of
\be
-  a_1 {\cdot} \eperp  \frac{i\partial_1}{|q|} \to  |b| a_1
  {\cdot} \frac{\partial}{\partial a}
 \partial_{x_2} \int \frac{dx_1}{x_1}\partial_{x_1}\,;
\ee
however, this is a bit too cumbersome. Instead, we will introduce a slightly modified transformed function for this contribution. Thus, the Fourier-transformed triangle coefficient  $c_{\bigtriangleup,2}$ becomes
\begin{align}
 \chi_{\mathrm{2PM},2}&:= -i \left(\frac{\kappa}{2} \right)^{4} \int d^{D}\!\mu \, e^{i q\cdot b} c_{\bigtriangleup,2} \mathcal{I}_{\bigtriangleup} \nn\\
 &=-\frac{G^2 \pi m_1 m_2^2 \sigma^3}{2  \sinh^3 \zeta}
 \big(1 {+} i \tanh\zeta  \partial_2 \big)^3  \bigg[ \frac{a {\cdot} v_2 }{ |b||a_1|} \Big( 1  {+} 
  \frac{i \partial_2}{\tanh \zeta }\Big) {\cal J}_{-2,0}\Big(\frac{\tilde\tau_+^2}{2|b|^2},\frac{a\star b}{|b|^2},\frac{(\star a)^2}{2|b|^2}\Big) \nn \\& \hskip55mm
  + \frac{a_1}{ |a_1|}
  \cdot \frac{\partial}{\partial a}\widetilde {\cal J}_{-2,0}\Big(\frac{\tilde\tau_+}{|b|},\frac{a\star b}{|b|^2},\frac{(\star a)^2}{2|b|^2}\Big) \bigg] \nn \\
  & \hskip55mm
  + (|a_1|\to -|a_1|)\,,
\end{align}
where we assume that $\frac{\partial}{\partial a}$ only acts on the second and third arguments, so $\frac{\partial\tilde\tau_+}{\partial a}\to 0$. And the slightly modified transformed Bessel-like function is
\begin{align}
\tilde {\cal J}_{n,k}({\bf x}):= i^{n+k}\!\sum_{l,m,j=-1}^{\infty} &  \frac{(2 l {-} 1)!!(2 m + 2 j - 1)!! (2 (2m {+}  j {+} l{-}k)-n {-} 1)!!}{ (2 l {+} 2 m {+}  2 j)!!}\nn \\
&\times \frac{(2l{-}n-3)!!}{(2l{-}n-1)!}\frac{x_1^{2l-n-1} x_2^{2m-k+1} x_3^j}{(2m{-}k+1)! j!}\,,
\end{align}
where $n$ again has to be even. We will not attempt to Fourier transform the third triangle coefficient, $c_{\bigtriangleup,3}$, instead we will present some explicitly spin-multipole expanded results. 

\subsubsection*{Scattering angle results}
The eikonal for generic spin configurations is provided in the ancillary files up to $\mathcal{O}(S^{11})$. We print here the full results for the aligned-spin scattering angle up to $\mathcal{O}(S^{11})$. Note that all dissipative terms, \ie those proportional to $\eta$, drops out when specializing to the case of aligned spin. The results are
\begin{align}
\label{eq:covscattangle}
    \theta^{S^0}_{\mathrm{2PM}}=& -\frac{3 \pi G^2 m_1^2 m_2   }{4 \sqrt{\sigma^2-1}|b|^2}\Big[ 5\sigma^2-1\Big] + \left(1 \leftrightarrow 2\right)\,, \nn \\
    \theta^{S^1}_{\mathrm{2PM}}=&\frac{\pi G^2 m_1^2 m_2   }{2 \left(\sigma^2-1\right)|b|^3}\Big[\sigma\left(5\sigma^2-3\right) \left(4 |a_1|+3 |a_2|\right) \Big] + \left(1 \leftrightarrow 2\right) \nn \,,\\
     \theta^{S^2}_{\mathrm{2PM}}=& -\frac{3 \pi G^2 m_1^2 m_2 }{16 |b|^4 \sqrt{\sigma^2-1}^3} \Big[\left(95 \sigma^4 - 102 \sigma^2 + 15\right)|a_1|^2 + 8\left(20\sigma^4 - 21 \sigma^2 + 3\right)|a_1||a_2| \nn \\
     &+ 4 \left(15\sigma^4 -15 \sigma^2 + 2\right)|a_2|^2\Big]+ \left(1 \leftrightarrow 2\right) \,,\nn\\
 \theta^{S^3}_{\mathrm{2PM}}=&\frac{3  \pi G^2  m_1^2 m_2 \sigma}{4 |b|^5\left(\sigma^2-1\right)}  \Big[ 4\left(9\sigma^2-5\right)|a_1|^3 + \left(95\sigma^2-51\right)|a_1|^2 |a_2| \nn \\
     &+ 40\left(2\sigma^2 -1\right)|a_1| |a_2|^2 + 4\left(5\sigma^2-2\right)|a_2|^3  \Big] + \left(1 \leftrightarrow 2\right)\,,\nn\\
  \theta^{S^4}_{\mathrm{2PM}}=&   -\frac{5 \pi G^2 m_1^2 m_2 }{32 |b|^6 \sqrt{\sigma^2-1}^3}\Big[\left(239\sigma^4-250\sigma^2+35\right)|a_1|^4+24\left(36 \sigma^4 -37 \sigma^2+5\right)|a_1|^3 |a_2| \nn \\
  &+12\left(95 \sigma^4 - 95 \sigma^2 +12\right)|a_1|^2|a_2|^2 + 32\left(20 \sigma^4-19\sigma^2+2\right)|a_1||a_2|^3 \nn \\ 
  &+ 24\sigma^2\left(5\sigma^2-4\right)|a_2|^4\Big]+ \left(1 \leftrightarrow 2\right)\nn \\ 
  \theta^{S^5}_{\mathrm{2PM}}=& -\frac{3  \pi G^2  m_1^2 m_2 \sigma}{64 |b|^7\left(\sigma^2-1\right)}\Big[-80  \left(13 \sigma ^2-7 \right)|a_1|^5- 20  \left(239 \sigma ^2-125 \right)    |a_1|^4|a_2| \nn \\
  &-4320  \left(2 \sigma ^2-1 \right)    |a_1|^3|a_2|^2  -80  \left(95 \sigma ^2-44 \right)  |a_1|^2|a_2|^3  -640  \left(5 \sigma ^2-2 \right)  |a_1||a_2|^4 \nn \\
  &+ 3  \left(7 \sigma ^4-178 \sigma ^2+51 \right)    |a_2|^5 \Big]+ \left(1 \leftrightarrow 2\right)\,,\nn\\
  \theta^{S^6}_{\mathrm{2PM}}=& -\frac{7\pi G^2 m_1^2 m_2 }{512|b|^8 \sqrt{\sigma^2-1}^3}\Big[  30  \left(149 \sigma ^4-154 \sigma ^2+21 \right)|a_1|^6+480  \left(52 \sigma ^4-53 \sigma ^2+7 \right) |a_1|^5 |a_2|\nn \\
     &+240  \left(239 \sigma
   ^4-239 \sigma ^2+30 \right) |a_1|^4 |a_2|^2+1920  \left(36 \sigma ^4-35 \sigma ^2+4 \right) |a_1|^3
   |a_2|^3\nn \\
     &+480  \left(95 \sigma ^4-88 \sigma
   ^2+8 \right) |a_1|^2 |a_2|^4
   -12  \left(21 \sigma ^6-1333 \sigma ^4+1123 \sigma ^2-51 \right)|a_1| |a_2|^5\nn \\
     &+\left(105 \sigma ^8-574 \sigma ^6+2984 \sigma
   ^4-2026 \sigma ^2-9\right) |a_2|^6    \Big]+ \left(1 \leftrightarrow 2\right)\,,\nn\\
    \theta^{S^7}_{\mathrm{2PM}}=& \frac{ \pi G^2  m_1^2 m_2 \sigma}{64 |b|^9 \left(\sigma^2-1\right)}\Big[  280 \left(17 \sigma ^2-9\right)|a_1|^7+210 \left(149 \sigma ^2-77\right) |a_1|^6  |a_2|\nn \\
     &+43680 \left(2 \sigma ^2-1\right) |a_1|^5  |a_2|^2+560 \left(239
   \sigma ^2-114\right) |a_1|^4
    |a_2|^3+13440 \left(9 \sigma ^2-4\right) |a_1|^3  |a_2|^4\nn \\
     &-42 \left(15 \sigma ^4-1552 \sigma ^2+617\right) |a_1|^2  |a_2|^5+28 \left(15
   \sigma ^6-74 \sigma ^4+747 \sigma ^2-248\right) |a_1|
    |a_2|^6\nn \\
     &+3 \left(85 \sigma ^6-353 \sigma ^4+1123 \sigma ^2-295\right) |a_2|^7      \Big]+ \left(1 \leftrightarrow 2\right)\,,\nn\\
    \theta^{S^8}_{\mathrm{2PM}}=& -\frac{9\pi G^2 m_1^2 m_2 }{16384 |b|^{10}\sqrt{\sigma^2-1}^3} \Big[ 224 \left(719 \sigma ^4-738 \sigma ^2+99\right) |a_1|^8\nn \\
     &+17920 \left(68 \sigma ^4-69 \sigma ^2+9\right)|a_1|^7 |a_2|+8960 \left(447
   \sigma ^4-447 \sigma ^2+56\right)|a_1|^6 |a_2|^2\nn \\
     &+143360 \left(52 \sigma ^4-51 \sigma ^2+6\right)|a_1|^5
   |a_2|^3+35840 \left(239 \sigma ^4-228
   \sigma ^2+24\right)|a_1|^4 |a_2|^4\nn \\
     &-448 \left(87 \sigma ^6-14037 \sigma ^4+12813 \sigma ^2-1103\right)|a_1|^3 |a_2|^5\nn \\
     &+112 \left(345 \sigma ^8-1874
   \sigma ^6+27544 \sigma ^4-22926 \sigma ^2+1391\right)|a_1|^2
   |a_2|^6\nn \\
     &+64 \left(795 \sigma ^8-3618 \sigma ^6+15956 \sigma ^4-11358
   \sigma ^2+465\right)|a_1| |a_2|^7\nn \\
     &+(3465 \sigma ^{10}+1095 \sigma ^8-43798 \sigma ^6+148062 \sigma ^4-92355 \sigma ^2+1451)|a_2|^8    \Big]+ (1 \leftrightarrow 2)\,,\nn\\
     \theta^{S^9}_{\mathrm{2PM}}=&\frac{5 \pi G^2  m_1^2 m_2 \sigma}{16384|b|^{11}(\sigma^2-1)}\Big[ 16128 \left(21 \sigma ^2-11\right)|a_1|^9+4032 \left(719 \sigma ^2-369\right)|a_1|^8 |a_2|\nn \\
     &+5483520 \left(2 \sigma ^2-1\right)|a_1|^7 |a_2|^2+161280
   \left(149 \sigma ^2-72\right)|a_1|^6
   |a_2|^3\nn \\
     &+2580480 \left(13 \sigma ^2-6\right)|a_1|^5 |a_2|^4-2016 \left(69 \sigma ^4-15434 \sigma
   ^2+6685\right)|a_1|^4 |a_2|^5\nn \\
     &+2688 \left(65 \sigma ^6-320 \sigma ^4+7357 \sigma ^2-2902\right)|a_1|^3
   |a_2|^6\nn \\
     &+288 \left(1215 \sigma ^6-4815 \sigma
   ^4+30361 \sigma ^2-10521\right)|a_1|^2 |a_2|^7\nn \\
     &+72 \left(385 \sigma ^8+1940 \sigma ^6-10338 \sigma ^4+34020 \sigma
   ^2-10327\right)|a_1| |a_2|^8\nn \\
     &+(19635 \sigma ^8-16300 \sigma ^6-98446 \sigma ^4+298388 \sigma ^2-82317)|a_2|^9   \Big]+ (1 \leftrightarrow 2)\,,\nn\\
     \theta^{S^{10}}_{\mathrm{2PM}}=&-\frac{11\pi G^2 m_1^2 m_2 }{131072|b|^{12}}\Big[ 1344 \left(1055 \sigma ^4-1078 \sigma ^2+143\right)|a_1|^{10}\nn \\
     &+161280 \left(84 \sigma ^4-85 \sigma ^2+11\right)|a_1|^9 |a_2|+80640
   \left(719 \sigma ^4-719 \sigma ^2+90\right)|a_1|^8 |a_2|^2\nn \\
     &+2150400 \left(68 \sigma ^4-67 \sigma ^2+8\right)|a_1|^7
   |a_2|^3+1612800 \left(149
   \sigma ^4-144 \sigma ^2+16\right)|a_1|^6 |a_2|^4\nn \\
     &-8064 \left(111 \sigma ^6-33547 \sigma ^4+31537 \sigma ^2-3141\right)|a_1|^5 |a_2|^5\nn \\
     &+3360
   \left(415 \sigma ^8-2250 \sigma ^6+64960 \sigma ^4-57966 \sigma ^2+4921\right)|a_1|^4
   |a_2|^6\nn \\
     &+3840 \left(985 \sigma ^8-4370 \sigma
   ^6+34274 \sigma ^4-27794 \sigma ^2+1945\right)|a_1|^3 |a_2|^7\nn \\
     &+180 \left(1925 \sigma ^{10}+14180 \sigma ^8-78306 \sigma ^6+316812
   \sigma ^4-227611 \sigma ^2+13320\right)|a_1|^2 |a_2|^8\nn \\
     &+20 \left(25795 \sigma ^{10}-12885 \sigma ^8-200106 \sigma ^6+721566 \sigma
   ^4-476729 \sigma ^2+22999\right)|a_1|
   |a_2|^9\nn \\
     &+(27027 \sigma ^{12}+41580 \sigma ^{10}-114965 \sigma ^8-433952 \sigma ^6+1605769
   \sigma ^4-997356 \sigma ^2\nn \\
     &+33177)|a_2|^{10}    \Big]+ (1 \leftrightarrow 2)\,,\nn\\
     \theta^{S^{11}}_{\mathrm{2PM}}=&\frac{3 \pi G^2  m_1^2 m_2 \sigma}{32768|b|^{13}(\sigma^2-1)}\Big[  59136 \left(25 \sigma ^2-13\right) |a_1|^{11}+14784 \left(1055 \sigma ^2-539\right)|a_1|^{10} |a_2|\nn \\
     &+37255680 \left(2 \sigma
   ^2-1\right)|a_1|^9 |a_2|^2+295680 \left(719 \sigma ^2-350\right)|a_1|^8
   |a_2|^3\nn \\
     &+23654400 \left(17 \sigma ^2-8\right)|a_1|^7 |a_2|^4-44352 \left(31 \sigma
   ^4-11980 \sigma ^2+5397\right)|a_1|^6 |a_2|^5\nn \\
     &+29568 \left(85 \sigma ^6-418 \sigma ^4+17209 \sigma ^2-7300\right)|a_1|^5
   |a_2|^6\nn \\
     &+5280 \left(1625
   \sigma ^6-6333 \sigma ^4+68887 \sigma ^2-26883\right)|a_1|^4 |a_2|^7\nn \\
     &+2640 \left(315 \sigma ^8+3515 \sigma ^6-15027 \sigma ^4+73433
   \sigma ^2-25948\right)|a_1|^3 |a_2|^8\nn \\
     &+110 \left(16975 \sigma ^8+20420 \sigma ^6-195206 \sigma ^4+650532 \sigma ^2-210481\right)|a_1|^2
   |a_2|^9\nn \\
     &+44
   \left(2457 \sigma ^{10}+20818 \sigma ^8-19494 \sigma ^6-113556 \sigma ^4+355669 \sigma ^2-108806\right)|a_1| |a_2|^{10}\nn \\
     &+(82719 \sigma
   ^{10}-23961 \sigma ^8-90386 \sigma ^6-530730 \sigma ^4+1542995 \sigma ^2-448413)|a_2|^{11}       \Big]\nn \\
     &+ (1 \leftrightarrow 2)\,.
\end{align}

\subsection{Comparison to literature and canonical spin}
\label{sec:comparison}

The results presented in previous sections pass various consistency checks with the available literature. For instance, our triangle and box coefficients agree with \rcite{Chen:2021kxt} up to $\cO(S^4)$. Our eikonal and scattering angle, computed using the covariant spin supplementary condition (SSC), agree with that of \rcite{Bautista:2023szu} up to $\cO(S^6)$. We also compared the impulse, up to $\cO(S^2)$ in the covariant SSC, against \rcites{Liu:2021zxr,Jakobsen:2022zsx,FebresCordero:2022jts}, and up to $\cO(S^5)$ in the canonical SSC against \rcite{Bautista:2023szu}, and we find agreement in both cases. At $\cO(S^6)$ we find some minor disagreement with~\rcite{Bautista:2023szu} for contributions corresponding to vector boxes. Since the all-order-in-spin formula \eqref{eq:ImpulseEperp}, coming from the vector box, appears to be robust, we expect that the issue is to be found elsewhere.

We now elaborate on the canonical SSC comparison, which requires some extra work. We need to convert our result from covariant to canonical SSC, also known as \textit{Newton-Wigner SSC}~\cite{Pryce:1935ibt,Pryce:1948pf,Newton:1949cq}. The details of the conversion are discussed in appendix~\ref{app:SSC}, here we give a brief version. In order to line up notation with \rcite{Bautista:2023szu}, the incoming black-hole momenta are taken to be in the center-of-mass frame $p_1^\mu = -(E_1,\vec{p})$ and $p_2^\mu = -(E_2,-\vec{p})$, and the covariant spin vectors $S_i^\mu$ are then
\begin{equation}
\label{eq:covScom}
        S_i^\mu = \left( \vec{v}_i \cdot \vec{s}_i , \vec{s}_i + \frac{\vec{v}_i \cdot \vec{s}_i}{\gamma_i+1} \vec{v}_i \right) .
\end{equation}

To rewrite our results in canonical SSC, we need to transform the spin vectors $S_i^\mu$ and the impact parameter $b^\mu$. As discussed in appendix~\ref{app:SSC}, the canonical spin vectors in the center-of-mass frame are simply
\begin{equation}
        S_{i,\text{can}}^\mu = \left( 0 , \vec{s}_i \right) .
\end{equation}
Since $\vec{S}_{i,\text{can}} = \vec{s}_i$, it is enough to use \eqn{eq:covScom} to express our results in terms of the canonical spin three-vector. On the other hand, the required shift of the impact parameter is given by \eqn{eq:covtocan}, which can be rewritten in terms of three-vectors as
\begin{equation}
\label{eq:bcantocov}
    \vec{b} = \vec{b}_{\text{can}} - \sum_{i=1}^2 \frac{\vec{p}\times \vec{s}_i}{m_i(E_i+m_i)} ,
\end{equation}
where we used the center-of-mass-frame identities $b^\mu = (0,\vec{b})$ and $b^\mu_{\text{can}} = (0,\vec{b}_{\text{can}})$.

Let us see how this works in a simple example. We start from the scattering angle given in \eqn{eq:covscattangle}, up to linear order in the spin $S_1^\mu$. For simplicity, we assume $S_2^\mu = 0$. In the center-of-mass frame this is equal to 
\begin{align}
\label{eq:covangles0s1}
\begin{split}
    \theta^{S^0}_{\mathrm{2PM}}=& -\frac{3\pi G^2 m_1 m_2 (m_1+m_2)  (5\sigma^2-1) }{4 \sqrt{\sigma^2-1}|\vec{b}|^2}\,, \\
    \theta^{S^1}_{\mathrm{2PM}}=&\frac{\pi G^2 m_2 \sigma(4m_1+3m_2)(5\sigma^2-3) |\vec{s}_1|   }{2 (\sigma^2-1)|\vec{b}|^3} \,.\\
\end{split}
\end{align}
As discussed, this is already expressed in terms of the canonical spin three-vector $\vec{s}_1$. Note that the scattering angle obeys aligned-spin kinematics, namely $\vec{s}_i\cdot\vec{b}=\vec{s}_i\cdot\vec{p}=0$ and $\vec{s}_i = \tfrac{|\vec{s}_i|}{|\vec{L}|}\vec{L}$, where $\vec{L} = \vec{b}\times\vec{p}$. To convert to canonical impact parameter $\vec{b}_{\text{can}}$, we use \eqn{eq:bcantocov} and get
\begin{equation}
\label{bcovtocanaligned}
    |\vec{b}|^2 = \left(\vec{b}_{\text{can}} -\frac{\vec{p}\times \vec{s}_1}{m_1(E_1+m_1)} \right)^2 = |\vec{b}_{\text{can}}|^2 \left(1 -\frac{|\vec{s}_1||\vec{p}|}{m_1(E_1+m_1)|\vec{b}_{\text{can}}|} \right)^2 ,
\end{equation}
where we used that $\vec{p}\times \vec{s}_1 = (|\vec{s}_1||\vec{p}|/|\vec{b}_{\text{can}}|) \vec{b}_{\text{can}}$ for aligned-spin kinematics. Therefore, substituting \eqn{bcovtocanaligned} into \eqn{eq:covangles0s1}, expanding in the spin $\vec{s}_1$ and isolating the linear contribution, we can derive the $\cO(\vec{s}_1)$ contribution to the canonical scattering angle, $\theta^{S^1}_{\mathrm{2PM},\text{can}}$, given by 
\begin{equation}
    \theta^{S^1}_{\mathrm{2PM},\text{can}}= 
    -\frac{3 \pi G^2 m_2 (m_1+m_2) (5\sigma^2-1) |\vec{s}_1||\vec{p}|}{2 \sqrt{\sigma^2-1}|\vec{b}_{\text{can}}|^3 (E_1+m_1)}
    + \frac{\pi G^2 m_2 (4m_1+3m_2)  \sigma\left(5\sigma^2-3\right) |\vec{s}_1|}{2 \left(\sigma^2-1\right)|\vec{b}_{\text{can}}|^3}
    \,.
\end{equation}

This result can be checked against the canonical impulse $\Delta p^\mu_{\mathrm{2PM},\text{can}}$ given in \rcite{Bautista:2023szu}, via the relation
\begin{equation}
      \theta_{\mathrm{2PM},\text{can}} = - \frac{\vec{b}_{\text{can}} \cdot \Delta \vec{p}_{\mathrm{2PM},\text{can}}}{|\vec{L}_{\text{can}}|} ,
\end{equation}
valid for aligned-spin kinematics at 2PM order. We have performed this check and find agreement up to $\cO(S^5)$, and for triangle contributions up to $\cO(S^6)$. The full canonical impulse up to $\cO(S^{11})$ can be extracted from our results by applying the same conversion between covariant and canonical SSC demonstrated above.
\section{Conclusion}
\label{sec:conclusion}
Using the framework of quantum higher-spin theory, a proposal for the Kerr Compton amplitude for any spin was given in \rcite{Cangemi:2023bpe}, which classically agrees with explicit black-hole perturbation calculations~\cite{Bautista:2022wjf,private,Bautista:2023sdf} for certain choices of near-zone/far-zone splittings. While there exist contact-term ambiguities due to the appearance of transcendental functions starting at ${\cal O}(S^5)$, which makes the notion of tree level unclear, one may expect that the simple all-order result of \rcite{Cangemi:2023bpe} captures a substantial part of the Kerr far-zone dynamics.  

The Kerr Compton amplitude can be used to extract observables for binary Kerr black-hole scattering at second-post-Minkowskian order, which we explore in this work. We employed on-shell unitarity methods to compute the relevant classical 2-to-2 one-loop integrand. For simplicity, we took the classical limit (infinite-spin limit) already at tree level, and the unitarity cuts only employed classical building blocks that are entire functions of the spin vector. From the classical integrand, we extracted the scalar box, vector box and scalar triangle coefficients to all orders in spin. We find simple novel formulae for these, specifically the box coefficients are exponential functions in spin and rapidity, where the individual helicity contributions give obvious imprints. The triangle coefficients are given as simple derivatives/integrations applied to the Bessel function $J_0$ of the first kind. We define explicit Bessel-like functions to make manifest the spin-multipole expansion to all orders in spin. With appropriate definitions, the triangle coefficients can be split into three contributions (originating from the pole term, subleading pole term and contact term of the Compton amplitude) that we presented as one-line expressions. 

The tensor-reduced one-loop integrand serves as input to compute classical observables such as the impulse and the closely related scattering angle and eikonal phase, all of which we explore. For the classical impulse, we use the KMOC formalism, which expresses it in terms of Fourier-transformed momentum-weighted amplitudes and cuts in impact-parameter space. The amplitude and cut contributions can be massaged such that the cancellation of hyper-classical iteration terms vanish and the remaining finite terms are split into three contributions based on their origin: scalar box, vector box and scalar triangle, all of which give classical contributions. 

While it is often implied that the scalar triangle captures the full 2PM results, the scalar box and vector box give tangible, albeit simple, contributions to the classical impulse. Even the same-helicity Compton amplitudes, once fed into the cuts, give a non-zero contribution to the impulse. While the box contributions are kinematically constrained to be related to the 1PM impulse, working them out in full generality at 2PM is instructive. We give closed-form all-order-in-spin expressions for the parallel 2PM impulse (from the scalar box) and also for the transverse spin-induced contributions coming from the vector box. For impulse contributions coming from triangle integrals, we give spin-expanded expressions up to $\cO(S^{11})$.

For the eikonal and scattering angle, we give certain Fourier-transformed triangle integrals to all orders in spin, specifically those contributions that originate from the simple entire functions of the Compton amplitude that are associated with the pole term and subleading-pole term. We defer the remaining all-order results of the genuine contact terms to future work. For the aligned-spin scattering angle we explicitly print out the results up to $\cO(S^{11})$, and the remaining results are given in the ancillary file. Specifically, we provide results up to $\cO(S^{11})$ for the triangle, scalar box, and tensor box, as well as the eikonal, scattering angle and impulse at 1PM and 2PM.

We have compared our computations to fixed-order results for low-spin multipoles in the literature and find convincing agreement. Specifically, we focused on \rcite{Bautista:2023szu} which gives the impulse up to $\cO(S^6)$ in canonical SSC gauge, which required nontrivial conversion of our observables that are otherwise expressed in the covariant SSC gauge. We reproduce the impulse up to $\cO(S^5)$, once taking into account the choice $\alpha=0$, $\eta\neq0$ that we employ for the contact terms in the Compton amplitude, and find a minor disagreement at $\cO(S^6)$ for pieces coming from the vector box, which are simple iteration pieces. 
For reference, we have included some practical details of the canonical SSC conversion.

There are several avenues that can be explored in further work. Firstly, we were able to get surprisingly simple all-order-in-spin formulae for the one-loop triangle coefficients, and while for certain of these closed formulae we also computed the classical observables in terms of the eikonal phase, more work is required to sufficiently simplify and present the remaining all-order-in-spin contributions. 
Secondly, our results are given for the case $\alpha = 0$, where $\alpha$ is a tag that was introduced in~\rcite{Bautista:2022wjf} to mark certain conspicuous contributions. The quantum higher-spin Compton amplitude seems to land naturally on $\alpha = 0$~\cite{Cangemi:2023bpe}, but it would also be desirable to obtain all-order-in-spin results for contributions that may represent $\alpha \neq 0$. This requires more knowledge of such terms to higher orders $S^{\gg8}$ ({\it c.f.}~\rcite{Bautista:2023sdf}) such that suitable candidate entire functions in spin can be studied. 
Lastly, there are important classical observables that we have not explored in this work, such as the 2PM spin kick, which can be extracted from the same 1-loop amplitude discussed in this work. The higher-spin Compton amplitude can also be used for computing the leading-order waveform, which can be obtained from the five-point tree-level amplitude where a graviton external state is included.

\begin{acknowledgments}

We thank Fabian Bautista, Maor Ben-Shahar, Marco Chiodaroli, Gang Chen, Paolo Di Vecchia, Harald Ita, Alex Ochirov, Johannes Schlenk, Oliver Schlotterer, Evgeny Skvortsov, Fei Teng, Tianheng Wang, Mao Zeng for useful discussions and related collaborations. The research is supported by the Knut and Alice Wallenberg Foundation under grants KAW 2018.0116 (From Scattering Amplitudes to Gravitational Waves) and KAW 2018.0162. 
L.B. acknowledges support from the Swiss National Science Foundation (SNSF) under
the grant 200020 192092. 
The work of P.P. was supported by the Science and Technology Facilities Council (STFC) Consolidated Grants ST/T000686/1 “Amplitudes, Strings \& Duality” and ST/X00063X/1 “Amplitudes, Strings \& Duality”; no new data were generated or analyzed during this study.
We appreciate the hospitality and support from the Munich Institute for Astro-, Particle and BioPhysics (MIAPbP), which is funded by the Deutsche Forschungsgemeinschaft (DFG, German Research Foundation) under Germany's Excellence Strategy – EXC-2094 – 390783311.
H.J. thanks the Galilei Galilei Institute for Theoretical Physics for the hospitality and the INFN for partial support during the completion of this work. 

\end{acknowledgments}

\appendix

\section{Fourier integrals}\label{ap:FourierTransform}
The types of integrals required for the Fourier transform to impact parameter space are
\begin{align}
     \mathcal{I}_{\alpha}&= \int d^{D}\!\mu \, e^{i q\cdot b} |q|^{\alpha},\,\\
     \mathcal{I}_{\alpha}^{\mu_1 \cdots \mu_k} &= \int d^{D}\!\mu \, e^{i q\cdot b} q^{\mu_1} \cdots q^{\mu_k} |q|^{\alpha} \label{eq:gentensorint}
\end{align}
A priori, the $q^{\mu}$ momentum has components parallel to the velocities,  
\begin{align}
    q^{\mu} = x_1 v_1^\mu + x_2 v_2^\mu + \qperp^{\mu}
\end{align}
such that the on-shell measure decomposes into
\begin{equation}
    d^{D}\, \mu := \frac{1}{4 m_1 m_2 \sqrt{\sigma^2-1}}  \hat{d}^{D-2} \qperp d x_1 d x_2 \delta(x_1) \delta(x_2)\,,
\end{equation} 
where $\hat{d}^{D-2} \qperp = (2 \pi)^{D-2} d^{D-2}  \qperp$. Note that the parallel components of $q^{\mu}$ in \eqn{eq:gentensorint} integrate zero due to the delta functions in the measure such that
\begin{equation}
    \mathcal{I}_{\alpha}^{\mu_1 \cdots \mu_k} = \frac{1}{4 m_1 m_2 \sqrt{\sigma^2-1}}\int \hat{d}^{D-2} \qperp \, e^{i \qperp\cdot b} \qperp^{\mu_1} \cdots \qperp^{\mu_k} |\qperp|^{\alpha}
\end{equation}

In the scalar case, the remaining $D-2$ dimensional Fourier transform evaluates to
\begin{align}
    \mathcal{I}_{\alpha} = \frac{2^\alpha \pi^{\frac{2-D}{2}} }{4  m_1  m_2 \sqrt{\sigma^2-1}} \frac{\Gamma(\frac{\alpha+D-2}{2})}{\Gamma(\frac{-\alpha}{2})} \left(\frac{1}{|b|}\right)^{\alpha+D-2}\,,
\end{align}
see \rcite{Herrmann:2021tct} for details. Note that in $D=4$ the integral diverges for even $\alpha<0$ and needs to be regulated. Using the regulation scheme $D=4-2 \epsilon$, as used for the loop integrals, the leading orders of the scalar integral are
\begin{align} \label{eq:scalarIntDiv}
    \mathcal{I}_{\alpha=-2r} \Big|_{D=4-2\epsilon} &= \frac{(-1)^{r}}{4^{r+1} \pi m_1  m_2 \sqrt{\sigma^2-1}} \frac{|b|^{2r-2}}{\Gamma(r)^2 } \left(\frac{1}{\epsilon} - 2 \log |b| - \psi(r) + \mathcal{O}(\epsilon) \right) \nn\\
    &= \frac{(-1)^{r}}{4^{r+1} \pi m_1  m_2 \sqrt{\sigma^2-1}} \frac{|b|^{2r-2}}{\Gamma(r)^2} \left(\frac{1}{\epsilon |b|^{2\epsilon}} - \psi(r)\right)+ \mathcal{O}(\epsilon) \,,
\end{align} where $\psi(r):= \Gamma'(r)/\Gamma$ is the digamma function. In the special case $r=1$, a single $b$-derivative suffices to kill the infrared divergence
\begin{align}\label{eq:1derivI}
    \frac{\partial}{\partial b^{\mu}}\mathcal{I}_{\alpha=-2} \Big|_{D=4-2\epsilon} &=- \frac{1}{8\pi m_1  m_2 \sqrt{\sigma^2-1}}  \frac{\hat{b}^{\mu}}{|b|}+ \mathcal{O}(\epsilon)\,.
\end{align} 

The general tensor integrals $\mathcal{I}^{\mu_1 \dots \mu_k}_{\alpha}$ can be generated by taking derivatives with respect to $b^{\mu}$ of the scalar integral,
\begin{equation}
    \mathcal{I}_{\alpha}^{\mu_1 \cdots \mu_k} =  (-i)^k\left(\Pi^{\mu_1 \nu_1} \frac{\partial}{\partial b_{\nu_1}}\right)\cdots \left(\Pi^{\mu_k \nu_k} \frac{\partial}{\partial b_{\nu_k}}\right)\mathcal{I}_{\alpha}\,.
\end{equation} Since the Fourier integral projects out the plane spanned by $v_1$ and $v_2$ all the scalar products after integration live in the $(D-2)$-dimensional subspace transverse to $v_1$ and $v_2$. We enforce this by including the projector
\begin{equation}
    \Pi^{\mu\nu} = g^{\mu\nu} - v_{1}^{\mu}\check v_{1}^{\nu} - v_{2}^{\mu}\check v_{2}^{\nu}\,, \qquad \check v_{1}^{\mu} := \frac{\sigma v_2^{\mu} - v_1^{\mu}}{\sigma^2 -1} \,, \qquad    \check v_{2}^{\mu} := \frac{\sigma v_1^{\mu} - v_2^{\mu}}{\sigma^2 -1}\,,
\end{equation}
where $\check v_{i}^{\mu}$ are the dual velocities, $ v_{i}\cdot \check v_{j}=\delta_{ij}$.
Note that the impact parameter is fully transverse and $|b| = \sqrt{\bperp^2} = \sqrt{-b^2}$. The general action of the derivatives on the scalar integral is captured by
\begin{align}\label{eq:generalbderivs}
     \left(\Pi \cdot \frac{\partial}{\partial b}\right)^{\mu_1}\cdots \left(\Pi \cdot \frac{\partial}{\partial b}\right)^{\mu_k}\frac{1}{|b|^{\beta}} 
     =& \sum_{n=0}^{\lfloor k/2 \rfloor} \Big\{ \frac{(\beta + 2(k-n-1))!!}{(\beta-2)!!}\frac{2^{-n} }{n! (k-2n)!} \frac{1}{|b|^{\beta+2(k-n)}} \nn\\
     &\sum_{\vec{\sigma}\in \mathrm{perm}(\vec{\mu})}\Big[ \prod_{i=1}^{n} \Pi^{\sigma(2i-1)\sigma(2i)}\prod_{i=2 n +1}^{k} b^{\sigma(i)}\Big]\Big\}\,.
\end{align}
In practice, the majority of the integrals that show up in the computation of the impulse and eikonal simplify because of the simple tensor structure of the integrands. For example, the integrals that appear in the eikonal phase are of the form $\mathcal{I}^{ (\star a)^k}_{\alpha} := (\star a)_{\mu_1}\dots (\star a)_{\mu_k} \mathcal{I}^{\mu_1 \dots \mu_k}$. Note that since $(\star a)_{\mu}$ is already transverse to $v_1$, $v_2$ the action of the projector is trivial and the combinatorics in \eqn{eq:generalbderivs} simplify to 
\begin{align}
     \left( (\star a)^\mu \frac{\partial}{\partial b^{\mu}} \right)^{k}\frac{1}{|b|^{\beta}} 
     = \sum_{n=0}^{\lfloor k/2 \rfloor}\frac{(\beta + 2(k-n-1))!! \, k!}{2^{n}(\beta -2)!! \, (k-2n)!\, n!}\frac{(b\star a)^{k-2n} (\star a)^{2n}}{|b|^{\beta+2(k-n)}}\,, 
\end{align} 
Taking $\beta = \alpha +D-2$, and working in $D=4$, we find that the integrals $\mathcal{I}_{\alpha}^{(\star a)^k}$ vanish for even $\alpha \ge 0$. 

For even $\alpha <0$ the integrals are in general divergent and need to be regulated. However, the case $\alpha = - 2$, relevant for the $1$PM eikonal and the parallel contribution to $2$PM impulse $\Delta p^{\mu}_{\parallel}$, is only divergent for $k=0$, such that it is finite for $k>0$,
\begin{equation}
    \mathcal{I}_{D=4}^{(\star a)^k} = \begin{cases}
    -\frac{\mathcal{N}}{4} \left( \frac{1}{\epsilon} - 2 \log |b| - \psi(1)    \right) + \mathcal{O}(\epsilon) &\mathrm{ for }\,\, k=0\,, \\
    (-1)^{k+1} i^k\mathcal{N} \sum_{n=0}^{\lfloor k/2 \rfloor}\frac{2^{k-2n-2}(k-n-1)!\, k!}{(k-2n)! \, n!}\frac{(b\star a)^{k-2n} (\star a)^{2n}}{|b|^{2(k-n)}} &\mathrm{ for }\,\, k>0\,,
    \end{cases}
\end{equation} where $\mathcal{N} = \frac{1}{4 \pi m_1 m_2 \sqrt{\sigma^2-1}}$. As shown in \eqn{eq:1derivI}, the $b$-derivatives cancel the infrared divergence in the scalar integral $\mathcal{I}_{-2}$.

For odd $\alpha=2 r-1$,relevant for $2$PM computations, it is finite in $D=4$ and it can be written on the simple form
\begin{align}
\mathcal{I}_{D=4}^{(\star a)^k} = (-1)^{k+r} i^k \frac{\mathcal{N}}{2}  \sum_{n=0}^{\lfloor k/2 \rfloor} \frac{ (2k+ 2r - 2n -1)!! \, (2 r-1)!! \, k! }{2^{n} (k - 2 n)! \, n! } \frac{(b\star a)^{k-2n} (\star a)^{2n}}{|b|^{2(k-n+r)+1}}\,.
\end{align}

When computing the impulse via the KMOC framework, the relevant Fourier integrals have a free Lorentz index and are of the form
\begin{equation}
    \mathcal{I}^{\nu (\star a)^k}_{\alpha} := (\star a)_{\mu_1}\dots (\star a)_{\mu_k} \mathcal{I}^{\nu \mu_1 \dots \mu_k}\,,
\end{equation} such that the $b$-derivatives simplify to
\begin{align}
     \left( (\star a)^\mu \frac{\partial}{\partial b^{\mu}} \right)^{k} \left(\Pi \cdot \frac{\partial}{\partial b}\right)^{\nu}\frac{1}{|b|^{\beta}} 
     = \sum_{n=0}^{\lfloor (k+1)/2 \rfloor} \Big\{ &\frac{(\beta + 2(k-n))!!\, k!}{2^{n}(\beta -2)!! \,(k+1-2n)!\, n!}\frac{1}{|b|^{\beta+2(k+1-n)}}\nn \\
     \times \Big[ &(k+1 -2n) b^{\nu} (b\star a)^{k-2n} (\star a)^{2 n} \nn \\
     + &(2n)  (\star a)^{\nu} (b\star a)^{k+1-2n} (\star a)^{2n-2} \Big]\Big\}.
\end{align}

This relation is valid in generic dimension, which is necessary since the integrals for the leading-order momentum must be regularized. For example, the transverse contribution to the $2$PM impulse, $\Delta p_{e_{\perp}}^{\nu}$, requires computing the integral $\epsilon^{-1} (\star a)_{\mu_1}\dots(\star a)_{\mu_k} \mathcal{I}_{\alpha = -2 \epsilon}^{\nu (\star a)^k}$, where the dependence on the regulator $\epsilon$ is generated by the dimensional reduction of the loop integral, $D= 4-2\epsilon$. We will see that the impulse is finite so long as we regularize the Fourier integrals using the same prescription. At leading order in $\epsilon \to 0$, the integral is finite and can be reduced to
\begin{align}
\epsilon^{-1} (\star a)_{\mu_1}\dots(\star a)_{\mu_k} \mathcal{I}_{-2 \epsilon}^{\nu \mu_1 \dots \mu_k} = \mathcal{N} \!\! \sum_{n=0}^{\lfloor (k+1)/2 \rfloor} &\Big\{ \frac{2^{k - 2 n+1} (k-n+1)!}{n! (k-2n+1)!}\frac{1}{|b|^{2(k+2-n)}} \nn    \\
\Big[ &k! (k+1 -2n) (b\star a)^{k-2n} (\star a)^{2 n}b^{\nu}  \nn \\
+ &k! (2n)   (b\star a)^{k+1-2n} (\star a)^{2n-2}(\star a)^{\nu} \Big]\Big\}\,.
\end{align}
The all-order expression in \eqn{eq:ImpulseEperp} is generated by exponentiation of the above integral since $\tilde{c}_{\Box}$ introduces the exponential $e^{i q \star a}$.
\section{Spin Supplementary Condition}\label{app:SSC}
In this section, we discuss the redundancy in the definition of spin variables, and how to fix it by means of a \textit{spin supplementary condition} (SSC). We follow the approaches outlined in~\rcites{Jakobsen:2022zsx,Bautista:2023szu}.

Let us consider the motion of two bodies at positions $b_1^\mu$ and $b_2^\mu$ moving with momenta $p_1$ and $p_2$. In the center-of-mass frame, we have
\begin{align}
\label{eq:COMp}
    \begin{split}
        &p_1^\mu = -(E_1,\vec{p})= m_1 v_1\, ,\\
        &p_2^\mu = -(E_2,-\vec{p})=m_2 v_2\, .
    \end{split}
\end{align}
The unit vector $\hat{P}^\mu=(p_1+p_2)^\mu/E$,  
with $E = E_1+E_2$, picks out the center-of-mass frame, which becomes the statement $\hat{P}^\mu = (-1,\vec{0})$.

The total angular momentum of the system is described by a tensor $J^{\mu\nu}$ and it is a well-defined conserved quantity. Splitting this tensor into an orbital angular momentum $L^{\mu\nu}$ and intrinsic angular momentum of each body, $S_i^{\mu\nu}$ can be done as
\begin{equation}
    J^{\mu\nu} = L^{\mu\nu} + S_1^{\mu\nu} + S_2^{\mu\nu}\, ,
\end{equation}
where $L^{\mu\nu}$ is
\begin{equation}
    L^{\mu\nu} = 2b_1^{[\mu}p_1^{\nu]} + 2b_2^{[\mu}p_2^{\nu]}\,.
\end{equation}

However, the split is potentially ambiguous. If we redefine the notion of the ``center'' of each body via $b_i^\mu \to b_i^\mu + \delta b_i^\mu$, invariance of $J^{\mu\nu}$ demands that the intrinsic spin vectors transform as
\begin{equation}
\label{eq:SSCtransf}
    \delta S_i^{\mu\nu} = - 2\delta b_i^{[\mu} p_i^{\nu]}\, .
\end{equation}
This is effectively a gauge symmetry of the system, which needs to be accounted for by a gauge-fixing condition. Such a condition is known as the SSC, and the one we use in this work is the \textit{covariant} SSC, defined by
\begin{equation}
    p_{i\mu} S_i^{\mu\nu} = 0\,.
\end{equation}

Working in covariant SSC, it is customary to define spin vectors $J^\mu$, $L^\mu$ and $S_i^\mu$ as
\begin{align}
\label{eq:spinvecs}
\begin{split}
    &J^\mu = \frac{1}{2} \epsilon^{\mu\nu\rho\sigma} J_{\nu\rho} \hat{P}_{\sigma}\, ,\\
    &L^\mu = \frac{1}{2} \epsilon^{\mu\nu\rho\sigma} L_{\nu\rho} \hat{P}_{\sigma}\, ,\\
    &S_i^\mu = \frac{1}{2} \epsilon^{\mu\nu\rho\sigma} S_{i\nu\rho} v_{i\sigma}\, .
\end{split}
\end{align}
The orbital spin vector $L^\mu$ can be written in terms of the impact parameter $b^\mu = b_2^\mu - b_1^\mu$ through $L^\mu = - \epsilon^{\mu\nu\rho\sigma} b_\nu p_{1\rho} p_{2\sigma} / E$. The inverse relations that expresses the spin tensors in terms of spin vectors are similar, for instance
\begin{equation}
    S_i^{\mu\nu} = \epsilon^{\mu\nu\rho\sigma} v_{i\rho} S_{i\sigma}\,.
\end{equation}
The spin vector $S_i^\mu$ thus takes the natural form in the rest frame of particle $i$, namely
\begin{equation}
    S_i^\mu \big|_{v_i^\mu = (-1,\vec{0})} = (0, \vec{s}_i)\,,
\end{equation}
where $\vec{s}_i$ is the body's spin three-vector, and the frame exhibited is indicated by the condition it satisfies. 

It is useful to quote the explicit form of the covariant spin vectors $S_i^\mu$ in the center-of-mass frame of the system, namely
\begin{equation}
\label{}
    S_i^\mu \big|_{\hat{P}^\mu = (-1,\vec{0})} = \left( \vec{v}_i \cdot \vec{s}_i , \vec{s}_i + \frac{\vec{v}_i \cdot \vec{s}_i}{\gamma_i+1} \vec{v}_i \right)\, .
\end{equation}
where $\gamma_i = \hat{P}\cdot v_i= \cosh \zeta_i$ are the Lorentz factors, and $\zeta_i$ are the rapidities. 

The choice of covariant spin vectors has a drawback: it is clear from the definitions~\eqref{eq:spinvecs} that the intrinsic and orbital spin vectors do not sum to the total spin vector, namely
\begin{equation}
    J^\mu \neq L^\mu + S_1^\mu + S_2^\mu\, .
\end{equation}
A related issue is that the covariant spin three-vectors $\vec{s}_i$ do not obey canonical Poisson-bracket relations, which creates some difficulties when working with Hamiltonians, see \eg \rcite{Vines:2016unv}. Resolving this leads to the gauge-fixing choice known as \textit{canonical} (or \textit{Newton-Wigner}) SSC, namely
\begin{equation}
    (\hat{P}+v_i)_\mu\, S_{i,\text{can}}^{\mu\nu} = 0\, .
\end{equation}
In this case, while the spin vectors $J^\mu_{\text{can}}$ and $L^\mu_{\text{can}}$ are defined in the same manner as in \eqn{eq:spinvecs}, the intrinsic spin vectors $S_{i,\text{can}}^\mu$ are now defined as
\begin{equation}
    S_{i,\text{can}}^\mu = \frac{1}{2} \epsilon^{\mu\nu\rho\sigma} S_{i,\text{can} \nu\rho} \hat{P}_{\sigma}\, ,
\end{equation}
where the inverse transformation is now given by
\begin{equation}
    S_{i,\text{can}}^{\mu\nu} = \frac{1}{\gamma_i+1}\epsilon^{\mu\nu\rho\sigma} (\hat{P}+v_i)_\rho S_{i,\text{can} \sigma}\, ,
\end{equation}
This new choice satisfies $J^\mu = L^\mu_{\text{can}}+S^\mu_{1,\text{can}}+S^\mu_{2,\text{can}}$, and it can be shown that the three-vectors $\vec{S}_{i,\text{can}}$ satisfy canonical Poisson brackets (see \eg\cite{Vines:2016unv}). The transition from covariant to canonical SSC is realized via the transformation~\eqref{eq:SSCtransf}, where
\begin{equation}
    \delta b_i^\mu = -\frac{\hat{P}\cdot S_i^\mu}{E_i+m_i}\, .
\end{equation}

This gives rise to the following relations between quantities in covariant and canonical SSC:
\begin{align}
\label{eq:cantocov}
\begin{split}
    &b^\mu_{\text{can}} = b^\mu - \frac{\hat{P}\cdot S_2^\mu}{E_2+m_2} + \frac{\hat{P}\cdot S_1^\mu}{E_1+m_1} ,\\
    &S_{i,\text{can}}^{\mu\nu} = S_i^{\mu\nu} + \frac{2}{\gamma_i+1} \hat{P}\cdot S_i^{[\mu} v_i^{\nu]} ,\\
    &L^{\mu\nu}_{\text{can}} = L^{\mu\nu} - \sum_{i=1}^2 \frac{2}{\gamma_i+1} \hat{P}\cdot S_i^{[\mu} v_i^{\nu]}\, ,
\end{split}
\end{align}
or in terms of spin vectors,
\begin{align}
\begin{split}
    &S_{i,\text{can}}^\mu = S_i^\mu -\frac{\hat{P}\cdot S_i}{\gamma_i+1} (\hat{P}^\mu+v_i^\mu) ,\\
    &L^\mu_{\text{can}} = L^\mu + \sum_{i=1}^2 \left[ (\gamma_i-1)S_i^\mu + \frac{\hat{P}\cdot S_i}{\gamma_i+1} (\hat{P}^\mu - \gamma_i v_i^\mu) \right]\, ,
\end{split}
\end{align}
where we used $m_i \gamma_i = E_i$. One can also compute the explicit form for the canonical spin vectors $S_{i,\text{can}}^\mu$ in the center-of-mass frame, and it takes the expected form
\begin{equation}
    S_{i,\text{can}}^\mu \big|_{\hat{P}^\mu=(-1,\vec{0})} = \left( 0, \vec{s}_i \right)\, .
\end{equation}
It is also useful to derive some of the inverse transformations, expressing quantities in covariant SSC in terms of their canonical counterparts. They are
\begin{align}
\label{eq:covtocan}
\begin{split}
    &b^\mu = b^\mu_{\text{can}} + \frac{\hat{P}\cdot S_{2,\text{can}}^\mu}{m_2} - \frac{\hat{P}\cdot S_{1,\text{can}}^\mu}{m_1}\, ,\\
    &S_{i}^{\mu\nu} = S_{i,\text{can}}^{\mu\nu} + \frac{2}{\gamma_i+1} v_i\cdot S_{i,\text{can}}^{[\mu} (\hat{P}+v_i)^{\nu]}\, ,\\
    &S_{i}^\mu = S_{i,\text{can}}^\mu -\frac{v_i\cdot S_{i,\text{can}}}{\gamma_i+1} (\hat{P}^\mu+v_i^\mu)\, .
\end{split}
\end{align}

\bibliographystyle{JHEP}
\bibliography{references}

\providecommand{\href}[2]{#2}\begingroup\raggedright\begin{thebibliography}{100}

\bibitem{LIGOScientific:2016aoc}
{\scshape LIGO Scientific, Virgo} collaboration, \emph{{Observation of
  Gravitational Waves from a Binary Black Hole Merger}},
  \href{https://doi.org/10.1103/PhysRevLett.116.061102}{\emph{Phys. Rev. Lett.}
  {\bfseries 116} (2016) 061102}
  [\href{https://arxiv.org/abs/1602.03837}{{\ttfamily 1602.03837}}].

\bibitem{LIGOScientific:2017vwq}
{\scshape LIGO Scientific, Virgo} collaboration, \emph{{GW170817: Observation
  of Gravitational Waves from a Binary Neutron Star Inspiral}},
  \href{https://doi.org/10.1103/PhysRevLett.119.161101}{\emph{Phys. Rev. Lett.}
  {\bfseries 119} (2017) 161101}
  [\href{https://arxiv.org/abs/1710.05832}{{\ttfamily 1710.05832}}].

\bibitem{Punturo:2010zz}
M.~Punturo et~al., \emph{{The Einstein Telescope: A third-generation
  gravitational wave observatory}},
  \href{https://doi.org/10.1088/0264-9381/27/19/194002}{\emph{Class. Quant.
  Grav.} {\bfseries 27} (2010) 194002}.

\bibitem{LISA:2017pwj}
{\scshape LISA} collaboration, \emph{{Laser Interferometer Space Antenna}},
  \href{https://arxiv.org/abs/1702.00786}{{\ttfamily 1702.00786}}.

\bibitem{Reitze:2019iox}
D.~Reitze et~al., \emph{{Cosmic Explorer: The U.S. Contribution to
  Gravitational-Wave Astronomy beyond LIGO}}, {\emph{Bull. Am. Astron. Soc.}
  {\bfseries 51} (2019) 035}
  [\href{https://arxiv.org/abs/1907.04833}{{\ttfamily 1907.04833}}].

\bibitem{Buonanno:1998gg}
A.~Buonanno and T.~Damour, \emph{{Effective one-body approach to general
  relativistic two-body dynamics}},
  \href{https://doi.org/10.1103/PhysRevD.59.084006}{\emph{Phys. Rev.}
  {\bfseries D59} (1999) 084006}
  [\href{https://arxiv.org/abs/gr-qc/9811091}{{\ttfamily gr-qc/9811091}}].

\bibitem{Damour:2001tu}
T.~Damour, \emph{{Coalescence of two spinning black holes: an effective
  one-body approach}},
  \href{https://doi.org/10.1103/PhysRevD.64.124013}{\emph{Phys. Rev. D}
  {\bfseries 64} (2001) 124013}
  [\href{https://arxiv.org/abs/gr-qc/0103018}{{\ttfamily gr-qc/0103018}}].

\bibitem{Blanchet:2013haa}
L.~Blanchet, \emph{{Post-Newtonian Theory for Gravitational Waves}},
  \href{https://doi.org/10.12942/lrr-2014-2}{\emph{Living Rev. Rel.} {\bfseries
  17} (2014) 2} [\href{https://arxiv.org/abs/1310.1528}{{\ttfamily
  1310.1528}}].

\bibitem{Damour:2017zjx}
T.~Damour, \emph{{High-energy gravitational scattering and the general
  relativistic two-body problem}},
  \href{https://doi.org/10.1103/PhysRevD.97.044038}{\emph{Phys. Rev.}
  {\bfseries D97} (2018) 044038}
  [\href{https://arxiv.org/abs/1710.10599}{{\ttfamily 1710.10599}}].

\bibitem{Bjerrum-Bohr:2018xdl}
N.E.J.~Bjerrum-Bohr, P.H.~Damgaard, G.~Festuccia, L.~Planté and P.~Vanhove,
  \emph{{General Relativity from Scattering Amplitudes}},
  \href{https://doi.org/10.1103/PhysRevLett.121.171601}{\emph{Phys. Rev. Lett.}
  {\bfseries 121} (2018) 171601}
  [\href{https://arxiv.org/abs/1806.04920}{{\ttfamily 1806.04920}}].

\bibitem{Cheung:2018wkq}
C.~Cheung, I.Z.~Rothstein and M.P.~Solon, \emph{{From Scattering Amplitudes to
  Classical Potentials in the Post-Minkowskian Expansion}},
  \href{https://doi.org/10.1103/PhysRevLett.121.251101}{\emph{Phys. Rev. Lett.}
  {\bfseries 121} (2018) 251101}
  [\href{https://arxiv.org/abs/1808.02489}{{\ttfamily 1808.02489}}].

\bibitem{Bern:2019nnu}
Z.~Bern, C.~Cheung, R.~Roiban, C.-H.~Shen, M.P.~Solon and M.~Zeng,
  \emph{{Scattering Amplitudes and the Conservative Hamiltonian for Binary
  Systems at Third Post-Minkowskian Order}},
  \href{https://doi.org/10.1103/PhysRevLett.122.201603}{\emph{Phys. Rev. Lett.}
  {\bfseries 122} (2019) 201603}
  [\href{https://arxiv.org/abs/1901.04424}{{\ttfamily 1901.04424}}].

\bibitem{Bern:2019crd}
Z.~Bern, C.~Cheung, R.~Roiban, C.-H.~Shen, M.P.~Solon and M.~Zeng, \emph{{Black
  Hole Binary Dynamics from the Double Copy and Effective Theory}},
  \href{https://doi.org/10.1007/JHEP10(2019)206}{\emph{JHEP} {\bfseries 10}
  (2019) 206} [\href{https://arxiv.org/abs/1908.01493}{{\ttfamily
  1908.01493}}].

\bibitem{Damour:2019lcq}
T.~Damour, \emph{{Classical and quantum scattering in post-Minkowskian
  gravity}}, \href{https://doi.org/10.1103/PhysRevD.102.024060}{\emph{Phys.
  Rev. D} {\bfseries 102} (2020) 024060}
  [\href{https://arxiv.org/abs/1912.02139}{{\ttfamily 1912.02139}}].

\bibitem{Cheung:2020gyp}
C.~Cheung and M.P.~Solon, \emph{{Classical gravitational scattering at $
  \mathcal{O} $(G$^{3}$) from Feynman diagrams}},
  \href{https://doi.org/10.1007/JHEP06(2020)144}{\emph{JHEP} {\bfseries 06}
  (2020) 144} [\href{https://arxiv.org/abs/2003.08351}{{\ttfamily
  2003.08351}}].

\bibitem{Kalin:2020fhe}
G.~K\"alin, Z.~Liu and R.A.~Porto, \emph{{Conservative Dynamics of Binary
  Systems to Third Post-Minkowskian Order from the Effective Field Theory
  Approach}}, \href{https://doi.org/10.1103/PhysRevLett.125.261103}{\emph{Phys.
  Rev. Lett.} {\bfseries 125} (2020) 261103}
  [\href{https://arxiv.org/abs/2007.04977}{{\ttfamily 2007.04977}}].

\bibitem{Bjerrum-Bohr:2021din}
N.E.J.~Bjerrum-Bohr, P.H.~Damgaard, L.~Plant\'e and P.~Vanhove, \emph{{The
  amplitude for classical gravitational scattering at third Post-Minkowskian
  order}}, \href{https://doi.org/10.1007/JHEP08(2021)172}{\emph{JHEP}
  {\bfseries 08} (2021) 172}
  [\href{https://arxiv.org/abs/2105.05218}{{\ttfamily 2105.05218}}].

\bibitem{Brandhuber:2021eyq}
A.~Brandhuber, G.~Chen, G.~Travaglini and C.~Wen, \emph{{Classical
  gravitational scattering from a gauge-invariant double copy}},
  \href{https://doi.org/10.1007/JHEP10(2021)118}{\emph{JHEP} {\bfseries 10}
  (2021) 118} [\href{https://arxiv.org/abs/2108.04216}{{\ttfamily
  2108.04216}}].

\bibitem{Jakobsen:2022psy}
G.U.~Jakobsen, G.~Mogull, J.~Plefka and B.~Sauer, \emph{{All things retarded:
  radiation-reaction in worldline quantum field theory}},
  \href{https://doi.org/10.1007/JHEP10(2022)128}{\emph{JHEP} {\bfseries 10}
  (2022) 128} [\href{https://arxiv.org/abs/2207.00569}{{\ttfamily
  2207.00569}}].

\bibitem{Akpinar:2024meg}
D.~Akpinar, F.~Febres~Cordero, M.~Kraus, M.S.~Ruf and M.~Zeng, \emph{{Spinning
  Black Hole Scattering at $\mathcal{O}(G^3 S^2)$: Casimir Terms, Radial Action
  and Hidden Symmetry}},  \href{https://arxiv.org/abs/2407.19005}{{\ttfamily
  2407.19005}}.

\bibitem{Bern:2021dqo}
Z.~Bern, J.~Parra-Martinez, R.~Roiban, M.S.~Ruf, C.-H.~Shen, M.P.~Solon et~al.,
  \emph{{Scattering Amplitudes and Conservative Binary Dynamics at ${\cal
  O}(G^4)$}}, \href{https://doi.org/10.1103/PhysRevLett.126.171601}{\emph{Phys.
  Rev. Lett.} {\bfseries 126} (2021) 171601}
  [\href{https://arxiv.org/abs/2101.07254}{{\ttfamily 2101.07254}}].

\bibitem{Bern:2021yeh}
Z.~Bern, J.~Parra-Martinez, R.~Roiban, M.S.~Ruf, C.-H.~Shen, M.P.~Solon et~al.,
  \emph{{Scattering Amplitudes, the Tail Effect, and Conservative Binary
  Dynamics at O(G4)}},
  \href{https://doi.org/10.1103/PhysRevLett.128.161103}{\emph{Phys. Rev. Lett.}
  {\bfseries 128} (2022) 161103}
  [\href{https://arxiv.org/abs/2112.10750}{{\ttfamily 2112.10750}}].

\bibitem{Dlapa:2021npj}
C.~Dlapa, G.~K\"alin, Z.~Liu and R.A.~Porto, \emph{{Dynamics of binary systems
  to fourth Post-Minkowskian order from the effective field theory approach}},
  \href{https://doi.org/10.1016/j.physletb.2022.137203}{\emph{Phys. Lett. B}
  {\bfseries 831} (2022) 137203}
  [\href{https://arxiv.org/abs/2106.08276}{{\ttfamily 2106.08276}}].

\bibitem{Dlapa:2021vgp}
C.~Dlapa, G.~K\"alin, Z.~Liu and R.A.~Porto, \emph{{Conservative Dynamics of
  Binary Systems at Fourth Post-Minkowskian Order in the Large-Eccentricity
  Expansion}},
  \href{https://doi.org/10.1103/PhysRevLett.128.161104}{\emph{Phys. Rev. Lett.}
  {\bfseries 128} (2022) 161104}
  [\href{https://arxiv.org/abs/2112.11296}{{\ttfamily 2112.11296}}].

\bibitem{Khalil:2022ylj}
M.~Khalil, A.~Buonanno, J.~Steinhoff and J.~Vines, \emph{{Energetics and
  scattering of gravitational two-body systems at fourth post-Minkowskian
  order}}, \href{https://doi.org/10.1103/PhysRevD.106.024042}{\emph{Phys. Rev.
  D} {\bfseries 106} (2022) 024042}
  [\href{https://arxiv.org/abs/2204.05047}{{\ttfamily 2204.05047}}].

\bibitem{Jakobsen:2023pvx}
G.U.~Jakobsen, G.~Mogull, J.~Plefka and B.~Sauer, \emph{{Tidal effects and
  renormalization at fourth post-Minkowskian order}},
  \href{https://doi.org/10.1103/PhysRevD.109.L041504}{\emph{Phys. Rev. D}
  {\bfseries 109} (2024) L041504}
  [\href{https://arxiv.org/abs/2312.00719}{{\ttfamily 2312.00719}}].

\bibitem{Jakobsen:2023hig}
G.U.~Jakobsen, G.~Mogull, J.~Plefka and B.~Sauer, \emph{{Dissipative Scattering
  of Spinning Black Holes at Fourth Post-Minkowskian Order}},
  \href{https://doi.org/10.1103/PhysRevLett.131.241402}{\emph{Phys. Rev. Lett.}
  {\bfseries 131} (2023) 241402}
  [\href{https://arxiv.org/abs/2308.11514}{{\ttfamily 2308.11514}}].

\bibitem{Damgaard:2023ttc}
P.H.~Damgaard, E.R.~Hansen, L.~Plant\'e and P.~Vanhove, \emph{{Classical
  observables from the exponential representation of the gravitational
  S-matrix}}, \href{https://doi.org/10.1007/JHEP09(2023)183}{\emph{JHEP}
  {\bfseries 09} (2023) 183}
  [\href{https://arxiv.org/abs/2307.04746}{{\ttfamily 2307.04746}}].

\bibitem{Dlapa:2024cje}
C.~Dlapa, G.~K\"alin, Z.~Liu and R.A.~Porto, \emph{{Local in Time Conservative
  Binary Dynamics at Fourth Post-Minkowskian Order}},
  \href{https://doi.org/10.1103/PhysRevLett.132.221401}{\emph{Phys. Rev. Lett.}
  {\bfseries 132} (2024) 221401}
  [\href{https://arxiv.org/abs/2403.04853}{{\ttfamily 2403.04853}}].

\bibitem{Driesse:2024xad}
M.~Driesse, G.U.~Jakobsen, G.~Mogull, J.~Plefka, B.~Sauer and J.~Usovitsch,
  \emph{{Conservative Black Hole Scattering at Fifth Post-Minkowskian and First
  Self-Force Order}},
  \href{https://doi.org/10.1103/PhysRevLett.132.241402}{\emph{Phys. Rev. Lett.}
  {\bfseries 132} (2024) 241402}
  [\href{https://arxiv.org/abs/2403.07781}{{\ttfamily 2403.07781}}].

\bibitem{Kalin:2019rwq}
G.~K\"alin and R.A.~Porto, \emph{{From Boundary Data to Bound States}},
  \href{https://doi.org/10.1007/JHEP01(2020)072}{\emph{JHEP} {\bfseries 01}
  (2020) 072} [\href{https://arxiv.org/abs/1910.03008}{{\ttfamily
  1910.03008}}].

\bibitem{Kalin:2019inp}
G.~K\"alin and R.A.~Porto, \emph{{From boundary data to bound states. Part II.
  Scattering angle to dynamical invariants (with twist)}},
  \href{https://doi.org/10.1007/JHEP02(2020)120}{\emph{JHEP} {\bfseries 02}
  (2020) 120} [\href{https://arxiv.org/abs/1911.09130}{{\ttfamily
  1911.09130}}].

\bibitem{Cho:2021arx}
G.~Cho, G.~K\"alin and R.A.~Porto, \emph{{From boundary data to bound states.
  Part III. Radiative effects}},
  \href{https://doi.org/10.1007/JHEP04(2022)154}{\emph{JHEP} {\bfseries 04}
  (2022) 154} [\href{https://arxiv.org/abs/2112.03976}{{\ttfamily
  2112.03976}}].

\bibitem{Adamo:2022ooq}
T.~Adamo and R.~Gonzo, \emph{{Bethe-Salpeter equation for classical
  gravitational bound states}},
  \href{https://doi.org/10.1007/JHEP05(2023)088}{\emph{JHEP} {\bfseries 05}
  (2023) 088} [\href{https://arxiv.org/abs/2212.13269}{{\ttfamily
  2212.13269}}].

\bibitem{Gonzo:2023goe}
R.~Gonzo and C.~Shi, \emph{{Boundary to bound dictionary for generic Kerr
  orbits}}, \href{https://doi.org/10.1103/PhysRevD.108.084065}{\emph{Phys. Rev.
  D} {\bfseries 108} (2023) 084065}
  [\href{https://arxiv.org/abs/2304.06066}{{\ttfamily 2304.06066}}].

\bibitem{Siemonsen:2017yux}
N.~Siemonsen, J.~Steinhoff and J.~Vines, \emph{{Gravitational waves from
  spinning binary black holes at the leading post-Newtonian orders at all
  orders in spin}},
  \href{https://doi.org/10.1103/PhysRevD.97.124046}{\emph{Phys. Rev. D}
  {\bfseries 97} (2018) 124046}
  [\href{https://arxiv.org/abs/1712.08603}{{\ttfamily 1712.08603}}].

\bibitem{Foffa:2019yfl}
S.~Foffa, R.A.~Porto, I.~Rothstein and R.~Sturani, \emph{{Conservative dynamics
  of binary systems to fourth Post-Newtonian order in the EFT approach II:
  Renormalized Lagrangian}},
  \href{https://doi.org/10.1103/PhysRevD.100.024048}{\emph{Phys. Rev. D}
  {\bfseries 100} (2019) 024048}
  [\href{https://arxiv.org/abs/1903.05118}{{\ttfamily 1903.05118}}].

\bibitem{Blumlein:2019zku}
J.~Bl\"umlein, A.~Maier and P.~Marquard, \emph{{Five-Loop Static Contribution
  to the Gravitational Interaction Potential of Two Point Masses}},
  \href{https://doi.org/10.1016/j.physletb.2019.135100}{\emph{Phys. Lett. B}
  {\bfseries 800} (2020) 135100}
  [\href{https://arxiv.org/abs/1902.11180}{{\ttfamily 1902.11180}}].

\bibitem{Blumlein:2020znm}
J.~Bl\"umlein, A.~Maier, P.~Marquard and G.~Sch\"afer, \emph{{Testing binary
  dynamics in gravity at the sixth post-Newtonian level}},
  \href{https://doi.org/10.1016/j.physletb.2020.135496}{\emph{Phys. Lett. B}
  {\bfseries 807} (2020) 135496}
  [\href{https://arxiv.org/abs/2003.07145}{{\ttfamily 2003.07145}}].

\bibitem{Levi:2019kgk}
M.~Levi, S.~Mougiakakos and M.~Vieira, \emph{{Gravitational cubic-in-spin
  interaction at the next-to-leading post-Newtonian order}},
  \href{https://doi.org/10.1007/JHEP01(2021)036}{\emph{JHEP} {\bfseries 01}
  (2021) 036} [\href{https://arxiv.org/abs/1912.06276}{{\ttfamily
  1912.06276}}].

\bibitem{Levi:2020lfn}
M.~Levi and F.~Teng, \emph{{NLO gravitational quartic-in-spin interaction}},
  \href{https://doi.org/10.1007/JHEP01(2021)066}{\emph{JHEP} {\bfseries 01}
  (2021) 066} [\href{https://arxiv.org/abs/2008.12280}{{\ttfamily
  2008.12280}}].

\bibitem{Levi:2022rrq}
M.~Levi and Z.~Yin, \emph{{Completing the fifth PN precision frontier via the
  EFT of spinning gravitating objects}},
  \href{https://doi.org/10.1007/JHEP04(2023)079}{\emph{JHEP} {\bfseries 04}
  (2023) 079} [\href{https://arxiv.org/abs/2211.14018}{{\ttfamily
  2211.14018}}].

\bibitem{Levi:2022dqm}
M.~Levi, R.~Morales and Z.~Yin, \emph{{From the EFT of spinning gravitating
  objects to Poincar\'e and gauge invariance at the 4.5PN precision frontier}},
  \href{https://doi.org/10.1007/JHEP09(2023)090}{\emph{JHEP} {\bfseries 09}
  (2023) 090} [\href{https://arxiv.org/abs/2210.17538}{{\ttfamily
  2210.17538}}].

\bibitem{Mandal:2022ufb}
M.K.~Mandal, P.~Mastrolia, R.~Patil and J.~Steinhoff, \emph{{Gravitational
  quadratic-in-spin Hamiltonian at NNNLO in the post-Newtonian framework}},
  \href{https://doi.org/10.1007/JHEP07(2023)128}{\emph{JHEP} {\bfseries 07}
  (2023) 128} [\href{https://arxiv.org/abs/2210.09176}{{\ttfamily
  2210.09176}}].

\bibitem{Mandal:2022nty}
M.K.~Mandal, P.~Mastrolia, R.~Patil and J.~Steinhoff, \emph{{Gravitational
  spin-orbit Hamiltonian at NNNLO in the post-Newtonian framework}},
  \href{https://doi.org/10.1007/JHEP03(2023)130}{\emph{JHEP} {\bfseries 03}
  (2023) 130} [\href{https://arxiv.org/abs/2209.00611}{{\ttfamily
  2209.00611}}].

\bibitem{Placidi:2024yld}
A.~Placidi, P.~Rettegno and A.~Nagar, \emph{{Gravitational spin-orbit coupling
  through the third-subleading post-Newtonian order: Exploring spin-gauge
  flexibility}}, \href{https://doi.org/10.1103/PhysRevD.109.084065}{\emph{Phys.
  Rev. D} {\bfseries 109} (2024) 084065}
  [\href{https://arxiv.org/abs/2401.12290}{{\ttfamily 2401.12290}}].

\bibitem{Khalil:2021fpm}
M.~Khalil, \emph{{Gravitational spin-orbit dynamics at the fifth-and-a-half
  post-Newtonian order}},
  \href{https://doi.org/10.1103/PhysRevD.104.124015}{\emph{Phys. Rev. D}
  {\bfseries 104} (2021) 124015}
  [\href{https://arxiv.org/abs/2110.12813}{{\ttfamily 2110.12813}}].

\bibitem{Antonelli:2020ybz}
A.~Antonelli, C.~Kavanagh, M.~Khalil, J.~Steinhoff and J.~Vines,
  \emph{{Gravitational spin-orbit and aligned spin$_1$-spin$_2$ couplings
  through third-subleading post-Newtonian orders}},
  \href{https://doi.org/10.1103/PhysRevD.102.124024}{\emph{Phys. Rev. D}
  {\bfseries 102} (2020) 124024}
  [\href{https://arxiv.org/abs/2010.02018}{{\ttfamily 2010.02018}}].

\bibitem{Antonelli:2020aeb}
A.~Antonelli, C.~Kavanagh, M.~Khalil, J.~Steinhoff and J.~Vines,
  \emph{{Gravitational spin-orbit coupling through third-subleading
  post-Newtonian order: from first-order self-force to arbitrary mass ratios}},
  \href{https://doi.org/10.1103/PhysRevLett.125.011103}{\emph{Phys. Rev. Lett.}
  {\bfseries 125} (2020) 011103}
  [\href{https://arxiv.org/abs/2003.11391}{{\ttfamily 2003.11391}}].

\bibitem{Cho:2021mqw}
G.~Cho, B.~Pardo and R.A.~Porto, \emph{{Gravitational radiation from
  inspiralling compact objects: Spin-spin effects completed at the
  next-to-leading post-Newtonian order}},
  \href{https://doi.org/10.1103/PhysRevD.104.024037}{\emph{Phys. Rev. D}
  {\bfseries 104} (2021) 024037}
  [\href{https://arxiv.org/abs/2103.14612}{{\ttfamily 2103.14612}}].

\bibitem{Bini:2023mdz}
D.~Bini, A.~Geralico and P.~Rettegno, \emph{{Spin-orbit contribution to
  radiative losses for spinning binaries with aligned spins}},
  \href{https://doi.org/10.1103/PhysRevD.108.064049}{\emph{Phys. Rev. D}
  {\bfseries 108} (2023) 064049}
  [\href{https://arxiv.org/abs/2307.12670}{{\ttfamily 2307.12670}}].

\bibitem{Goldberger:2004jt}
W.D.~Goldberger and I.Z.~Rothstein, \emph{{An Effective field theory of gravity
  for extended objects}},
  \href{https://doi.org/10.1103/PhysRevD.73.104029}{\emph{Phys. Rev.}
  {\bfseries D73} (2006) 104029}
  [\href{https://arxiv.org/abs/hep-th/0409156}{{\ttfamily hep-th/0409156}}].

\bibitem{Goldberger:2007hy}
W.D.~Goldberger, \emph{Les houches lectures on effective field theories and
  gravitational radiation},  in \emph{Les Houches Summer School - Session 86:
  Particle Physics and Cosmology: The Fabric of Spacetime}, 1, 2007
  [\href{https://arxiv.org/abs/hep-ph/0701129}{{\ttfamily hep-ph/0701129}}].

\bibitem{Kol:2007bc}
B.~Kol and M.~Smolkin, \emph{{Non-Relativistic Gravitation: From Newton to
  Einstein and Back}},
  \href{https://doi.org/10.1088/0264-9381/25/14/145011}{\emph{Class. Quant.
  Grav.} {\bfseries 25} (2008) 145011}
  [\href{https://arxiv.org/abs/0712.4116}{{\ttfamily 0712.4116}}].

\bibitem{Goldberger:2009qd}
W.D.~Goldberger and A.~Ross, \emph{{Gravitational radiative corrections from
  effective field theory}},
  \href{https://doi.org/10.1103/PhysRevD.81.124015}{\emph{Phys. Rev.}
  {\bfseries D81} (2010) 124015}
  [\href{https://arxiv.org/abs/0912.4254}{{\ttfamily 0912.4254}}].

\bibitem{Foffa:2013qca}
S.~Foffa and R.~Sturani, \emph{{Effective field theory methods to model compact
  binaries}}, \href{https://doi.org/10.1088/0264-9381/31/4/043001}{\emph{Class.
  Quant. Grav.} {\bfseries 31} (2014) 043001}
  [\href{https://arxiv.org/abs/1309.3474}{{\ttfamily 1309.3474}}].

\bibitem{Foffa:2016rgu}
S.~Foffa, P.~Mastrolia, R.~Sturani and C.~Sturm, \emph{{Effective field theory
  approach to the gravitational two-body dynamics, at fourth post-Newtonian
  order and quintic in the Newton constant}},
  \href{https://doi.org/10.1103/PhysRevD.95.104009}{\emph{Phys. Rev.}
  {\bfseries D95} (2017) 104009}
  [\href{https://arxiv.org/abs/1612.00482}{{\ttfamily 1612.00482}}].

\bibitem{Kalin:2020mvi}
G.~K\"alin and R.A.~Porto, \emph{{Post-Minkowskian Effective Field Theory for
  Conservative Binary Dynamics}},
  \href{https://doi.org/10.1007/JHEP11(2020)106}{\emph{JHEP} {\bfseries 11}
  (2020) 106} [\href{https://arxiv.org/abs/2006.01184}{{\ttfamily
  2006.01184}}].

\bibitem{Porto:2005ac}
R.A.~Porto, \emph{{Post-Newtonian corrections to the motion of spinning bodies
  in NRGR}}, \href{https://doi.org/10.1103/PhysRevD.73.104031}{\emph{Phys.
  Rev.} {\bfseries D73} (2006) 104031}
  [\href{https://arxiv.org/abs/gr-qc/0511061}{{\ttfamily gr-qc/0511061}}].

\bibitem{Porto:2006bt}
R.A.~Porto and I.Z.~Rothstein, \emph{{The Hyperfine Einstein-Infeld-Hoffmann
  potential}}, \href{https://doi.org/10.1103/PhysRevLett.97.021101}{\emph{Phys.
  Rev. Lett.} {\bfseries 97} (2006) 021101}
  [\href{https://arxiv.org/abs/gr-qc/0604099}{{\ttfamily gr-qc/0604099}}].

\bibitem{Porto:2008tb}
R.A.~Porto and I.Z.~Rothstein, \emph{{Spin(1)Spin(2) Effects in the Motion of
  Inspiralling Compact Binaries at Third Order in the Post-Newtonian
  Expansion}}, \href{https://doi.org/10.1103/PhysRevD.78.044012}{\emph{Phys.
  Rev. D} {\bfseries 78} (2008) 044012}
  [\href{https://arxiv.org/abs/0802.0720}{{\ttfamily 0802.0720}}].

\bibitem{Porto:2008jj}
R.A.~Porto and I.Z.~Rothstein, \emph{{Next to Leading Order Spin(1)Spin(1)
  Effects in the Motion of Inspiralling Compact Binaries}},
  \href{https://doi.org/10.1103/PhysRevD.78.044013}{\emph{Phys. Rev. D}
  {\bfseries 78} (2008) 044013}
  [\href{https://arxiv.org/abs/0804.0260}{{\ttfamily 0804.0260}}].

\bibitem{Levi:2008nh}
M.~Levi, \emph{{Next to Leading Order gravitational Spin1-Spin2 coupling with
  Kaluza-Klein reduction}},
  \href{https://doi.org/10.1103/PhysRevD.82.064029}{\emph{Phys. Rev. D}
  {\bfseries 82} (2010) 064029}
  [\href{https://arxiv.org/abs/0802.1508}{{\ttfamily 0802.1508}}].

\bibitem{Porto:2010tr}
R.A.~Porto, \emph{{Next to leading order spin-orbit effects in the motion of
  inspiralling compact binaries}},
  \href{https://doi.org/10.1088/0264-9381/27/20/205001}{\emph{Class. Quant.
  Grav.} {\bfseries 27} (2010) 205001}
  [\href{https://arxiv.org/abs/1005.5730}{{\ttfamily 1005.5730}}].

\bibitem{Porto:2010zg}
R.A.~Porto, A.~Ross and I.Z.~Rothstein, \emph{{Spin induced multipole moments
  for the gravitational wave flux from binary inspirals to third Post-Newtonian
  order}}, \href{https://doi.org/10.1088/1475-7516/2011/03/009}{\emph{JCAP}
  {\bfseries 03} (2011) 009} [\href{https://arxiv.org/abs/1007.1312}{{\ttfamily
  1007.1312}}].

\bibitem{Levi:2010zu}
M.~Levi, \emph{{Next to Leading Order gravitational Spin-Orbit coupling in an
  Effective Field Theory approach}},
  \href{https://doi.org/10.1103/PhysRevD.82.104004}{\emph{Phys. Rev. D}
  {\bfseries 82} (2010) 104004}
  [\href{https://arxiv.org/abs/1006.4139}{{\ttfamily 1006.4139}}].

\bibitem{Levi:2011eq}
M.~Levi, \emph{{Binary dynamics from spin1-spin2 coupling at fourth
  post-Newtonian order}},
  \href{https://doi.org/10.1103/PhysRevD.85.064043}{\emph{Phys. Rev. D}
  {\bfseries 85} (2012) 064043}
  [\href{https://arxiv.org/abs/1107.4322}{{\ttfamily 1107.4322}}].

\bibitem{Porto:2012as}
R.A.~Porto, A.~Ross and I.Z.~Rothstein, \emph{{Spin induced multipole moments
  for the gravitational wave amplitude from binary inspirals to 2.5
  Post-Newtonian order}},
  \href{https://doi.org/10.1088/1475-7516/2012/09/028}{\emph{JCAP} {\bfseries
  09} (2012) 028} [\href{https://arxiv.org/abs/1203.2962}{{\ttfamily
  1203.2962}}].

\bibitem{Levi:2014gsa}
M.~Levi and J.~Steinhoff, \emph{{Leading order finite size effects with spins
  for inspiralling compact binaries}},
  \href{https://doi.org/10.1007/JHEP06(2015)059}{\emph{JHEP} {\bfseries 06}
  (2015) 059} [\href{https://arxiv.org/abs/1410.2601}{{\ttfamily 1410.2601}}].

\bibitem{Levi:2014sba}
M.~Levi and J.~Steinhoff, \emph{{Equivalence of ADM Hamiltonian and Effective
  Field Theory approaches at next-to-next-to-leading order spin1-spin2 coupling
  of binary inspirals}},
  \href{https://doi.org/10.1088/1475-7516/2014/12/003}{\emph{JCAP} {\bfseries
  12} (2014) 003} [\href{https://arxiv.org/abs/1408.5762}{{\ttfamily
  1408.5762}}].

\bibitem{Levi:2015msa}
M.~Levi and J.~Steinhoff, \emph{{Spinning gravitating objects in the effective
  field theory in the post-Newtonian scheme}},
  \href{https://doi.org/10.1007/JHEP09(2015)219}{\emph{JHEP} {\bfseries 09}
  (2015) 219} [\href{https://arxiv.org/abs/1501.04956}{{\ttfamily
  1501.04956}}].

\bibitem{Levi:2015uxa}
M.~Levi and J.~Steinhoff, \emph{{Next-to-next-to-leading order gravitational
  spin-orbit coupling via the effective field theory for spinning objects in
  the post-Newtonian scheme}},
  \href{https://doi.org/10.1088/1475-7516/2016/01/011}{\emph{JCAP} {\bfseries
  01} (2016) 011} [\href{https://arxiv.org/abs/1506.05056}{{\ttfamily
  1506.05056}}].

\bibitem{Levi:2015ixa}
M.~Levi and J.~Steinhoff, \emph{{Next-to-next-to-leading order gravitational
  spin-squared potential via the effective field theory for spinning objects in
  the post-Newtonian scheme}},
  \href{https://doi.org/10.1088/1475-7516/2016/01/008}{\emph{JCAP} {\bfseries
  1601} (2016) 008} [\href{https://arxiv.org/abs/1506.05794}{{\ttfamily
  1506.05794}}].

\bibitem{Levi:2016ofk}
M.~Levi and J.~Steinhoff, \emph{{Complete conservative dynamics for
  inspiralling compact binaries with spins at fourth post-Newtonian order}},
  \href{https://arxiv.org/abs/1607.04252}{{\ttfamily 1607.04252}}.

\bibitem{Maia:2017yok}
N.T.~Maia, C.R.~Galley, A.K.~Leibovich and R.A.~Porto, \emph{{Radiation
  reaction for spinning bodies in effective field theory II: Spin-spin
  effects}}, \href{https://doi.org/10.1103/PhysRevD.96.084065}{\emph{Phys. Rev.
  D} {\bfseries 96} (2017) 084065}
  [\href{https://arxiv.org/abs/1705.07938}{{\ttfamily 1705.07938}}].

\bibitem{Maia:2017gxn}
N.T.~Maia, C.R.~Galley, A.K.~Leibovich and R.A.~Porto, \emph{{Radiation
  reaction for spinning bodies in effective field theory I: Spin-orbit
  effects}}, \href{https://doi.org/10.1103/PhysRevD.96.084064}{\emph{Phys. Rev.
  D} {\bfseries 96} (2017) 084064}
  [\href{https://arxiv.org/abs/1705.07934}{{\ttfamily 1705.07934}}].

\bibitem{Levi:2020kvb}
M.~Levi, A.J.~Mcleod and M.~Von~Hippel, \emph{{N$^{3}$LO gravitational
  spin-orbit coupling at order G$^{4}$}},
  \href{https://doi.org/10.1007/JHEP07(2021)115}{\emph{JHEP} {\bfseries 07}
  (2021) 115} [\href{https://arxiv.org/abs/2003.02827}{{\ttfamily
  2003.02827}}].

\bibitem{Levi:2020uwu}
M.~Levi, A.J.~Mcleod and M.~Von~Hippel, \emph{{N$^{3}$LO gravitational
  quadratic-in-spin interactions at G$^{4}$}},
  \href{https://doi.org/10.1007/JHEP07(2021)116}{\emph{JHEP} {\bfseries 07}
  (2021) 116} [\href{https://arxiv.org/abs/2003.07890}{{\ttfamily
  2003.07890}}].

\bibitem{Liu:2021zxr}
Z.~Liu, R.A.~Porto and Z.~Yang, \emph{{Spin Effects in the Effective Field
  Theory Approach to Post-Minkowskian Conservative Dynamics}},
  \href{https://doi.org/10.1007/JHEP06(2021)012}{\emph{JHEP} {\bfseries 06}
  (2021) 012} [\href{https://arxiv.org/abs/2102.10059}{{\ttfamily
  2102.10059}}].

\bibitem{Maybee:2019jus}
B.~Maybee, D.~O'Connell and J.~Vines, \emph{{Observables and amplitudes for
  spinning particles and black holes}},
  \href{https://doi.org/10.1007/JHEP12(2019)156}{\emph{JHEP} {\bfseries 12}
  (2019) 156} [\href{https://arxiv.org/abs/1906.09260}{{\ttfamily
  1906.09260}}].

\bibitem{Cristofoli:2021vyo}
A.~Cristofoli, R.~Gonzo, D.A.~Kosower and D.~O'Connell, \emph{{Waveforms from
  amplitudes}}, \href{https://doi.org/10.1103/PhysRevD.106.056007}{\emph{Phys.
  Rev. D} {\bfseries 106} (2022) 056007}
  [\href{https://arxiv.org/abs/2107.10193}{{\ttfamily 2107.10193}}].

\bibitem{Cristofoli:2021jas}
A.~Cristofoli, R.~Gonzo, N.~Moynihan, D.~O'Connell, A.~Ross, M.~Sergola et~al.,
  \emph{{The uncertainty principle and classical amplitudes}},
  \href{https://doi.org/10.1007/JHEP06(2024)181}{\emph{JHEP} {\bfseries 06}
  (2024) 181} [\href{https://arxiv.org/abs/2112.07556}{{\ttfamily
  2112.07556}}].

\bibitem{Cristofoli:2022phh}
A.~Cristofoli, A.~Elkhidir, A.~Ilderton and D.~O'Connell, \emph{{Large gauge
  effects and the structure of amplitudes}},
  \href{https://doi.org/10.1007/JHEP06(2023)204}{\emph{JHEP} {\bfseries 06}
  (2023) 204} [\href{https://arxiv.org/abs/2211.16438}{{\ttfamily
  2211.16438}}].

\bibitem{Adamo:2022rmp}
T.~Adamo, A.~Cristofoli and A.~Ilderton, \emph{{Classical physics from
  amplitudes on curved backgrounds}},
  \href{https://doi.org/10.1007/JHEP08(2022)281}{\emph{JHEP} {\bfseries 08}
  (2022) 281} [\href{https://arxiv.org/abs/2203.13785}{{\ttfamily
  2203.13785}}].

\bibitem{DiVecchia:2023frv}
P.~Di~Vecchia, C.~Heissenberg, R.~Russo and G.~Veneziano, \emph{{The
  gravitational eikonal: From particle, string and brane collisions to
  black-hole encounters}},
  \href{https://doi.org/10.1016/j.physrep.2024.06.002}{\emph{Phys. Rept.}
  {\bfseries 1083} (2024) 1}
  [\href{https://arxiv.org/abs/2306.16488}{{\ttfamily 2306.16488}}].

\bibitem{KoemansCollado:2019ggb}
A.~Koemans~Collado, P.~Di~Vecchia and R.~Russo, \emph{{Revisiting the second
  post-Minkowskian eikonal and the dynamics of binary black holes}},
  \href{https://doi.org/10.1103/PhysRevD.100.066028}{\emph{Phys. Rev.}
  {\bfseries D100} (2019) 066028}
  [\href{https://arxiv.org/abs/1904.02667}{{\ttfamily 1904.02667}}].

\bibitem{DiVecchia:2021bdo}
P.~Di~Vecchia, C.~Heissenberg, R.~Russo and G.~Veneziano, \emph{{The eikonal
  approach to gravitational scattering and radiation at $ \mathcal{O}
  $(G$^{3}$)}}, \href{https://doi.org/10.1007/JHEP07(2021)169}{\emph{JHEP}
  {\bfseries 07} (2021) 169}
  [\href{https://arxiv.org/abs/2104.03256}{{\ttfamily 2104.03256}}].

\bibitem{Heissenberg:2021tzo}
C.~Heissenberg, \emph{{Infrared divergences and the eikonal exponentiation}},
  \href{https://doi.org/10.1103/PhysRevD.104.046016}{\emph{Phys. Rev. D}
  {\bfseries 104} (2021) 046016}
  [\href{https://arxiv.org/abs/2105.04594}{{\ttfamily 2105.04594}}].

\bibitem{Haddad:2021znf}
K.~Haddad, \emph{{Exponentiation of the leading eikonal phase with spin}},
  \href{https://doi.org/10.1103/PhysRevD.105.026004}{\emph{Phys. Rev. D}
  {\bfseries 105} (2022) 026004}
  [\href{https://arxiv.org/abs/2109.04427}{{\ttfamily 2109.04427}}].

\bibitem{Adamo:2021rfq}
T.~Adamo, A.~Cristofoli and P.~Tourkine, \emph{{Eikonal amplitudes from curved
  backgrounds}},
  \href{https://doi.org/10.21468/SciPostPhys.13.2.032}{\emph{SciPost Phys.}
  {\bfseries 13} (2022) 032}
  [\href{https://arxiv.org/abs/2112.09113}{{\ttfamily 2112.09113}}].

\bibitem{DiVecchia:2022piu}
P.~Di~Vecchia, C.~Heissenberg, R.~Russo and G.~Veneziano, \emph{{Classical
  gravitational observables from the Eikonal operator}},
  \href{https://doi.org/10.1016/j.physletb.2023.138049}{\emph{Phys. Lett. B}
  {\bfseries 843} (2023) 138049}
  [\href{https://arxiv.org/abs/2210.12118}{{\ttfamily 2210.12118}}].

\bibitem{Bellazzini:2022wzv}
B.~Bellazzini, G.~Isabella and M.M.~Riva, \emph{{Classical vs quantum eikonal
  scattering and its causal structure}},
  \href{https://doi.org/10.1007/JHEP04(2023)023}{\emph{JHEP} {\bfseries 04}
  (2023) 023} [\href{https://arxiv.org/abs/2211.00085}{{\ttfamily
  2211.00085}}].

\bibitem{Luna:2023uwd}
A.~Luna, N.~Moynihan, D.~O'Connell and A.~Ross, \emph{{Observables from the
  Spinning Eikonal}},  \href{https://arxiv.org/abs/2312.09960}{{\ttfamily
  2312.09960}}.

\bibitem{Gatica:2023iws}
J.P.~Gatica, \emph{{The Eikonal Phase and Spinning Observables}},
  \href{https://arxiv.org/abs/2312.04680}{{\ttfamily 2312.04680}}.

\bibitem{Fernandes:2024xqr}
K.~Fernandes and F.-L.~Lin, \emph{{Next-to-eikonal corrected double graviton
  dressing and gravitational wave observables at $
  \mathcal{O}\left({G}^2\right) $}},
  \href{https://doi.org/10.1007/JHEP06(2024)015}{\emph{JHEP} {\bfseries 06}
  (2024) 015} [\href{https://arxiv.org/abs/2401.03900}{{\ttfamily
  2401.03900}}].

\bibitem{Du:2024rkf}
Y.~Du, S.~Ajith, R.~Rajagopal and D.~Vaman, \emph{{Worldline Proof of Eikonal
  Exponentiation}},  \href{https://arxiv.org/abs/2409.12895}{{\ttfamily
  2409.12895}}.

\bibitem{Mogull:2020sak}
G.~Mogull, J.~Plefka and J.~Steinhoff, \emph{{Classical black hole scattering
  from a worldline quantum field theory}},
  \href{https://doi.org/10.1007/JHEP02(2021)048}{\emph{JHEP} {\bfseries 02}
  (2021) 048} [\href{https://arxiv.org/abs/2010.02865}{{\ttfamily
  2010.02865}}].

\bibitem{Jakobsen:2021smu}
G.U.~Jakobsen, G.~Mogull, J.~Plefka and J.~Steinhoff, \emph{{Classical
  Gravitational Bremsstrahlung from a Worldline Quantum Field Theory}},
  \href{https://doi.org/10.1103/PhysRevLett.126.201103}{\emph{Phys. Rev. Lett.}
  {\bfseries 126} (2021) 201103}
  [\href{https://arxiv.org/abs/2101.12688}{{\ttfamily 2101.12688}}].

\bibitem{Jakobsen:2021lvp}
G.U.~Jakobsen, G.~Mogull, J.~Plefka and J.~Steinhoff, \emph{{Gravitational
  Bremsstrahlung and Hidden Supersymmetry of Spinning Bodies}},
  \href{https://doi.org/10.1103/PhysRevLett.128.011101}{\emph{Phys. Rev. Lett.}
  {\bfseries 128} (2022) 011101}
  [\href{https://arxiv.org/abs/2106.10256}{{\ttfamily 2106.10256}}].

\bibitem{Jakobsen:2021zvh}
G.U.~Jakobsen, G.~Mogull, J.~Plefka and J.~Steinhoff, \emph{{SUSY in the sky
  with gravitons}}, \href{https://doi.org/10.1007/JHEP01(2022)027}{\emph{JHEP}
  {\bfseries 01} (2022) 027}
  [\href{https://arxiv.org/abs/2109.04465}{{\ttfamily 2109.04465}}].

\bibitem{Comberiati:2022cpm}
F.~Comberiati and C.~Shi, \emph{{Classical Double Copy of Spinning Worldline
  Quantum Field Theory}},
  \href{https://doi.org/10.1007/JHEP04(2023)008}{\emph{JHEP} {\bfseries 04}
  (2023) 008} [\href{https://arxiv.org/abs/2212.13855}{{\ttfamily
  2212.13855}}].

\bibitem{Ben-Shahar:2023djm}
M.~Ben-Shahar, \emph{{Scattering of spinning compact objects from a worldline
  EFT}},  \href{https://arxiv.org/abs/2311.01430}{{\ttfamily 2311.01430}}.

\bibitem{Damgaard:2023vnx}
P.H.~Damgaard, E.R.~Hansen, L.~Plant\'e and P.~Vanhove, \emph{{The relation
  between KMOC and worldline formalisms for classical gravity}},
  \href{https://doi.org/10.1007/JHEP09(2023)059}{\emph{JHEP} {\bfseries 09}
  (2023) 059} [\href{https://arxiv.org/abs/2306.11454}{{\ttfamily
  2306.11454}}].

\bibitem{Damgaard:2019lfh}
P.H.~Damgaard, K.~Haddad and A.~Helset, \emph{{Heavy Black Hole Effective
  Theory}}, \href{https://doi.org/10.1007/JHEP11(2019)070}{\emph{JHEP}
  {\bfseries 11} (2019) 070}
  [\href{https://arxiv.org/abs/1908.10308}{{\ttfamily 1908.10308}}].

\bibitem{Aoude:2020onz}
R.~Aoude, K.~Haddad and A.~Helset, \emph{{On-shell heavy particle effective
  theories}}, \href{https://doi.org/10.1007/JHEP05(2020)051}{\emph{JHEP}
  {\bfseries 05} (2020) 051}
  [\href{https://arxiv.org/abs/2001.09164}{{\ttfamily 2001.09164}}].

\bibitem{Bern:2008qj}
Z.~Bern, J.~Carrasco and H.~Johansson, \emph{{New Relations for Gauge-Theory
  Amplitudes}},
  \href{https://doi.org/10.1103/PhysRevD.78.085011}{\emph{Phys.Rev.} {\bfseries
  D78} (2008) 085011} [\href{https://arxiv.org/abs/0805.3993}{{\ttfamily
  0805.3993}}].

\bibitem{Bern:2010ue}
Z.~Bern, J.J.M.~Carrasco and H.~Johansson, \emph{{Perturbative Quantum Gravity
  as a Double Copy of Gauge Theory}},
  \href{https://doi.org/10.1103/PhysRevLett.105.061602}{\emph{Phys.Rev.Lett.}
  {\bfseries 105} (2010) 061602}
  [\href{https://arxiv.org/abs/1004.0476}{{\ttfamily 1004.0476}}].

\bibitem{Bern:2019prr}
Z.~Bern, J.J.~Carrasco, M.~Chiodaroli, H.~Johansson and R.~Roiban, \emph{{The
  Duality Between Color and Kinematics and its Applications}},
  \href{https://arxiv.org/abs/1909.01358}{{\ttfamily 1909.01358}}.

\bibitem{Adamo:2022dcm}
T.~Adamo, J.J.M.~Carrasco, M.~Carrillo-Gonz\'alez, M.~Chiodaroli, H.~Elvang,
  H.~Johansson et~al., \emph{{Snowmass White Paper: the Double Copy and its
  Applications}},  in \emph{{Snowmass 2021}}, 4, 2022
  [\href{https://arxiv.org/abs/2204.06547}{{\ttfamily 2204.06547}}].

\bibitem{Bern:2022wqg}
Z.~Bern, J.J.~Carrasco, M.~Chiodaroli, H.~Johansson and R.~Roiban, \emph{{The
  SAGEX review on scattering amplitudes Chapter 2: An invitation to
  color-kinematics duality and the double copy}},
  \href{https://doi.org/10.1088/1751-8121/ac93cf}{\emph{J. Phys. A} {\bfseries
  55} (2022) 443003} [\href{https://arxiv.org/abs/2203.13013}{{\ttfamily
  2203.13013}}].

\bibitem{Luna:2016due}
A.~Luna, R.~Monteiro, I.~Nicholson, D.~O'Connell and C.D.~White, \emph{{The
  double copy: Bremsstrahlung and accelerating black holes}},
  \href{https://doi.org/10.1007/JHEP06(2016)023}{\emph{JHEP} {\bfseries 06}
  (2016) 023} [\href{https://arxiv.org/abs/1603.05737}{{\ttfamily
  1603.05737}}].

\bibitem{Luna:2017dtq}
A.~Luna, I.~Nicholson, D.~O'Connell and C.D.~White, \emph{{Inelastic Black Hole
  Scattering from Charged Scalar Amplitudes}},
  \href{https://doi.org/10.1007/JHEP03(2018)044}{\emph{JHEP} {\bfseries 03}
  (2018) 044} [\href{https://arxiv.org/abs/1711.03901}{{\ttfamily
  1711.03901}}].

\bibitem{Shen:2018ebu}
C.-H.~Shen, \emph{{Gravitational Radiation from Color-Kinematics Duality}},
  \href{https://doi.org/10.1007/JHEP11(2018)162}{\emph{JHEP} {\bfseries 11}
  (2018) 162} [\href{https://arxiv.org/abs/1806.07388}{{\ttfamily
  1806.07388}}].

\bibitem{Li:2018qap}
J.~Li and S.G.~Prabhu, \emph{{Gravitational radiation from the classical
  spinning double copy}},
  \href{https://doi.org/10.1103/PhysRevD.97.105019}{\emph{Phys. Rev.}
  {\bfseries D97} (2018) 105019}
  [\href{https://arxiv.org/abs/1803.02405}{{\ttfamily 1803.02405}}].

\bibitem{Goldberger:2019xef}
W.D.~Goldberger and J.~Li, \emph{{Strings, extended objects, and the classical
  double copy}}, \href{https://doi.org/10.1007/JHEP02(2020)092}{\emph{JHEP}
  {\bfseries 02} (2020) 092}
  [\href{https://arxiv.org/abs/1912.01650}{{\ttfamily 1912.01650}}].

\bibitem{Plefka:2019wyg}
J.~Plefka, C.~Shi and T.~Wang, \emph{{Double copy of massive scalar QCD}},
  \href{https://doi.org/10.1103/PhysRevD.101.066004}{\emph{Phys. Rev. D}
  {\bfseries 101} (2020) 066004}
  [\href{https://arxiv.org/abs/1911.06785}{{\ttfamily 1911.06785}}].

\bibitem{Bautista:2019tdr}
Y.F.~Bautista and A.~Guevara, \emph{{From Scattering Amplitudes to Classical
  Physics: Universality, Double Copy and Soft Theorems}},
  \href{https://arxiv.org/abs/1903.12419}{{\ttfamily 1903.12419}}.

\bibitem{Kim:2019jwm}
K.~Kim, K.~Lee, R.~Monteiro, I.~Nicholson and D.~Peinador~Veiga, \emph{{The
  Classical Double Copy of a Point Charge}},
  \href{https://doi.org/10.1007/JHEP02(2020)046}{\emph{JHEP} {\bfseries 02}
  (2020) 046} [\href{https://arxiv.org/abs/1912.02177}{{\ttfamily
  1912.02177}}].

\bibitem{Monteiro:2020plf}
R.~Monteiro, D.~O'Connell, D.~Peinador~Veiga and M.~Sergola, \emph{{Classical
  solutions and their double copy in split signature}},
  \href{https://doi.org/10.1007/JHEP05(2021)268}{\emph{JHEP} {\bfseries 05}
  (2021) 268} [\href{https://arxiv.org/abs/2012.11190}{{\ttfamily
  2012.11190}}].

\bibitem{Haddad:2020tvs}
K.~Haddad and A.~Helset, \emph{{The double copy for heavy particles}},
  \href{https://doi.org/10.1103/PhysRevLett.125.181603}{\emph{Phys. Rev. Lett.}
  {\bfseries 125} (2020) 181603}
  [\href{https://arxiv.org/abs/2005.13897}{{\ttfamily 2005.13897}}].

\bibitem{Carrasco:2020ywq}
J.J.M.~Carrasco and I.A.~Vazquez-Holm, \emph{{Loop-Level Double-Copy for
  Massive Quantum Particles}},
  \href{https://doi.org/10.1103/PhysRevD.103.045002}{\emph{Phys. Rev. D}
  {\bfseries 103} (2021) 045002}
  [\href{https://arxiv.org/abs/2010.13435}{{\ttfamily 2010.13435}}].

\bibitem{Carrasco:2021bmu}
J.J.M.~Carrasco and I.A.~Vazquez-Holm, \emph{{Extracting Einstein from the
  loop-level double-copy}},
  \href{https://doi.org/10.1007/JHEP11(2021)088}{\emph{JHEP} {\bfseries 11}
  (2021) 088} [\href{https://arxiv.org/abs/2108.06798}{{\ttfamily
  2108.06798}}].

\bibitem{Brandhuber:2021kpo}
A.~Brandhuber, G.~Chen, G.~Travaglini and C.~Wen, \emph{{A new gauge-invariant
  double copy for heavy-mass effective theory}},
  \href{https://doi.org/10.1007/JHEP07(2021)047}{\emph{JHEP} {\bfseries 07}
  (2021) 047} [\href{https://arxiv.org/abs/2104.11206}{{\ttfamily
  2104.11206}}].

\bibitem{Gonzo:2021drq}
R.~Gonzo and C.~Shi, \emph{{Geodesics from classical double copy}},
  \href{https://doi.org/10.1103/PhysRevD.104.105012}{\emph{Phys. Rev. D}
  {\bfseries 104} (2021) 105012}
  [\href{https://arxiv.org/abs/2109.01072}{{\ttfamily 2109.01072}}].

\bibitem{Shi:2021qsb}
C.~Shi and J.~Plefka, \emph{{Classical double copy of worldline quantum field
  theory}}, \href{https://doi.org/10.1103/PhysRevD.105.026007}{\emph{Phys. Rev.
  D} {\bfseries 105} (2022) 026007}
  [\href{https://arxiv.org/abs/2109.10345}{{\ttfamily 2109.10345}}].

\bibitem{Almeida:2020mrg}
G.L.~Almeida, S.~Foffa and R.~Sturani, \emph{{Classical Gravitational
  Self-Energy from Double Copy}},
  \href{https://doi.org/10.1007/JHEP11(2020)165}{\emph{JHEP} {\bfseries 11}
  (2020) 165} [\href{https://arxiv.org/abs/2008.06195}{{\ttfamily
  2008.06195}}].

\bibitem{Wang:2022ntx}
T.~Wang, \emph{{Binary dynamics from worldline QFT for scalar QED}},
  \href{https://doi.org/10.1103/PhysRevD.107.085011}{\emph{Phys. Rev. D}
  {\bfseries 107} (2023) 085011}
  [\href{https://arxiv.org/abs/2205.15753}{{\ttfamily 2205.15753}}].

\bibitem{CarrilloGonzalez:2022mxx}
M.~Carrillo~Gonz\'alez, A.~Momeni and J.~Rumbutis, \emph{{Cotton double copy
  for gravitational waves}},
  \href{https://doi.org/10.1103/PhysRevD.106.025006}{\emph{Phys. Rev. D}
  {\bfseries 106} (2022) 025006}
  [\href{https://arxiv.org/abs/2202.10476}{{\ttfamily 2202.10476}}].

\bibitem{Saha:2019tub}
A.P.~Saha, B.~Sahoo and A.~Sen, \emph{{Proof of the classical soft graviton
  theorem in $D$ = 4}},
  \href{https://doi.org/10.1007/JHEP06(2020)153}{\emph{JHEP} {\bfseries 06}
  (2020) 153} [\href{https://arxiv.org/abs/1912.06413}{{\ttfamily
  1912.06413}}].

\bibitem{Manu:2020zxl}
A.~Manu, D.~Ghosh, A.~Laddha and P.V.~Athira, \emph{{Soft radiation from
  scattering amplitudes revisited}},
  \href{https://doi.org/10.1007/JHEP05(2021)056}{\emph{JHEP} {\bfseries 05}
  (2021) 056} [\href{https://arxiv.org/abs/2007.02077}{{\ttfamily
  2007.02077}}].

\bibitem{DiVecchia:2021ndb}
P.~Di~Vecchia, C.~Heissenberg, R.~Russo and G.~Veneziano, \emph{{Radiation
  Reaction from Soft Theorems}},
  \href{https://doi.org/10.1016/j.physletb.2021.136379}{\emph{Phys. Lett. B}
  {\bfseries 818} (2021) 136379}
  [\href{https://arxiv.org/abs/2101.05772}{{\ttfamily 2101.05772}}].

\bibitem{Alessio:2022kwv}
F.~Alessio and P.~Di~Vecchia, \emph{{Radiation reaction for spinning black-hole
  scattering}},
  \href{https://doi.org/10.1016/j.physletb.2022.137258}{\emph{Phys. Lett. B}
  {\bfseries 832} (2022) 137258}
  [\href{https://arxiv.org/abs/2203.13272}{{\ttfamily 2203.13272}}].

\bibitem{A:2022wsk}
M.~A. and D.~Ghosh, \emph{{Classical spinning soft factors from gauge theory
  amplitudes}},  \href{https://arxiv.org/abs/2210.07561}{{\ttfamily
  2210.07561}}.

\bibitem{Alessio:2024wmz}
F.~Alessio and P.~Di~Vecchia, \emph{{2PM waveform from loop corrected soft
  theorems}},  \href{https://arxiv.org/abs/2402.06533}{{\ttfamily 2402.06533}}.

\bibitem{Elkhidir:2024izo}
A.~Elkhidir, D.~O'Connell and R.~Roiban, \emph{{Supertranslations from
  Scattering Amplitudes}},  \href{https://arxiv.org/abs/2408.15961}{{\ttfamily
  2408.15961}}.

\bibitem{Akhtar:2024lkk}
S.~Akhtar, \emph{{On the classical limit of the (sub)$^{n}$-leading soft
  graviton theorems in $D = 4$ without deflection}},
  \href{https://arxiv.org/abs/2409.12898}{{\ttfamily 2409.12898}}.

\bibitem{Alessio:2024onn}
F.~Alessio, P.~Di~Vecchia and C.~Heissenberg, \emph{{Logarithmic soft theorems
  and soft spectra}},  \href{https://arxiv.org/abs/2407.04128}{{\ttfamily
  2407.04128}}.

\bibitem{Cachazo:2017jef}
F.~Cachazo and A.~Guevara, \emph{{Leading Singularities and Classical
  Gravitational Scattering}},
  \href{https://arxiv.org/abs/1705.10262}{{\ttfamily 1705.10262}}.

\bibitem{Guevara:2017csg}
A.~Guevara, \emph{{Holomorphic Classical Limit for Spin Effects in
  Gravitational and Electromagnetic Scattering}},
  \href{https://doi.org/10.1007/JHEP04(2019)033}{\emph{JHEP} {\bfseries 04}
  (2019) 033} [\href{https://arxiv.org/abs/1706.02314}{{\ttfamily
  1706.02314}}].

\bibitem{Kim:2023vgb}
J.-H.~Kim, \emph{{Asymptotic Spinspacetime}},
  \href{https://arxiv.org/abs/2309.11886}{{\ttfamily 2309.11886}}.

\bibitem{Kim:2023aff}
J.-H.~Kim and S.~Lee, \emph{{Symplectic Perturbation Theory in Massive
  Ambitwistor Space: A Zig-Zag Theory of Massive Spinning Particles}},
  \href{https://arxiv.org/abs/2301.06203}{{\ttfamily 2301.06203}}.

\bibitem{Kim:2024grz}
J.-H.~Kim, J.-W.~Kim and S.~Lee, \emph{{Massive twistor worldline in
  electromagnetic fields}},
  \href{https://doi.org/10.1007/JHEP08(2024)080}{\emph{JHEP} {\bfseries 08}
  (2024) 080} [\href{https://arxiv.org/abs/2405.17056}{{\ttfamily
  2405.17056}}].

\bibitem{Porto:2016pyg}
R.A.~Porto, \emph{{The effective field theorist’s approach to gravitational
  dynamics}}, \href{https://doi.org/10.1016/j.physrep.2016.04.003}{\emph{Phys.
  Rept.} {\bfseries 633} (2016) 1}
  [\href{https://arxiv.org/abs/1601.04914}{{\ttfamily 1601.04914}}].

\bibitem{Barack:2018yly}
L.~Barack et~al., \emph{{Black holes, gravitational waves and fundamental
  physics: a roadmap}},
  \href{https://doi.org/10.1088/1361-6382/ab0587}{\emph{Class. Quant. Grav.}
  {\bfseries 36} (2019) 143001}
  [\href{https://arxiv.org/abs/1806.05195}{{\ttfamily 1806.05195}}].

\bibitem{Levi:2018nxp}
M.~Levi, \emph{{Effective Field Theories of Post-Newtonian Gravity: A
  comprehensive review}},
  \href{https://doi.org/10.1088/1361-6633/ab12bc}{\emph{Rept. Prog. Phys.}
  {\bfseries 83} (2020) 075901}
  [\href{https://arxiv.org/abs/1807.01699}{{\ttfamily 1807.01699}}].

\bibitem{Buonanno:2022pgc}
A.~Buonanno, M.~Khalil, D.~O'Connell, R.~Roiban, M.P.~Solon and M.~Zeng,
  \emph{{Snowmass White Paper: Gravitational Waves and Scattering Amplitudes}},
   in \emph{{Snowmass 2021}}, 4, 2022
  [\href{https://arxiv.org/abs/2204.05194}{{\ttfamily 2204.05194}}].

\bibitem{Kosower:2022yvp}
D.A.~Kosower, R.~Monteiro and D.~O'Connell, \emph{{The SAGEX review on
  scattering amplitudes Chapter 14: Classical gravity from scattering
  amplitudes}}, \href{https://doi.org/10.1088/1751-8121/ac8846}{\emph{J. Phys.
  A} {\bfseries 55} (2022) 443015}
  [\href{https://arxiv.org/abs/2203.13025}{{\ttfamily 2203.13025}}].

\bibitem{Bjerrum-Bohr:2022blt}
N.E.J.~Bjerrum-Bohr, P.H.~Damgaard, L.~Plante and P.~Vanhove, \emph{{The SAGEX
  review on scattering amplitudes Chapter 13: Post-Minkowskian expansion from
  scattering amplitudes}},
  \href{https://doi.org/10.1088/1751-8121/ac7a78}{\emph{J. Phys. A} {\bfseries
  55} (2022) 443014} [\href{https://arxiv.org/abs/2203.13024}{{\ttfamily
  2203.13024}}].

\bibitem{Goldberger:2022rqf}
W.D.~Goldberger, \emph{{Effective Field Theory for Compact Binary Dynamics}},
  \href{https://arxiv.org/abs/2212.06677}{{\ttfamily 2212.06677}}.

\bibitem{Bjerrum-Bohr:2022ows}
N.E.J.~Bjerrum-Bohr, L.~Plant\'e and P.~Vanhove, \emph{{Effective Field Theory
  and Applications: Weak Field Observables from Scattering Amplitudes in
  Quantum Field Theory}},  \href{https://arxiv.org/abs/2212.08957}{{\ttfamily
  2212.08957}}.

\bibitem{Guevara:2018wpp}
A.~Guevara, A.~Ochirov and J.~Vines, \emph{{Scattering of Spinning Black Holes
  from Exponentiated Soft Factors}},
  \href{https://doi.org/10.1007/JHEP09(2019)056}{\emph{JHEP} {\bfseries 09}
  (2019) 056} [\href{https://arxiv.org/abs/1812.06895}{{\ttfamily
  1812.06895}}].

\bibitem{Bini:2018zxp}
D.~Bini and A.~Geralico, \emph{{High-energy hyperbolic scattering by neutron
  stars and black holes}},
  \href{https://doi.org/10.1103/PhysRevD.98.024049}{\emph{Phys. Rev. D}
  {\bfseries 98} (2018) 024049}
  [\href{https://arxiv.org/abs/1806.02085}{{\ttfamily 1806.02085}}].

\bibitem{Chung:2019duq}
M.-Z.~Chung, Y.-T.~Huang and J.-W.~Kim, \emph{{Classical potential for general
  spinning bodies}}, \href{https://doi.org/10.1007/JHEP09(2020)074}{\emph{JHEP}
  {\bfseries 09} (2020) 074}
  [\href{https://arxiv.org/abs/1908.08463}{{\ttfamily 1908.08463}}].

\bibitem{Bjerrum-Bohr:2019kec}
N.E.J.~Bjerrum-Bohr, A.~Cristofoli and P.H.~Damgaard, \emph{{Post-Minkowskian
  Scattering Angle in Einstein Gravity}},
  \href{https://doi.org/10.1007/JHEP08(2020)038}{\emph{JHEP} {\bfseries 08}
  (2020) 038} [\href{https://arxiv.org/abs/1910.09366}{{\ttfamily
  1910.09366}}].

\bibitem{Cristofoli:2019neg}
A.~Cristofoli, N.E.J.~Bjerrum-Bohr, P.H.~Damgaard and P.~Vanhove, \emph{{On
  Post-Minkowskian Hamiltonians in General Relativity}},
  \href{https://doi.org/10.1103/PhysRevD.100.084040}{\emph{Phys. Rev.}
  {\bfseries D100} (2019) 084040}
  [\href{https://arxiv.org/abs/1906.01579}{{\ttfamily 1906.01579}}].

\bibitem{Goldberger:2020wbx}
W.D.~Goldberger and I.Z.~Rothstein, \emph{{Horizon radiation reaction forces}},
  \href{https://doi.org/10.1007/JHEP10(2020)026}{\emph{JHEP} {\bfseries 10}
  (2020) 026} [\href{https://arxiv.org/abs/2007.00731}{{\ttfamily
  2007.00731}}].

\bibitem{Kol:2021jjc}
U.~Kol, D.~O'connell and O.~Telem, \emph{{The radial action from probe
  amplitudes to all orders}},
  \href{https://doi.org/10.1007/JHEP03(2022)141}{\emph{JHEP} {\bfseries 03}
  (2022) 141} [\href{https://arxiv.org/abs/2109.12092}{{\ttfamily
  2109.12092}}].

\bibitem{Chen:2021kxt}
W.-M.~Chen, M.-Z.~Chung, Y.-t.~Huang and J.-W.~Kim, \emph{{The 2PM Hamiltonian
  for binary Kerr to quartic in spin}},
  \href{https://doi.org/10.1007/JHEP08(2022)148}{\emph{JHEP} {\bfseries 08}
  (2022) 148} [\href{https://arxiv.org/abs/2111.13639}{{\ttfamily
  2111.13639}}].

\bibitem{Bjerrum-Bohr:2021wwt}
N.E.J.~Bjerrum-Bohr, L.~Plant\'e and P.~Vanhove, \emph{{Post-Minkowskian radial
  action from soft limits and velocity cuts}},
  \href{https://doi.org/10.1007/JHEP03(2022)071}{\emph{JHEP} {\bfseries 03}
  (2022) 071} [\href{https://arxiv.org/abs/2111.02976}{{\ttfamily
  2111.02976}}].

\bibitem{Herrmann:2021tct}
E.~Herrmann, J.~Parra-Martinez, M.S.~Ruf and M.~Zeng, \emph{{Radiative
  classical gravitational observables at $ \mathcal{O} $(G$^{3}$) from
  scattering amplitudes}},
  \href{https://doi.org/10.1007/JHEP10(2021)148}{\emph{JHEP} {\bfseries 10}
  (2021) 148} [\href{https://arxiv.org/abs/2104.03957}{{\ttfamily
  2104.03957}}].

\bibitem{Jakobsen:2022fcj}
G.U.~Jakobsen and G.~Mogull, \emph{{Conservative and Radiative Dynamics of
  Spinning Bodies at Third Post-Minkowskian Order Using Worldline Quantum Field
  Theory}}, \href{https://doi.org/10.1103/PhysRevLett.128.141102}{\emph{Phys.
  Rev. Lett.} {\bfseries 128} (2022) 141102}
  [\href{https://arxiv.org/abs/2201.07778}{{\ttfamily 2201.07778}}].

\bibitem{Jakobsen:2022zsx}
G.U.~Jakobsen and G.~Mogull, \emph{{Linear response, Hamiltonian, and radiative
  spinning two-body dynamics}},
  \href{https://doi.org/10.1103/PhysRevD.107.044033}{\emph{Phys. Rev. D}
  {\bfseries 107} (2023) 044033}
  [\href{https://arxiv.org/abs/2210.06451}{{\ttfamily 2210.06451}}].

\bibitem{Aoude:2022thd}
R.~Aoude, K.~Haddad and A.~Helset, \emph{{Classical Gravitational
  Spinning-Spinless Scattering at O(G2S\ensuremath{\infty})}},
  \href{https://doi.org/10.1103/PhysRevLett.129.141102}{\emph{Phys. Rev. Lett.}
  {\bfseries 129} (2022) 141102}
  [\href{https://arxiv.org/abs/2205.02809}{{\ttfamily 2205.02809}}].

\bibitem{Menezes:2022tcs}
G.~Menezes and M.~Sergola, \emph{{NLO deflections for spinning particles and
  Kerr black holes}},
  \href{https://doi.org/10.1007/JHEP10(2022)105}{\emph{JHEP} {\bfseries 10}
  (2022) 105} [\href{https://arxiv.org/abs/2205.11701}{{\ttfamily
  2205.11701}}].

\bibitem{Damgaard:2022jem}
P.H.~Damgaard, J.~Hoogeveen, A.~Luna and J.~Vines, \emph{{Scattering angles in
  Kerr metrics}},
  \href{https://doi.org/10.1103/PhysRevD.106.124030}{\emph{Phys. Rev. D}
  {\bfseries 106} (2022) 124030}
  [\href{https://arxiv.org/abs/2208.11028}{{\ttfamily 2208.11028}}].

\bibitem{FebresCordero:2022jts}
F.~Febres~Cordero, M.~Kraus, G.~Lin, M.S.~Ruf and M.~Zeng, \emph{{Conservative
  Binary Dynamics with a Spinning Black Hole at O(G3) from Scattering
  Amplitudes}},
  \href{https://doi.org/10.1103/PhysRevLett.130.021601}{\emph{Phys. Rev. Lett.}
  {\bfseries 130} (2023) 021601}
  [\href{https://arxiv.org/abs/2205.07357}{{\ttfamily 2205.07357}}].

\bibitem{Alessio:2023kgf}
F.~Alessio, \emph{{Kerr binary dynamics from minimal coupling and double
  copy}},  \href{https://arxiv.org/abs/2303.12784}{{\ttfamily 2303.12784}}.

\bibitem{Bautista:2023szu}
Y.F.~Bautista, \emph{{Dynamics for super-extremal Kerr binary systems at
  O(G2)}}, \href{https://doi.org/10.1103/PhysRevD.108.084036}{\emph{Phys. Rev.
  D} {\bfseries 108} (2023) 084036}
  [\href{https://arxiv.org/abs/2304.04287}{{\ttfamily 2304.04287}}].

\bibitem{Gonzo:2024xjk}
R.~Gonzo, J.~Lewis and A.~Pound, \emph{{The first law of binary black hole
  scattering}},  \href{https://arxiv.org/abs/2409.03437}{{\ttfamily
  2409.03437}}.

\bibitem{Manohar:2022dea}
A.V.~Manohar, A.K.~Ridgway and C.-H.~Shen, \emph{{Radiated Angular Momentum and
  Dissipative Effects in Classical Scattering}},
  \href{https://doi.org/10.1103/PhysRevLett.129.121601}{\emph{Phys. Rev. Lett.}
  {\bfseries 129} (2022) 121601}
  [\href{https://arxiv.org/abs/2203.04283}{{\ttfamily 2203.04283}}].

\bibitem{Kalin:2022hph}
G.~K\"alin, J.~Neef and R.A.~Porto, \emph{{Radiation-reaction in the Effective
  Field Theory approach to Post-Minkowskian dynamics}},
  \href{https://doi.org/10.1007/JHEP01(2023)140}{\emph{JHEP} {\bfseries 01}
  (2023) 140} [\href{https://arxiv.org/abs/2207.00580}{{\ttfamily
  2207.00580}}].

\bibitem{Bini:2022enm}
D.~Bini, T.~Damour and A.~Geralico, \emph{{Radiated momentum and radiation
  reaction in gravitational two-body scattering including time-asymmetric
  effects}}, \href{https://doi.org/10.1103/PhysRevD.107.024012}{\emph{Phys.
  Rev. D} {\bfseries 107} (2023) 024012}
  [\href{https://arxiv.org/abs/2210.07165}{{\ttfamily 2210.07165}}].

\bibitem{Dlapa:2022lmu}
C.~Dlapa, G.~K\"alin, Z.~Liu, J.~Neef and R.A.~Porto, \emph{{Radiation Reaction
  and Gravitational Waves at Fourth Post-Minkowskian Order}},
  \href{https://doi.org/10.1103/PhysRevLett.130.101401}{\emph{Phys. Rev. Lett.}
  {\bfseries 130} (2023) 101401}
  [\href{https://arxiv.org/abs/2210.05541}{{\ttfamily 2210.05541}}].

\bibitem{Riva:2021vnj}
M.M.~Riva and F.~Vernizzi, \emph{{Radiated momentum in the post-Minkowskian
  worldline approach via reverse unitarity}},
  \href{https://doi.org/10.1007/JHEP11(2021)228}{\emph{JHEP} {\bfseries 11}
  (2021) 228} [\href{https://arxiv.org/abs/2110.10140}{{\ttfamily
  2110.10140}}].

\bibitem{Brandhuber:2023hhy}
A.~Brandhuber, G.R.~Brown, G.~Chen, S.~De~Angelis, J.~Gowdy and G.~Travaglini,
  \emph{{One-loop gravitational bremsstrahlung and waveforms from a heavy-mass
  effective field theory}},
  \href{https://doi.org/10.1007/JHEP06(2023)048}{\emph{JHEP} {\bfseries 06}
  (2023) 048} [\href{https://arxiv.org/abs/2303.06111}{{\ttfamily
  2303.06111}}].

\bibitem{Georgoudis:2023lgf}
A.~Georgoudis, C.~Heissenberg and I.~Vazquez-Holm, \emph{{Inelastic
  exponentiation and classical gravitational scattering at one loop}},
  \href{https://doi.org/10.1007/JHEP06(2023)126}{\emph{JHEP} {\bfseries 2023}
  (2023) 126} [\href{https://arxiv.org/abs/2303.07006}{{\ttfamily
  2303.07006}}].

\bibitem{Georgoudis:2023ozp}
A.~Georgoudis, C.~Heissenberg and I.~Vazquez-Holm, \emph{{Addendum to:
  Inelastic exponentiation and classical gravitational scattering at one
  loop}}, \href{https://doi.org/10.1007/JHEP02(2024)161}{\emph{JHEP} {\bfseries
  2024} (2024) 161} [\href{https://arxiv.org/abs/2312.14710}{{\ttfamily
  2312.14710}}].

\bibitem{Caron-Huot:2023vxl}
S.~Caron-Huot, M.~Giroux, H.S.~Hannesdottir and S.~Mizera, \emph{{What can be
  measured asymptotically?}},
  \href{https://doi.org/10.1007/JHEP01(2024)139}{\emph{JHEP} {\bfseries 01}
  (2024) 139} [\href{https://arxiv.org/abs/2308.02125}{{\ttfamily
  2308.02125}}].

\bibitem{Caron-Huot:2023ikn}
S.~Caron-Huot, M.~Giroux, H.S.~Hannesdottir and S.~Mizera, \emph{{Crossing
  beyond scattering amplitudes}},
  \href{https://doi.org/10.1007/JHEP04(2024)060}{\emph{JHEP} {\bfseries 04}
  (2024) 060} [\href{https://arxiv.org/abs/2310.12199}{{\ttfamily
  2310.12199}}].

\bibitem{Bini:2023fiz}
D.~Bini, T.~Damour and A.~Geralico, \emph{{Comparing one-loop gravitational
  bremsstrahlung amplitudes to the multipolar-post-Minkowskian waveform}},
  \href{https://doi.org/10.1103/PhysRevD.108.124052}{\emph{Phys. Rev. D}
  {\bfseries 108} (2023) 124052}
  [\href{https://arxiv.org/abs/2309.14925}{{\ttfamily 2309.14925}}].

\bibitem{Georgoudis:2023eke}
A.~Georgoudis, C.~Heissenberg and R.~Russo, \emph{{An eikonal-inspired approach
  to the gravitational scattering waveform}},
  \href{https://doi.org/10.1007/JHEP03(2024)089}{\emph{JHEP} {\bfseries 03}
  (2024) 089} [\href{https://arxiv.org/abs/2312.07452}{{\ttfamily
  2312.07452}}].

\bibitem{Georgoudis:2024pdz}
A.~Georgoudis, C.~Heissenberg and R.~Russo, \emph{{Post-Newtonian multipoles
  from the next-to-leading post-Minkowskian gravitational waveform}},
  \href{https://doi.org/10.1103/PhysRevD.109.106020}{\emph{Phys. Rev. D}
  {\bfseries 109} (2024) 106020}
  [\href{https://arxiv.org/abs/2402.06361}{{\ttfamily 2402.06361}}].

\bibitem{Bini:2024rsy}
D.~Bini, T.~Damour, S.~De~Angelis, A.~Geralico, A.~Herderschee, R.~Roiban
  et~al., \emph{{Gravitational waveforms: A tale of two formalisms}},
  \href{https://doi.org/10.1103/PhysRevD.109.125008}{\emph{Phys. Rev. D}
  {\bfseries 109} (2024) 125008}
  [\href{https://arxiv.org/abs/2402.06604}{{\ttfamily 2402.06604}}].

\bibitem{Buonanno:2024byg}
A.~Buonanno, G.~Mogull, R.~Patil and L.~Pompili, \emph{{Post-Minkowskian Theory
  Meets the Spinning Effective-One-Body Approach for Bound-Orbit Waveforms}},
  \href{https://arxiv.org/abs/2405.19181}{{\ttfamily 2405.19181}}.

\bibitem{Mougiakakos:2021ckm}
S.~Mougiakakos, M.M.~Riva and F.~Vernizzi, \emph{{Gravitational Bremsstrahlung
  in the post-Minkowskian effective field theory}},
  \href{https://doi.org/10.1103/PhysRevD.104.024041}{\emph{Phys. Rev. D}
  {\bfseries 104} (2021) 024041}
  [\href{https://arxiv.org/abs/2102.08339}{{\ttfamily 2102.08339}}].

\bibitem{Burger:2019wkq}
D.J.~Burger, W.T.~Emond and N.~Moynihan, \emph{{Rotating Black Holes in Cubic
  Gravity}}, \href{https://doi.org/10.1103/PhysRevD.101.084009}{\emph{Phys.
  Rev. D} {\bfseries 101} (2020) 084009}
  [\href{https://arxiv.org/abs/1910.11618}{{\ttfamily 1910.11618}}].

\bibitem{Emond:2019crr}
W.T.~Emond and N.~Moynihan, \emph{{Scattering Amplitudes, Black Holes and
  Leading Singularities in Cubic Theories of Gravity}},
  \href{https://doi.org/10.1007/JHEP12(2019)019}{\emph{JHEP} {\bfseries 12}
  (2019) 019} [\href{https://arxiv.org/abs/1905.08213}{{\ttfamily
  1905.08213}}].

\bibitem{Cristofoli:2019ewu}
A.~Cristofoli, \emph{{Post-Minkowskian Hamiltonians in Modified Theories of
  Gravity}}, \href{https://doi.org/10.1016/j.physletb.2019.135095}{\emph{Phys.
  Lett. B} {\bfseries 800} (2020) 135095}
  [\href{https://arxiv.org/abs/1906.05209}{{\ttfamily 1906.05209}}].

\bibitem{Brandhuber:2019qpg}
A.~Brandhuber and G.~Travaglini, \emph{{On higher-derivative effects on the
  gravitational potential and particle bending}},
  \href{https://doi.org/10.1007/JHEP01(2020)010}{\emph{JHEP} {\bfseries 01}
  (2020) 010} [\href{https://arxiv.org/abs/1905.05657}{{\ttfamily
  1905.05657}}].

\bibitem{KoemansCollado:2019lnh}
A.~Koemans~Collado and S.~Thomas, \emph{{Eikonal Scattering in Kaluza-Klein
  Gravity}}, \href{https://doi.org/10.1007/JHEP04(2019)171}{\emph{JHEP}
  {\bfseries 04} (2019) 171}
  [\href{https://arxiv.org/abs/1901.05869}{{\ttfamily 1901.05869}}].

\bibitem{AccettulliHuber:2020dal}
M.~Accettulli~Huber, A.~Brandhuber, S.~De~Angelis and G.~Travaglini,
  \emph{{From amplitudes to gravitational radiation with cubic interactions and
  tidal effects}},
  \href{https://doi.org/10.1103/PhysRevD.103.045015}{\emph{Phys. Rev. D}
  {\bfseries 103} (2021) 045015}
  [\href{https://arxiv.org/abs/2012.06548}{{\ttfamily 2012.06548}}].

\bibitem{AccettulliHuber:2020oou}
M.~Accettulli~Huber, A.~Brandhuber, S.~De~Angelis and G.~Travaglini,
  \emph{{Eikonal phase matrix, deflection angle and time delay in effective
  field theories of gravity}},
  \href{https://doi.org/10.1103/PhysRevD.102.046014}{\emph{Phys. Rev. D}
  {\bfseries 102} (2020) 046014}
  [\href{https://arxiv.org/abs/2006.02375}{{\ttfamily 2006.02375}}].

\bibitem{Carrillo-Gonzalez:2021mqj}
M.~Carrillo-Gonz\'alez, C.~de~Rham and A.J.~Tolley, \emph{{Scattering
  amplitudes for binary systems beyond GR}},
  \href{https://doi.org/10.1007/JHEP11(2021)087}{\emph{JHEP} {\bfseries 11}
  (2021) 087} [\href{https://arxiv.org/abs/2107.11384}{{\ttfamily
  2107.11384}}].

\bibitem{Davis:2023zqv}
A.-C.~Davis and S.~Melville, \emph{{Orbital precession and hidden symmetries in
  scalar-tensor theories}},
  \href{https://doi.org/10.1088/1475-7516/2023/11/034}{\emph{JCAP} {\bfseries
  11} (2023) 034} [\href{https://arxiv.org/abs/2307.06331}{{\ttfamily
  2307.06331}}].

\bibitem{Bhattacharyya:2024aeq}
A.~Bhattacharyya, D.~Ghosh, S.~Ghosh and S.~Pal, \emph{{Observables from
  classical black hole scattering in Scalar-Tensor theory of gravity from
  worldline quantum field theory}},
  \href{https://doi.org/10.1007/JHEP04(2024)015}{\emph{JHEP} {\bfseries 04}
  (2024) 015} [\href{https://arxiv.org/abs/2401.05492}{{\ttfamily
  2401.05492}}].

\bibitem{Bhattacharyya:2024kxj}
A.~Bhattacharyya, D.~Ghosh, S.~Ghosh and S.~Pal, \emph{{Bootstrapping spinning
  two body problem in dynamical Chern-Simons gravity using worldline QFT}},
  \href{https://arxiv.org/abs/2407.07195}{{\ttfamily 2407.07195}}.

\bibitem{Fransen:2024fzs}
K.~Fransen and S.B.~Giddings, \emph{{Gravitational wave signatures of
  departures from classical black hole scattering}},
  \href{https://doi.org/10.1103/PhysRevD.110.064029}{\emph{Phys. Rev. D}
  {\bfseries 110} (2024) 064029}
  [\href{https://arxiv.org/abs/2405.05970}{{\ttfamily 2405.05970}}].

\bibitem{Brandhuber:2024bnz}
A.~Brandhuber, G.R.~Brown, P.~Pichini, G.~Travaglini and P.~Vives~Matasan,
  \emph{{Spinning binary dynamics in cubic effective field theories of
  gravity}}, \href{https://doi.org/10.1007/JHEP08(2024)188}{\emph{JHEP}
  {\bfseries 08} (2024) 188}
  [\href{https://arxiv.org/abs/2405.13826}{{\ttfamily 2405.13826}}].

\bibitem{Bautista:2024agp}
Y.F.~Bautista, M.~Khalil, M.~Sergola, C.~Kavanagh and J.~Vines,
  \emph{{Post-Newtonian observables for aligned-spin binaries to sixth order in
  spin from gravitational self-force and Compton amplitudes}},
  \href{https://arxiv.org/abs/2408.01871}{{\ttfamily 2408.01871}}.

\bibitem{Brandhuber:2024qdn}
A.~Brandhuber, G.R.~Brown, G.~Chen, G.~Travaglini and P.~Vives~Matasan,
  \emph{{Spinning waveforms in cubic effective field theories of gravity}},
  \href{https://arxiv.org/abs/2408.00587}{{\ttfamily 2408.00587}}.

\bibitem{Falkowski:2024bgb}
A.~Falkowski and P.~Marinellis, \emph{{Spinning waveforms of scalar radiation
  in quadratic modified gravity}},
  \href{https://arxiv.org/abs/2407.16457}{{\ttfamily 2407.16457}}.

\bibitem{Aoude:2020ygw}
R.~Aoude, K.~Haddad and A.~Helset, \emph{{Tidal effects for spinning
  particles}}, \href{https://doi.org/10.1007/JHEP03(2021)097}{\emph{JHEP}
  {\bfseries 03} (2021) 097}
  [\href{https://arxiv.org/abs/2012.05256}{{\ttfamily 2012.05256}}].

\bibitem{Kalin:2020lmz}
G.~K\"alin, Z.~Liu and R.A.~Porto, \emph{{Conservative Tidal Effects in Compact
  Binary Systems to Next-to-Leading Post-Minkowskian Order}},
  \href{https://doi.org/10.1103/PhysRevD.102.124025}{\emph{Phys. Rev. D}
  {\bfseries 102} (2020) 124025}
  [\href{https://arxiv.org/abs/2008.06047}{{\ttfamily 2008.06047}}].

\bibitem{Kim:2020dif}
J.-W.~Kim and M.~Shim, \emph{{Quantum corrections to tidal Love number for
  Schwarzschild black holes}},
  \href{https://doi.org/10.1103/PhysRevD.104.046022}{\emph{Phys. Rev. D}
  {\bfseries 104} (2021) 046022}
  [\href{https://arxiv.org/abs/2011.03337}{{\ttfamily 2011.03337}}].

\bibitem{Bern:2020uwk}
Z.~Bern, J.~Parra-Martinez, R.~Roiban, E.~Sawyer and C.-H.~Shen, \emph{{Leading
  Nonlinear Tidal Effects and Scattering Amplitudes}},
  \href{https://doi.org/10.1007/JHEP05(2021)188}{\emph{JHEP} {\bfseries 05}
  (2021) 188} [\href{https://arxiv.org/abs/2010.08559}{{\ttfamily
  2010.08559}}].

\bibitem{Haddad:2020que}
K.~Haddad and A.~Helset, \emph{{Tidal effects in quantum field theory}},
  \href{https://doi.org/10.1007/JHEP12(2020)024}{\emph{JHEP} {\bfseries 12}
  (2020) 024} [\href{https://arxiv.org/abs/2008.04920}{{\ttfamily
  2008.04920}}].

\bibitem{Cheung:2020sdj}
C.~Cheung and M.P.~Solon, \emph{{Tidal Effects in the Post-Minkowskian
  Expansion}},
  \href{https://doi.org/10.1103/PhysRevLett.125.191601}{\emph{Phys. Rev. Lett.}
  {\bfseries 125} (2020) 191601}
  [\href{https://arxiv.org/abs/2006.06665}{{\ttfamily 2006.06665}}].

\bibitem{Bini:2020flp}
D.~Bini, T.~Damour and A.~Geralico, \emph{{Scattering of tidally interacting
  bodies in post-Minkowskian gravity}},
  \href{https://doi.org/10.1103/PhysRevD.101.044039}{\emph{Phys. Rev. D}
  {\bfseries 101} (2020) 044039}
  [\href{https://arxiv.org/abs/2001.00352}{{\ttfamily 2001.00352}}].

\bibitem{Heissenberg:2022tsn}
C.~Heissenberg, \emph{{Angular Momentum Loss due to Tidal Effects in the
  Post-Minkowskian Expansion}},
  \href{https://doi.org/10.1103/PhysRevLett.131.011603}{\emph{Phys. Rev. Lett.}
  {\bfseries 131} (2023) 011603}
  [\href{https://arxiv.org/abs/2210.15689}{{\ttfamily 2210.15689}}].

\bibitem{Mougiakakos:2022sic}
S.~Mougiakakos, M.M.~Riva and F.~Vernizzi, \emph{{Gravitational Bremsstrahlung
  with Tidal Effects in the Post-Minkowskian Expansion}},
  \href{https://doi.org/10.1103/PhysRevLett.129.121101}{\emph{Phys. Rev. Lett.}
  {\bfseries 129} (2022) 121101}
  [\href{https://arxiv.org/abs/2204.06556}{{\ttfamily 2204.06556}}].

\bibitem{Adamo:2020qru}
T.~Adamo and A.~Ilderton, \emph{{Classical and quantum double copy of
  back-reaction}}, \href{https://doi.org/10.1007/JHEP09(2020)200}{\emph{JHEP}
  {\bfseries 09} (2020) 200}
  [\href{https://arxiv.org/abs/2005.05807}{{\ttfamily 2005.05807}}].

\bibitem{Cristofoli:2020hnk}
A.~Cristofoli, \emph{{Gravitational shock waves and scattering amplitudes}},
  \href{https://doi.org/10.1007/JHEP11(2020)160}{\emph{JHEP} {\bfseries 11}
  (2020) 160} [\href{https://arxiv.org/abs/2006.08283}{{\ttfamily
  2006.08283}}].

\bibitem{Kim:2020cvf}
J.-W.~Kim and M.~Shim, \emph{{Gravitational Dyonic Amplitude at One-Loop and
  its Inconsistency with the Classical Impulse}},
  \href{https://doi.org/10.1007/JHEP02(2021)217}{\emph{JHEP} {\bfseries 02}
  (2021) 217} [\href{https://arxiv.org/abs/2010.14347}{{\ttfamily
  2010.14347}}].

\bibitem{Crawley:2021auj}
E.~Crawley, A.~Guevara, N.~Miller and A.~Strominger, \emph{{Black holes in
  Klein space}}, \href{https://doi.org/10.1007/JHEP10(2022)135}{\emph{JHEP}
  {\bfseries 10} (2022) 135}
  [\href{https://arxiv.org/abs/2112.03954}{{\ttfamily 2112.03954}}].

\bibitem{Adamo:2022qci}
T.~Adamo, A.~Cristofoli, A.~Ilderton and S.~Klisch, \emph{{All Order
  Gravitational Waveforms from Scattering Amplitudes}},
  \href{https://doi.org/10.1103/PhysRevLett.131.011601}{\emph{Phys. Rev. Lett.}
  {\bfseries 131} (2023) 011601}
  [\href{https://arxiv.org/abs/2210.04696}{{\ttfamily 2210.04696}}].

\bibitem{Adamo:2022rob}
T.~Adamo, A.~Cristofoli and P.~Tourkine, \emph{{The ultrarelativistic limit of
  Kerr}}, \href{https://doi.org/10.1007/JHEP02(2023)107}{\emph{JHEP} {\bfseries
  02} (2023) 107} [\href{https://arxiv.org/abs/2209.05730}{{\ttfamily
  2209.05730}}].

\bibitem{Gonzo:2022tjm}
R.~Gonzo, T.~McLoughlin and A.~Puhm, \emph{{Celestial holography on Kerr-Schild
  backgrounds}}, \href{https://doi.org/10.1007/JHEP10(2022)073}{\emph{JHEP}
  {\bfseries 10} (2022) 073}
  [\href{https://arxiv.org/abs/2207.13719}{{\ttfamily 2207.13719}}].

\bibitem{Kosmopoulos:2023bwc}
D.~Kosmopoulos and M.P.~Solon, \emph{{Gravitational self force from scattering
  amplitudes in curved space}},
  \href{https://doi.org/10.1007/JHEP03(2024)125}{\emph{JHEP} {\bfseries 03}
  (2024) 125} [\href{https://arxiv.org/abs/2308.15304}{{\ttfamily
  2308.15304}}].

\bibitem{Jones:2023tgz}
C.R.T.~Jones, \emph{{Classical dynamics of vortex solitons from perturbative
  scattering amplitudes}},
  \href{https://doi.org/10.1007/JHEP11(2023)092}{\emph{JHEP} {\bfseries 11}
  (2023) 092} [\href{https://arxiv.org/abs/2305.08902}{{\ttfamily
  2305.08902}}].

\bibitem{Crawley:2023brz}
E.~Crawley, A.~Guevara, E.~Himwich and A.~Strominger, \emph{{Self-dual black
  holes in celestial holography}},
  \href{https://doi.org/10.1007/JHEP09(2023)109}{\emph{JHEP} {\bfseries 09}
  (2023) 109} [\href{https://arxiv.org/abs/2302.06661}{{\ttfamily
  2302.06661}}].

\bibitem{Guevara:2023wlr}
A.~Guevara and U.~Kol, \emph{{Self Dual Black Holes as the Hydrogen Atom}},
  \href{https://arxiv.org/abs/2311.07933}{{\ttfamily 2311.07933}}.

\bibitem{Adamo:2023fbj}
T.~Adamo, G.~Bogna, L.~Mason and A.~Sharma, \emph{{Scattering on self-dual
  Taub-NUT}}, \href{https://doi.org/10.1088/1361-6382/ad12ee}{\emph{Class.
  Quant. Grav.} {\bfseries 41} (2024) 015030}
  [\href{https://arxiv.org/abs/2309.03834}{{\ttfamily 2309.03834}}].

\bibitem{Adamo:2023cfp}
T.~Adamo, A.~Cristofoli, A.~Ilderton and S.~Klisch, \emph{{Scattering
  amplitudes for self-force}},
  \href{https://doi.org/10.1088/1361-6382/ad210f}{\emph{Class. Quant. Grav.}
  {\bfseries 41} (2024) 065006}
  [\href{https://arxiv.org/abs/2307.00431}{{\ttfamily 2307.00431}}].

\bibitem{Adamo:2024oxy}
T.~Adamo, R.~Gonzo and A.~Ilderton, \emph{{Gravitational bound waveforms from
  amplitudes}}, \href{https://doi.org/10.1007/JHEP05(2024)034}{\emph{JHEP}
  {\bfseries 05} (2024) 034}
  [\href{https://arxiv.org/abs/2402.00124}{{\ttfamily 2402.00124}}].

\bibitem{Abreu:2020lyk}
S.~Abreu, F.~Febres~Cordero, H.~Ita, M.~Jaquier, B.~Page, M.S.~Ruf et~al.,
  \emph{{Two-Loop Four-Graviton Scattering Amplitudes}},
  \href{https://doi.org/10.1103/PhysRevLett.124.211601}{\emph{Phys. Rev. Lett.}
  {\bfseries 124} (2020) 211601}
  [\href{https://arxiv.org/abs/2002.12374}{{\ttfamily 2002.12374}}].

\bibitem{Bern:2020gjj}
Z.~Bern, H.~Ita, J.~Parra-Martinez and M.S.~Ruf, \emph{{Universality in the
  classical limit of massless gravitational scattering}},
  \href{https://doi.org/10.1103/PhysRevLett.125.031601}{\emph{Phys. Rev. Lett.}
  {\bfseries 125} (2020) 031601}
  [\href{https://arxiv.org/abs/2002.02459}{{\ttfamily 2002.02459}}].

\bibitem{Caron-Huot:2018ape}
S.~Caron-Huot and Z.~Zahraee, \emph{{Integrability of Black Hole Orbits in
  Maximal Supergravity}},
  \href{https://doi.org/10.1007/JHEP07(2019)179}{\emph{JHEP} {\bfseries 07}
  (2019) 179} [\href{https://arxiv.org/abs/1810.04694}{{\ttfamily
  1810.04694}}].

\bibitem{KoemansCollado:2018hss}
A.~Koemans~Collado, P.~Di~Vecchia, R.~Russo and S.~Thomas, \emph{{The
  subleading eikonal in supergravity theories}},
  \href{https://doi.org/10.1007/JHEP10(2018)038}{\emph{JHEP} {\bfseries 10}
  (2018) 038} [\href{https://arxiv.org/abs/1807.04588}{{\ttfamily
  1807.04588}}].

\bibitem{Moynihan:2019bor}
N.~Moynihan, \emph{{Kerr-Newman from Minimal Coupling}},
  \href{https://doi.org/10.1007/JHEP01(2020)014}{\emph{JHEP} {\bfseries 01}
  (2020) 014} [\href{https://arxiv.org/abs/1909.05217}{{\ttfamily
  1909.05217}}].

\bibitem{Parra-Martinez:2020dzs}
J.~Parra-Martinez, M.S.~Ruf and M.~Zeng, \emph{{Extremal black hole scattering
  at $\mathcal{O}(G^3)$: graviton dominance, eikonal exponentiation, and
  differential equations}},
  \href{https://doi.org/10.1007/JHEP11(2020)023}{\emph{JHEP} {\bfseries 11}
  (2020) 023} [\href{https://arxiv.org/abs/2005.04236}{{\ttfamily
  2005.04236}}].

\bibitem{Cristofoli:2020uzm}
A.~Cristofoli, P.H.~Damgaard, P.~Di~Vecchia and C.~Heissenberg,
  \emph{{Second-order Post-Minkowskian scattering in arbitrary dimensions}},
  \href{https://doi.org/10.1007/JHEP07(2020)122}{\emph{JHEP} {\bfseries 07}
  (2020) 122} [\href{https://arxiv.org/abs/2003.10274}{{\ttfamily
  2003.10274}}].

\bibitem{Kim:2022iub}
J.-W.~Kim, \emph{{Quantum corrections to frame-dragging in scattering
  amplitudes}}, \href{https://doi.org/10.1103/PhysRevD.106.L081901}{\emph{Phys.
  Rev. D} {\bfseries 106} (2022) L081901}
  [\href{https://arxiv.org/abs/2207.04970}{{\ttfamily 2207.04970}}].

\bibitem{Chen:2021huj}
B.-T.~Chen, M.-Z.~Chung, Y.-t.~Huang and M.K.~Tam, \emph{{Minimal spin
  deflection of Kerr-Newman and supersymmetric black hole}},
  \href{https://doi.org/10.1007/JHEP10(2021)011}{\emph{JHEP} {\bfseries 10}
  (2021) 011} [\href{https://arxiv.org/abs/2106.12518}{{\ttfamily
  2106.12518}}].

\bibitem{Chung:2019yfs}
M.-Z.~Chung, Y.-T.~Huang and J.-W.~Kim, \emph{{Kerr-Newman stress-tensor from
  minimal coupling}},
  \href{https://doi.org/10.1007/JHEP12(2020)103}{\emph{JHEP} {\bfseries 12}
  (2020) 103} [\href{https://arxiv.org/abs/1911.12775}{{\ttfamily
  1911.12775}}].

\bibitem{Monteiro:2021ztt}
R.~Monteiro, S.~Nagy, D.~O'Connell, D.~Peinador~Veiga and M.~Sergola,
  \emph{{NS-NS spacetimes from amplitudes}},
  \href{https://doi.org/10.1007/JHEP06(2022)021}{\emph{JHEP} {\bfseries 06}
  (2022) 021} [\href{https://arxiv.org/abs/2112.08336}{{\ttfamily
  2112.08336}}].

\bibitem{Jones:2022aji}
C.R.T.~Jones and M.~Solon, \emph{{Scattering amplitudes and N-body
  post-Minkowskian Hamiltonians in general relativity and beyond}},
  \href{https://doi.org/10.1007/JHEP02(2023)105}{\emph{JHEP} {\bfseries 02}
  (2023) 105} [\href{https://arxiv.org/abs/2208.02281}{{\ttfamily
  2208.02281}}].

\bibitem{Emond:2020lwi}
W.T.~Emond, Y.-T.~Huang, U.~Kol, N.~Moynihan and D.~O'Connell,
  \emph{{Amplitudes from Coulomb to Kerr-Taub-NUT}},
  \href{https://arxiv.org/abs/2010.07861}{{\ttfamily 2010.07861}}.

\bibitem{Cangemi:2022abk}
L.~Cangemi and P.~Pichini, \emph{{Classical limit of higher-spin string
  amplitudes}}, \href{https://doi.org/10.1007/JHEP06(2023)167}{\emph{JHEP}
  {\bfseries 06} (2023) 167}
  [\href{https://arxiv.org/abs/2207.03947}{{\ttfamily 2207.03947}}].

\bibitem{Hoogeveen:2023bqa}
J.~Hoogeveen, \emph{{Charged test particle scattering and effective one-body
  metrics with spin}},
  \href{https://doi.org/10.1103/PhysRevD.108.024049}{\emph{Phys. Rev. D}
  {\bfseries 108} (2023) 024049}
  [\href{https://arxiv.org/abs/2303.00317}{{\ttfamily 2303.00317}}].

\bibitem{Bianchi:2023lrg}
M.~Bianchi, C.~Gambino and F.~Riccioni, \emph{{A Rutherford-like formula for
  scattering off Kerr-Newman BHs and subleading corrections}},
  \href{https://doi.org/10.1007/JHEP08(2023)188}{\emph{JHEP} {\bfseries 08}
  (2023) 188} [\href{https://arxiv.org/abs/2306.08969}{{\ttfamily
  2306.08969}}].

\bibitem{Gambino:2024uge}
C.~Gambino, P.~Pani and F.~Riccioni, \emph{{Rotating metrics and new multipole
  moments from scattering amplitudes in arbitrary dimensions}},
  \href{https://doi.org/10.1103/PhysRevD.109.124018}{\emph{Phys. Rev. D}
  {\bfseries 109} (2024) 124018}
  [\href{https://arxiv.org/abs/2403.16574}{{\ttfamily 2403.16574}}].

\bibitem{Saketh:2021sri}
M.V.S.~Saketh, J.~Vines, J.~Steinhoff and A.~Buonanno, \emph{{Conservative and
  radiative dynamics in classical relativistic scattering and bound systems}},
  \href{https://doi.org/10.1103/PhysRevResearch.4.013127}{\emph{Phys. Rev.
  Res.} {\bfseries 4} (2022) 013127}
  [\href{https://arxiv.org/abs/2109.05994}{{\ttfamily 2109.05994}}].

\bibitem{Akhtar:2024mbg}
S.~Akhtar, A.~Manna and A.~Manu, \emph{{Classical observables using
  exponentiated spin factors: electromagnetic scattering}},
  \href{https://doi.org/10.1007/JHEP05(2024)148}{\emph{JHEP} {\bfseries 05}
  (2024) 148} [\href{https://arxiv.org/abs/2401.15574}{{\ttfamily
  2401.15574}}].

\bibitem{Aoki:2024boe}
K.~Aoki, A.~Cristofoli and Y.-t.~Huang, \emph{{On-Shell Approach to Black Hole
  Mergers}},  \href{https://arxiv.org/abs/2410.13632}{{\ttfamily 2410.13632}}.

\bibitem{Rettegno:2023ghr}
P.~Rettegno, G.~Pratten, L.M.~Thomas, P.~Schmidt and T.~Damour,
  \emph{{Strong-field scattering of two spinning black holes: Numerical
  relativity versus post-Minkowskian gravity}},
  \href{https://doi.org/10.1103/PhysRevD.108.124016}{\emph{Phys. Rev. D}
  {\bfseries 108} (2023) 124016}
  [\href{https://arxiv.org/abs/2307.06999}{{\ttfamily 2307.06999}}].

\bibitem{Damour:2022ybd}
T.~Damour and P.~Rettegno, \emph{{Strong-field scattering of two black holes:
  Numerical relativity meets post-Minkowskian gravity}},
  \href{https://doi.org/10.1103/PhysRevD.107.064051}{\emph{Phys. Rev. D}
  {\bfseries 107} (2023) 064051}
  [\href{https://arxiv.org/abs/2211.01399}{{\ttfamily 2211.01399}}].

\bibitem{Vaidya:2014kza}
V.~Vaidya, \emph{{Gravitational spin Hamiltonians from the S matrix}},
  \href{https://doi.org/10.1103/PhysRevD.91.024017}{\emph{Phys. Rev.}
  {\bfseries D91} (2015) 024017}
  [\href{https://arxiv.org/abs/1410.5348}{{\ttfamily 1410.5348}}].

\bibitem{Vines:2017hyw}
J.~Vines, \emph{{Scattering of two spinning black holes in post-Minkowskian
  gravity, to all orders in spin, and effective-one-body mappings}},
  \href{https://doi.org/10.1088/1361-6382/aaa3a8}{\emph{Class. Quant. Grav.}
  {\bfseries 35} (2018) 084002}
  [\href{https://arxiv.org/abs/1709.06016}{{\ttfamily 1709.06016}}].

\bibitem{Arkani-Hamed:2017jhn}
N.~Arkani-Hamed, T.-C.~Huang and Y.-t.~Huang, \emph{{Scattering amplitudes for
  all masses and spins}},
  \href{https://doi.org/10.1007/JHEP11(2021)070}{\emph{JHEP} {\bfseries 11}
  (2021) 070} [\href{https://arxiv.org/abs/1709.04891}{{\ttfamily
  1709.04891}}].

\bibitem{Chung:2018kqs}
M.-Z.~Chung, Y.-T.~Huang, J.-W.~Kim and S.~Lee, \emph{{The simplest massive
  S-matrix: from minimal coupling to Black Holes}},
  \href{https://doi.org/10.1007/JHEP04(2019)156}{\emph{JHEP} {\bfseries 04}
  (2019) 156} [\href{https://arxiv.org/abs/1812.08752}{{\ttfamily
  1812.08752}}].

\bibitem{Guevara:2019fsj}
A.~Guevara, A.~Ochirov and J.~Vines, \emph{{Black-hole scattering with general
  spin directions from minimal-coupling amplitudes}},
  \href{https://doi.org/10.1103/PhysRevD.100.104024}{\emph{Phys. Rev.}
  {\bfseries D100} (2019) 104024}
  [\href{https://arxiv.org/abs/1906.10071}{{\ttfamily 1906.10071}}].

\bibitem{Arkani-Hamed:2019ymq}
N.~Arkani-Hamed, Y.-t.~Huang and D.~O'Connell, \emph{{Kerr black holes as
  elementary particles}},
  \href{https://doi.org/10.1007/JHEP01(2020)046}{\emph{JHEP} {\bfseries 01}
  (2020) 046} [\href{https://arxiv.org/abs/1906.10100}{{\ttfamily
  1906.10100}}].

\bibitem{Guevara:2020xjx}
A.~Guevara, B.~Maybee, A.~Ochirov, D.~O'Connell and J.~Vines, \emph{{A
  worldsheet for Kerr}},
  \href{https://doi.org/10.1007/JHEP03(2021)201}{\emph{JHEP} {\bfseries 03}
  (2021) 201} [\href{https://arxiv.org/abs/2012.11570}{{\ttfamily
  2012.11570}}].

\bibitem{Aoude:2020mlg}
R.~Aoude, M.-Z.~Chung, Y.-t.~Huang, C.S.~Machado and M.-K.~Tam, \emph{{Silence
  of Binary Kerr Black Holes}},
  \href{https://doi.org/10.1103/PhysRevLett.125.181602}{\emph{Phys. Rev. Lett.}
  {\bfseries 125} (2020) 181602}
  [\href{https://arxiv.org/abs/2007.09486}{{\ttfamily 2007.09486}}].

\bibitem{Johansson:2019dnu}
H.~Johansson and A.~Ochirov, \emph{{Double copy for massive quantum particles
  with spin}}, \href{https://doi.org/10.1007/JHEP09(2019)040}{\emph{JHEP}
  {\bfseries 09} (2019) 040}
  [\href{https://arxiv.org/abs/1906.12292}{{\ttfamily 1906.12292}}].

\bibitem{Falkowski:2020aso}
A.~Falkowski and C.S.~Machado, \emph{{Soft Matters, or the Recursions with
  Massive Spinors}}, \href{https://doi.org/10.1007/JHEP05(2021)238}{\emph{JHEP}
  {\bfseries 05} (2021) 238}
  [\href{https://arxiv.org/abs/2005.08981}{{\ttfamily 2005.08981}}].

\bibitem{Chen:2022clh}
W.-M.~Chen, M.-Z.~Chung, Y.-t.~Huang and J.-W.~Kim, \emph{{Gravitational
  Faraday effect from on-shell amplitudes}},
  \href{https://doi.org/10.1007/JHEP12(2022)058}{\emph{JHEP} {\bfseries 12}
  (2022) 058} [\href{https://arxiv.org/abs/2205.07305}{{\ttfamily
  2205.07305}}].

\bibitem{Bautista:2021wfy}
Y.F.~Bautista, A.~Guevara, C.~Kavanagh and J.~Vines, \emph{{Scattering in black
  hole backgrounds and higher-spin amplitudes. Part I}},
  \href{https://doi.org/10.1007/JHEP03(2023)136}{\emph{JHEP} {\bfseries 03}
  (2023) 136} [\href{https://arxiv.org/abs/2107.10179}{{\ttfamily
  2107.10179}}].

\bibitem{Bern:2022kto}
Z.~Bern, D.~Kosmopoulos, A.~Luna, R.~Roiban and F.~Teng, \emph{{Binary Dynamics
  through the Fifth Power of Spin at O(G2)}},
  \href{https://doi.org/10.1103/PhysRevLett.130.201402}{\emph{Phys. Rev. Lett.}
  {\bfseries 130} (2023) 201402}
  [\href{https://arxiv.org/abs/2203.06202}{{\ttfamily 2203.06202}}].

\bibitem{Aoude:2022trd}
R.~Aoude, K.~Haddad and A.~Helset, \emph{{Searching for Kerr in the 2PM
  amplitude}}, \href{https://doi.org/10.1007/JHEP07(2022)072}{\emph{JHEP}
  {\bfseries 07} (2022) 072}
  [\href{https://arxiv.org/abs/2203.06197}{{\ttfamily 2203.06197}}].

\bibitem{Saketh:2022wap}
M.V.S.~Saketh and J.~Vines, \emph{{Scattering of gravitational waves off
  spinning compact objects with an effective worldline theory}},
  \href{https://doi.org/10.1103/PhysRevD.106.124026}{\emph{Phys. Rev. D}
  {\bfseries 106} (2022) 124026}
  [\href{https://arxiv.org/abs/2208.03170}{{\ttfamily 2208.03170}}].

\bibitem{Bautista:2022wjf}
Y.F.~Bautista, A.~Guevara, C.~Kavanagh and J.~Vines, \emph{{Scattering in black
  hole backgrounds and higher-spin amplitudes. Part II}},
  \href{https://doi.org/10.1007/JHEP05(2023)211}{\emph{JHEP} {\bfseries 05}
  (2023) 211} [\href{https://arxiv.org/abs/2212.07965}{{\ttfamily
  2212.07965}}].

\bibitem{Bjerrum-Bohr:2023jau}
N.E.J.~Bjerrum-Bohr, G.~Chen and M.~Skowronek, \emph{{Classical spin
  gravitational Compton scattering}},
  \href{https://doi.org/10.1007/JHEP06(2023)170}{\emph{JHEP} {\bfseries 06}
  (2023) 170} [\href{https://arxiv.org/abs/2302.00498}{{\ttfamily
  2302.00498}}].

\bibitem{Bautista:2023sdf}
Y.F.~Bautista, G.~Bonelli, C.~Iossa, A.~Tanzini and Z.~Zhou, \emph{{Black Hole
  Perturbation Theory Meets CFT$_2$: Kerr Compton Amplitudes from
  Nekrasov-Shatashvili Functions}},
  \href{https://arxiv.org/abs/2312.05965}{{\ttfamily 2312.05965}}.

\bibitem{Bjerrum-Bohr:2023iey}
N.E.J.~Bjerrum-Bohr, G.~Chen and M.~Skowronek, \emph{{Covariant Compton
  Amplitudes in Gravity with Classical Spin}},
  \href{https://arxiv.org/abs/2309.11249}{{\ttfamily 2309.11249}}.

\bibitem{Scheopner:2023rzp}
T.~Scheopner and J.~Vines, \emph{{Dynamical Implications of the Kerr Multipole
  Moments for Spinning Black Holes}},
  \href{https://arxiv.org/abs/2311.18421}{{\ttfamily 2311.18421}}.

\bibitem{Aoude:2023vdk}
R.~Aoude, K.~Haddad and A.~Helset, \emph{{Classical gravitational scattering
  amplitude at O(G2S1\ensuremath{\infty}S2\ensuremath{\infty})}},
  \href{https://doi.org/10.1103/PhysRevD.108.024050}{\emph{Phys. Rev. D}
  {\bfseries 108} (2023) 024050}
  [\href{https://arxiv.org/abs/2304.13740}{{\ttfamily 2304.13740}}].

\bibitem{Haddad:2023ylx}
K.~Haddad, \emph{{Recursion in the classical limit and the neutron-star Compton
  amplitude}}, \href{https://doi.org/10.1007/JHEP05(2023)177}{\emph{JHEP}
  {\bfseries 05} (2023) 177}
  [\href{https://arxiv.org/abs/2303.02624}{{\ttfamily 2303.02624}}].

\bibitem{Azevedo:2024rrf}
T.~Azevedo, D.E.A.~Matamoros and G.~Menezes, \emph{{Compton scattering from
  superstrings}},  \href{https://arxiv.org/abs/2403.08899}{{\ttfamily
  2403.08899}}.

\bibitem{Kawai:1985xq}
H.~Kawai, D.~Lewellen and S.~Tye, \emph{{A Relation Between Tree Amplitudes of
  Closed and Open Strings}},
  \href{https://doi.org/10.1016/0550-3213(86)90362-7}{\emph{Nucl.Phys.}
  {\bfseries B269} (1986) 1}.

\bibitem{Bautista:2019evw}
Y.F.~Bautista and A.~Guevara, \emph{{On the double copy for spinning matter}},
  \href{https://doi.org/10.1007/JHEP11(2021)184}{\emph{JHEP} {\bfseries 11}
  (2021) 184} [\href{https://arxiv.org/abs/1908.11349}{{\ttfamily
  1908.11349}}].

\bibitem{Johansson:2015oia}
H.~Johansson and A.~Ochirov, \emph{{Color-Kinematics Duality for QCD
  Amplitudes}}, \href{https://doi.org/10.1007/JHEP01(2016)170}{\emph{JHEP}
  {\bfseries 01} (2016) 170}
  [\href{https://arxiv.org/abs/1507.00332}{{\ttfamily 1507.00332}}].

\bibitem{Chiodaroli:2021eug}
M.~Chiodaroli, H.~Johansson and P.~Pichini, \emph{{Compton black-hole
  scattering for s \ensuremath{\leq} 5/2}},
  \href{https://doi.org/10.1007/JHEP02(2022)156}{\emph{JHEP} {\bfseries 02}
  (2022) 156} [\href{https://arxiv.org/abs/2107.14779}{{\ttfamily
  2107.14779}}].

\bibitem{Guevara:2021yud}
A.~Guevara, \emph{{Reconstructing Classical Spacetimes from the S-Matrix in
  Twistor Space}},  \href{https://arxiv.org/abs/2112.05111}{{\ttfamily
  2112.05111}}.

\bibitem{Zinoviev:2001dt}
Y.M.~Zinoviev, \emph{{On massive high spin particles in AdS}},
  \href{https://arxiv.org/abs/hep-th/0108192}{{\ttfamily hep-th/0108192}}.

\bibitem{Cangemi:2022bew}
L.~Cangemi, M.~Chiodaroli, H.~Johansson, A.~Ochirov, P.~Pichini and
  E.~Skvortsov, \emph{{Kerr Black Holes From Massive Higher-Spin Gauge
  Symmetry}}, \href{https://doi.org/10.1103/PhysRevLett.131.221401}{\emph{Phys.
  Rev. Lett.} {\bfseries 131} (2023) 221401}
  [\href{https://arxiv.org/abs/2212.06120}{{\ttfamily 2212.06120}}].

\bibitem{Ochirov:2022nqz}
A.~Ochirov and E.~Skvortsov, \emph{{Chiral Approach to Massive Higher Spins}},
  \href{https://doi.org/10.1103/PhysRevLett.129.241601}{\emph{Phys. Rev. Lett.}
  {\bfseries 129} (2022) 241601}
  [\href{https://arxiv.org/abs/2207.14597}{{\ttfamily 2207.14597}}].

\bibitem{Cangemi:2023ysz}
L.~Cangemi, M.~Chiodaroli, H.~Johansson, A.~Ochirov, P.~Pichini and
  E.~Skvortsov, \emph{{From higher-spin gauge interactions to Compton
  amplitudes for root-Kerr}},
  \href{https://arxiv.org/abs/2311.14668}{{\ttfamily 2311.14668}}.

\bibitem{Cangemi:2023bpe}
L.~Cangemi, M.~Chiodaroli, H.~Johansson, A.~Ochirov, P.~Pichini and
  E.~Skvortsov, \emph{{Compton Amplitude for Rotating Black Hole from QFT}},
  \href{https://doi.org/10.1103/PhysRevLett.133.071601}{\emph{Phys. Rev. Lett.}
  {\bfseries 133} (2024) 071601}
  [\href{https://arxiv.org/abs/2312.14913}{{\ttfamily 2312.14913}}].

\bibitem{Aoude:2021oqj}
R.~Aoude and A.~Ochirov, \emph{{Classical observables from coherent-spin
  amplitudes}}, \href{https://doi.org/10.1007/JHEP10(2021)008}{\emph{JHEP}
  {\bfseries 10} (2021) 008}
  [\href{https://arxiv.org/abs/2108.01649}{{\ttfamily 2108.01649}}].

\bibitem{Correia:2024jgr}
M.~Correia and G.~Isabella, \emph{{The Born regime of gravitational
  amplitudes}},  \href{https://arxiv.org/abs/2406.13737}{{\ttfamily
  2406.13737}}.

\bibitem{Chen:2024mmm}
G.~Chen and T.~Wang, \emph{{Dynamics of Spinning Binary at 2PM}},
  \href{https://arxiv.org/abs/2406.09086}{{\ttfamily 2406.09086}}.

\bibitem{Jakobsen:2023ndj}
G.U.~Jakobsen, G.~Mogull, J.~Plefka, B.~Sauer and Y.~Xu, \emph{{Conservative
  Scattering of Spinning Black Holes at Fourth Post-Minkowskian Order}},
  \href{https://doi.org/10.1103/PhysRevLett.131.151401}{\emph{Phys. Rev. Lett.}
  {\bfseries 131} (2023) 151401}
  [\href{https://arxiv.org/abs/2306.01714}{{\ttfamily 2306.01714}}].

\bibitem{Bini:2018ywr}
D.~Bini and T.~Damour, \emph{{Gravitational spin-orbit coupling in binary
  systems at the second post-Minkowskian approximation}},
  \href{https://doi.org/10.1103/PhysRevD.98.044036}{\emph{Phys. Rev.}
  {\bfseries D98} (2018) 044036}
  [\href{https://arxiv.org/abs/1805.10809}{{\ttfamily 1805.10809}}].

\bibitem{Bern:2020buy}
Z.~Bern, A.~Luna, R.~Roiban, C.-H.~Shen and M.~Zeng, \emph{{Spinning black hole
  binary dynamics, scattering amplitudes, and effective field theory}},
  \href{https://doi.org/10.1103/PhysRevD.104.065014}{\emph{Phys. Rev. D}
  {\bfseries 104} (2021) 065014}
  [\href{https://arxiv.org/abs/2005.03071}{{\ttfamily 2005.03071}}].

\bibitem{Chung:2020rrz}
M.-Z.~Chung, Y.-t.~Huang, J.-W.~Kim and S.~Lee, \emph{{Complete Hamiltonian for
  spinning binary systems at first post-Minkowskian order}},
  \href{https://doi.org/10.1007/JHEP05(2020)105}{\emph{JHEP} {\bfseries 05}
  (2020) 105} [\href{https://arxiv.org/abs/2003.06600}{{\ttfamily
  2003.06600}}].

\bibitem{Kosmopoulos:2021zoq}
D.~Kosmopoulos and A.~Luna, \emph{{Quadratic-in-spin Hamiltonian at $
  \mathcal{O} $(G$^{2}$) from scattering amplitudes}},
  \href{https://doi.org/10.1007/JHEP07(2021)037}{\emph{JHEP} {\bfseries 07}
  (2021) 037} [\href{https://arxiv.org/abs/2102.10137}{{\ttfamily
  2102.10137}}].

\bibitem{Buonanno:2024vkx}
A.~Buonanno, G.U.~Jakobsen and G.~Mogull, \emph{{Post-Minkowskian theory meets
  the spinning effective-one-body approach for two-body scattering}},
  \href{https://doi.org/10.1103/PhysRevD.110.044038}{\emph{Phys. Rev. D}
  {\bfseries 110} (2024) 044038}
  [\href{https://arxiv.org/abs/2402.12342}{{\ttfamily 2402.12342}}].

\bibitem{Gonzo:2024zxo}
R.~Gonzo and C.~Shi, \emph{{Scattering and bound observables for spinning
  particles in Kerr spacetime with generic spin orientations}},
  \href{https://arxiv.org/abs/2405.09687}{{\ttfamily 2405.09687}}.

\bibitem{Bern:2023ity}
Z.~Bern, D.~Kosmopoulos, A.~Luna, R.~Roiban, T.~Scheopner, F.~Teng et~al.,
  \emph{{Quantum Field Theory, Worldline Theory, and Spin Magnitude Change in
  Orbital Evolution}},  \href{https://arxiv.org/abs/2308.14176}{{\ttfamily
  2308.14176}}.

\bibitem{Kim:2023drc}
J.-W.~Kim and J.~Steinhoff, \emph{{Spin supplementary condition in quantum
  field theory: covariant SSC and physical state projection}},
  \href{https://doi.org/10.1007/JHEP07(2023)042}{\emph{JHEP} {\bfseries 07}
  (2023) 042} [\href{https://arxiv.org/abs/2302.01944}{{\ttfamily
  2302.01944}}].

\bibitem{Alaverdian:2024spu}
M.~Alaverdian, Z.~Bern, D.~Kosmopoulos, A.~Luna, R.~Roiban, T.~Scheopner
  et~al., \emph{{Conservative Spin Magnitude Change in Orbital Evolution in
  General Relativity}},  \href{https://arxiv.org/abs/2407.10928}{{\ttfamily
  2407.10928}}.

\bibitem{Chen:2023qzo}
Y.-J.~Chen, T.~Hsieh, Y.-T.~Huang and J.-W.~Kim, \emph{{On-shell approach to
  (spinning) gravitational absorption processes}},
  \href{https://arxiv.org/abs/2312.04513}{{\ttfamily 2312.04513}}.

\bibitem{Aoude:2023fdm}
R.~Aoude and A.~Ochirov, \emph{{Gravitational partial-wave absorption from
  scattering amplitudes}},
  \href{https://doi.org/10.1007/JHEP12(2023)103}{\emph{JHEP} {\bfseries 12}
  (2023) 103} [\href{https://arxiv.org/abs/2307.07504}{{\ttfamily
  2307.07504}}].

\bibitem{Bautista:2021inx}
Y.F.~Bautista and N.~Siemonsen, \emph{{Post-Newtonian waveforms from spinning
  scattering amplitudes}},
  \href{https://doi.org/10.1007/JHEP01(2022)006}{\emph{JHEP} {\bfseries 01}
  (2022) 006} [\href{https://arxiv.org/abs/2110.12537}{{\ttfamily
  2110.12537}}].

\bibitem{Riva:2022fru}
M.M.~Riva, F.~Vernizzi and L.K.~Wong, \emph{{Gravitational bremsstrahlung from
  spinning binaries in the post-Minkowskian expansion}},
  \href{https://doi.org/10.1103/PhysRevD.106.044013}{\emph{Phys. Rev. D}
  {\bfseries 106} (2022) 044013}
  [\href{https://arxiv.org/abs/2205.15295}{{\ttfamily 2205.15295}}].

\bibitem{Heissenberg:2023uvo}
C.~Heissenberg, \emph{{Angular momentum loss due to spin-orbit effects in the
  post-Minkowskian expansion}},
  \href{https://doi.org/10.1103/PhysRevD.108.106003}{\emph{Phys. Rev. D}
  {\bfseries 108} (2023) 106003}
  [\href{https://arxiv.org/abs/2308.11470}{{\ttfamily 2308.11470}}].

\bibitem{DeAngelis:2023lvf}
S.~De~Angelis, R.~Gonzo and P.P.~Novichkov, \emph{{Spinning waveforms from KMOC
  at leading order}},  \href{https://arxiv.org/abs/2309.17429}{{\ttfamily
  2309.17429}}.

\bibitem{Brandhuber:2023hhl}
A.~Brandhuber, G.R.~Brown, G.~Chen, J.~Gowdy and G.~Travaglini, \emph{{Resummed
  spinning waveforms from five-point amplitudes}},
  \href{https://arxiv.org/abs/2310.04405}{{\ttfamily 2310.04405}}.

\bibitem{Aoude:2023dui}
R.~Aoude, K.~Haddad, C.~Heissenberg and A.~Helset, \emph{{Leading-order
  gravitational radiation to all spin orders}},
  \href{https://arxiv.org/abs/2310.05832}{{\ttfamily 2310.05832}}.

\bibitem{Bohnenblust:2023qmy}
L.~Bohnenblust, H.~Ita, M.~Kraus and J.~Schlenk, \emph{{Gravitational
  Bremsstrahlung in Black-Hole Scattering at $\mathcal{O}(G^3)$: Linear-in-Spin
  Effects}},  \href{https://arxiv.org/abs/2312.14859}{{\ttfamily 2312.14859}}.

\bibitem{Chen:2024bpf}
G.~Chen, J.-W.~Kim and T.~Wang, \emph{{Systematic integral evaluation for
  spin-resummed binary dynamics}},
  \href{https://arxiv.org/abs/2406.17658}{{\ttfamily 2406.17658}}.

\bibitem{Kosower:2018adc}
D.A.~Kosower, B.~Maybee and D.~O'Connell, \emph{{Amplitudes, Observables, and
  Classical Scattering}},
  \href{https://doi.org/10.1007/JHEP02(2019)137}{\emph{JHEP} {\bfseries 02}
  (2019) 137} [\href{https://arxiv.org/abs/1811.10950}{{\ttfamily
  1811.10950}}].

\bibitem{private}
Y.F.~Bautista, A.~Guevara, C.~Kavanagh and J.~Vines. private communication,
  2023.

\bibitem{Pryce:1935ibt}
M.H.L.~Pryce, \emph{{Commuting co-ordinates in the new field theory}},
  \href{https://doi.org/10.1098/rspa.1935.0094}{\emph{Proc. Roy. Soc. Lond. A}
  {\bfseries 150} (1935) 166}.

\bibitem{Pryce:1948pf}
M.H.L.~Pryce, \emph{{The Mass center in the restricted theory of relativity and
  its connection with the quantum theory of elementary particles}},
  \href{https://doi.org/10.1098/rspa.1948.0103}{\emph{Proc. Roy. Soc. Lond. A}
  {\bfseries 195} (1948) 62}.

\bibitem{Newton:1949cq}
T.D.~Newton and E.P.~Wigner, \emph{{Localized States for Elementary Systems}},
  \href{https://doi.org/10.1103/RevModPhys.21.400}{\emph{Rev. Mod. Phys.}
  {\bfseries 21} (1949) 400}.

\bibitem{Vines:2016unv}
J.~Vines, D.~Kunst, J.~Steinhoff and T.~Hinderer, \emph{{Canonical Hamiltonian
  for an extended test body in curved spacetime: To quadratic order in spin}},
  \href{https://doi.org/10.1103/PhysRevD.104.029902}{\emph{Phys. Rev. D}
  {\bfseries 93} (2016) 103008}
  [\href{https://arxiv.org/abs/1601.07529}{{\ttfamily 1601.07529}}].

\end{thebibliography}\endgroup

\end{document}